%% file: paper.tex
\documentclass[twocolumn,hyperpdf,amsmath,amssymb,aps,prd,10pt,superscriptaddress,nofootinbib,noeprint,preprintnumbers,floatfix]{revtex4-1}

\input{header}

\begin{document}

\preprint{arXiv:2009.10034}
\preprint{JLAB-THY-20-3249}

\title{Decays of an exotic $1^{-+}$ hybrid meson resonance in QCD}
\author{Antoni~J.~Woss}
\email{a.j.woss@damtp.cam.ac.uk}
\affiliation{DAMTP, University of Cambridge, Centre for Mathematical Sciences, Wilberforce Road, Cambridge, CB3 0WA, UK}
\author{Jozef~J.~Dudek}
\email{dudek@jlab.org}
\affiliation{\lsstyle Thomas Jefferson National Accelerator Facility, 12000 Jefferson Avenue, Newport News, VA 23606, USA}
\affiliation{Department of Physics, College of William and Mary, Williamsburg, VA 23187, USA}
\author{Robert~G.~Edwards}
\email{edwards@jlab.org}
\affiliation{\lsstyle Thomas Jefferson National Accelerator Facility, 12000 Jefferson Avenue, Newport News, VA 23606, USA}
\author{Christopher~E.~Thomas}
\email{c.e.thomas@damtp.cam.ac.uk}
\affiliation{DAMTP, University of Cambridge, Centre for Mathematical Sciences, Wilberforce Road, Cambridge, CB3 0WA, UK}
\author{David~J.~Wilson}
\email{d.j.wilson@damtp.cam.ac.uk}
\affiliation{DAMTP, University of Cambridge, Centre for Mathematical Sciences, Wilberforce Road, Cambridge, CB3 0WA, UK}

\collaboration{for the Hadron Spectrum Collaboration}
\date{22 December 2020}

\begin{abstract}
We present the first determination of the hadronic decays of the lightest exotic $J^{PC}=1^{-+}$ resonance in lattice QCD. Working with SU(3) flavor symmetry, where the up, down and strange quark masses approximately match the physical strange-quark mass giving $m_\pi \sim 700$~MeV, we compute finite-volume spectra on six lattice volumes which constrain a scattering system featuring eight coupled channels. Analytically continuing the scattering amplitudes into the complex energy plane, we find a pole singularity corresponding to a narrow resonance which shows relatively weak coupling to the open pseudoscalar--pseudoscalar, vector--pseudoscalar and vector--vector decay channels, but large couplings to at least one kinematically-closed axial-vector--pseudoscalar channel. Attempting a simple extrapolation of the couplings to physical light-quark mass suggests a broad $\pi_1$ resonance decaying dominantly through the $b_1 \pi$ mode with much smaller decays into $f_1 \pi$, $\rho \pi$, $\eta' \pi$ and $\eta \pi$. A large total width is potentially in agreement with the experimental $\pi_1(1564)$ candidate state, observed in $\eta \pi$, $\eta' \pi$, which we suggest may be heavily suppressed decay channels.
\end{abstract}

\maketitle

%%%%%%%%%%%%%%%%%%%%%%%%%%%%%%%%%%%%%%%%%%%%%%%%%%%%%%%%%%%%%%%%%%%%%%
\section{Introduction}\label{Intro}
\input{sections/intro}

\section{Resonances in lattice QCD}
\label{Method}
\input{sections/method}

\newpage

\section{Mesons with exact $\mathbf{SU(3)}$ flavor symmetry}
\label{SU3F}
\input{sections/SU3F}

\section{Lattice QCD spectra}
\label{lattice}
\input{sections/lattice}

\pagebreak
\section{Scattering Amplitudes}
\label{scattering}
\input{sections/scattering}

\section{Resonance Pole Singularities}
\label{poles_section}
\input{sections/poles}

\section{Interpretation}
\label{interpret}
\input{sections/interpret}

\section{Summary}
\label{summary}
\input{sections/summary}

%%%%%%%%%%%%%%%%%%%%%%%%%%%%%%%%%%%%%%%%%%%%%%%%%%%%%%%%%%%%%%%%%%%%%%%%%%%%%%%%%
\input{acknow}

%%%%%%%%%%%%%%%%%%%%%%%%%%%%%%%%%%%%%%%%%%%%%%%%%%%%%%%%%%%%%%%%%%%%%%%%%%%%%%%%%
\pagebreak
\clearpage
\appendix

\section{$\mathbf{SU(3)}$ Clebsch-Gordan Coefficients}
\label{su3sym}
\input{sections/SU3F_syms}

\section{$\mathbf{SU(3)}$ Bose symmetry}
\label{bosesym}
\input{sections/Bose_syms}
\section{Indistinguishable vector-vector $P$-waves in $T^-_1$}
\label{vvpwaves}
\input{sections/VV_pwaves}

\section{``Trapped" levels for factorized K-matrix poles}
\label{squeezed_levels}
\input{sections/squeezed_levels}

\section{Sensitivity to $\fOneOctet\etaOctet\{\threeSone\}$, $\hOneOctet\etaOctet\{\threeSone\}$ couplings}
\label{toy_model}
\input{sections/toy_model}

%%%%%%%%%%%%%%%%%%%%%%%%%%%%%%%%%%%%%%%%%%%%%%%%%%%%%%%%%%%%%%%%%%%%%%%%%%%%%%%%%
\bibliographystyle{apsrev4-1}
\bibliography{bib}

%%%%%%%%%%%%%%%%%%%%%%%%%%%%%%%%%%%%%%%%%%%%%%%%%%%%%%%%%%%%%%%%%%%%%%%%%%%%%%%%%

\end{document}

%% file: header.tex
\usepackage{graphicx, color}
\graphicspath{{./figures/}}

\usepackage[letterspace=-10]{microtype} 

\usepackage{bm, amsmath, amsfonts, amssymb,xfrac}
\usepackage{multirow, tabularx, dcolumn}
\usepackage{mathtools,leftidx, braket, slashed, cancel}
\usepackage{blkarray}
\usepackage[figures]{rotating}

\usepackage[utf8]{inputenc} 
\usepackage{hyperref}

\definecolor{jlab_red}{RGB}{192,39,45}
\definecolor{jlab_orange}{RGB}{249,102,0}
\definecolor{jlab_blue}{RGB}{47,122,121}
\definecolor{jlab_green}{RGB}{65,125,10}
\definecolor{jlab_gray}{gray}{0.6}

\newcommand{\cm}{\ensuremath{\mathsf{cm}}}
\newcommand{\SU}{\ensuremath{\mathrm{SU}}}
\newcommand{\SUF}{\ensuremath{\mathrm{SU}(3)_\mathrm{F}}}

\newcommand{\oneMP}{\ensuremath{1^{-(\!+\!)}}}

\newcommand{\etaOctet}{     \eta^{\scriptscriptstyle\mathbf{8}}}
\newcommand{\etaSinglet}{   \eta^{\scriptscriptstyle\mathbf{1}}}
\newcommand{\omegaOctet}{   \omega^{\scriptscriptstyle\mathbf{8}}}
\newcommand{\omegaSinglet}{ \omega^{\scriptscriptstyle\mathbf{1}}}
\newcommand{\fOneOctet}{    f_1^{\scriptscriptstyle\mathbf{8}}}
\newcommand{\fOneSinglet}{  f_1^{\scriptscriptstyle\mathbf{1}}}
\newcommand{\fZeroOctet}{    f_0^{\scriptscriptstyle\mathbf{8}}}
\newcommand{\fZeroSinglet}{  f_0^{\scriptscriptstyle\mathbf{1}}}
\newcommand{\fTwoOctet}{    f_2^{\scriptscriptstyle\mathbf{8}}}
\newcommand{\fTwoSinglet}{  f_2^{\scriptscriptstyle\mathbf{1}}}
\newcommand{\hOneOctet}{    h_1^{\scriptscriptstyle\mathbf{8}}}
\newcommand{\hOneSinglet}{  h_1^{\scriptscriptstyle\mathbf{1}}}
\newcommand{\etaOneOctet}{   \eta_1^{\scriptscriptstyle\mathbf{8}}}
\newcommand{\etaOneSinglet}{ \eta_1^{\scriptscriptstyle\mathbf{1}}}

\newcommand{\threeSone}{\prescript{3\!}{}{S}_1} 

\newcommand{\threePone}{\prescript{3\!}{}{P}_1}
  
\newcommand{\threeDone}{\prescript{3\!}{}{D}_1}

\newcommand{\threeFthree}{\prescript{3\!}{}{F}_3}

\newcommand{\onePone}{\prescript{1\!}{}{P}_1} 
\newcommand{\oneSzero}{\prescript{1\!}{}{S}_0} 
\newcommand{\fivePone}{\prescript{5\!}{}{P}_1} 
\newcommand{\fivePthree}{\prescript{5\!}{}{P}_3} 
\newcommand{\XPone}{\prescript{X\!}{}{P}_1}

\newcommand{\oneGfour}{\prescript{1\!}{}{G}_4}
\newcommand{\oneFthree}{\prescript{1\!}{}{F}_3}

\newcommand{\fiveFone}{\prescript{5\!}{}{F}_1}  
\newcommand{\fiveFthree}{\prescript{5\!}{}{F}_3}

\newcommand{\kbar}{\,\overline{\!K}}
\newcommand{\Kbar}{\,\overline{\!K}}

\newcommand{\KbarSuper}[1]{\,\overline{\!K} \vphantom{K}^{#1} }

\hypersetup{%
pdftitle = {Decays of an exotic 1-+ hybrid meson resonance in QCD},
pdfsubject = {},
pdfkeywords = {},
pdfauthor = {Hadron Spectrum Collaboration},
colorlinks = {true},
filecolor = {black},
linkcolor = {jlab_blue},
menucolor = {black},
citecolor = {jlab_green},
urlcolor = {jlab_green},
}{}

%% file: sections/intro.tex
The composition of hadrons has been the subject of experimental and theoretical studies for many decades.
Historically, the majority of mesons could be understood in a quark-model picture where they consist of a quark-antiquark pair ($q\bar{q}$).
There are some notable long-standing exceptions that do not appear to fit into this framework, such as the light scalar mesons, and more recently it has been challenged by the observation of a number of unexpected structures in the charm and bottom sectors.

In principle
mesons can contain constituent combinations beyond $q\bar{q}$, but whether QCD allows for such arrangements continues to motivate investigations in both theory and experiment.
One particular focus is on \emph{hybrid mesons} in which a quark-antiquark pair is coupled to an excitation of the gluonic field. Such states are an attractive target because the additional quantum numbers potentially supplied by the gluonic field allow for $J^{PC}$ combinations not allowed to a $q\bar{q}$ system. These \emph{exotic} $J^{PC} = 0^{--}, 0^{+-}, 1^{-+}, 2^{+-} \ldots$ serve as a smoking-gun signature that a novel state has been observed.

Suggestions that hybrid mesons are a feature of QCD are longstanding, but until recently predictions of their properties came only within models whose connection to QCD is not always clear~\cite{Horn:1977rq, Barnes:1982tx, Chanowitz:1982qj, Isgur:1985vy, Jaffe:1985qp, General:2006ed, Guo:2008yz}. While dynamical pictures like the flux-tube model, the bag model, and constituent gluon approaches generally agree that hybrids form part of the meson spectrum, some with exotic $J^{PC}$, they differ in details. A common feature is that typically a $J^{PC} = 1^{-+}$ state (labelled $\pi_1$ when the state has isospin--1) appears with a mass somewhere above 1.5 GeV. A particular challenge has been for these models to provide reliable predictions for the decay properties of hybrid mesons, which we expect to appear as resonances that can decay into several final states. Having some advance knowledge of which final states are more heavily populated in their decay is useful to experiments which perform amplitude analyses final-state by final-state. A folklore has developed, largely following from models in which the hybrid decay proceeds by the breaking of an oscillating tube of gluonic flux or through conversion of a constituent gluon to a $q\bar{q}$ pair~\cite{Tanimoto:1982eh, Iddir:1988jc, Ishida:1991mx, Close:1994hc, Swanson:1997wy, Page:1998gz}, where decays featuring only the lightest hadrons are suppressed, such as $\pi_1 \to \eta \pi, \eta' \pi, \rho \pi$, while decays which include a more excited hadron are prominent, such as $\pi_1 \to  b_1 \pi$. Whether these results are really a feature of QCD, or reflect the assumptions built into the flux-tube (a picture whose validity looks increasingly unlikely~\cite{Dudek:2011bn}) or constituent gluon pictures, has yet to be established. 

\vspace{3mm}
The experimental focus has remained largely on the $\pi_1$, and historically the picture has been quite confused~\cite{Meyer:2010ku,Meyer:2015eta}. Analyses have mostly considered the $\eta \pi$, $\eta' \pi$ and $\rho \pi \to \pi \pi \pi$ final states which have the lowest possible multiplicities. Recent data sets of unprecedented statistics from COMPASS provide our clearest picture~\cite{Adolph:2014rpp}: a broad bump in $\eta \pi$ peaking near 1400 MeV appears to match poorly with another bump in $\eta' \pi$ peaking near 1600 MeV. These results are similar to those observed in earlier experiments which were interpreted as \emph{two} resonances, $\pi_1(1400)$ and $\pi_1(1600)$, with there being some further evidence for the heavier resonance in the $\rho \pi$ final state.

A recent analysis of the COMPASS data by JPAC comes to a different conclusion~\cite{Rodas:2018owy}: the two bumps in $\eta \pi, \eta' \pi$ are actually due to a single resonance decaying into both final states. They proceed by parameterizating the production process and the scattering of the coupled-channel $\eta \pi, \eta' \pi$ system, respecting unitarity in these two channels. The scattering $t$-matrix is constrained for real values of the energy using experimental data. When the amplitude is considered for complex values of the energy, a single pole singularity is found which can be interpreted as \emph{one} resonance with a mass slightly below 1.6 GeV and a width of around 500 MeV. A combined analysis of COMPASS and Crystal Barrel data~\cite{Kopf:2020yoa} which appeared while this paper was in the final stages of preparation finds a very similar mass, but a slightly smaller width $\sim 388$ MeV.

Currently the GlueX experiment~\cite{AlekSejevs:2013mkl, Dobbs:2019sgr} is collecting large data sets using photoproduction in which they will search for hybrid mesons. Since the higher multiplicity final states suggested as preferred by the flux-tube picture, e.g.\ $\pi_1 \to b_1 \pi \to (\omega \pi) \pi \to \pi \pi \pi \pi \pi$, are much harder to analyze than those investigated in COMPASS, it would be of benefit to have some evidence within QCD that these channels are in fact dominant in the decays of hybrid mesons. It is to this task that we turn our attention in this paper, using the technique known as Lattice QCD.

\vspace{3mm}

Lattice QCD, which offers a first-principles numerical approach to QCD, has matured to the point where it has been able to make some fairly definitive statements about the excited spectrum of hadrons. In Refs.~\cite{Dudek:2009qf, Dudek:2010wm, Dudek:2011tt, Thomas:2011rh, Dudek:2013yja}, bases of composite operators built from fermion bilinears and up to three gauge-covariant derivatives were used to construct matrices of two-point correlation functions. Analyzing the time dependences of these matrices led to predictions for the spectrum of mesons with a wide range of $J^{PC}$. The spectra obtained, for several values of the light quark mass, show a strong qualitative similarity with the experimental meson spectrum, but also feature clear indications of exotic $J^{PC}$ states with notably a lightest $\pi_1$. A phenomenology was developed~\cite{Dudek:2011bn} based upon the observation that this state, along with states having $J^{PC} = 0^{-+}, 2^{-+}$ and $1^{--}$ at similar masses, have large matrix elements to be produced by operators of the form $\overline{\psi} \Gamma t^a \psi \, B^a$, which has the $q\bar{q}$ pair in a color-octet with the color neutralized by the chromomagnetic field operator, $B^a$. It was proposed that this large overlap signals that these states are hybrid mesons, and
they systematically appear roughly $1.3 - 1.4$ GeV above the lightest vector meson, even for quark masses corresponding to charmonium~\cite{Liu:2012ze, Cheung:2016bym}. The picture extends into the baryon spectrum~\cite{Dudek:2012ag}, where hybrid baryons can be identified, although in this case exotic quantum numbers are not possible.

While these calculations have provided us with the first picture of hybrid hadrons directly connected to QCD, the picture is clearly incomplete. These excited hadrons are not stable particles having a definite mass, rather they are unstable resonances which should appear as enhancements in the scattering of lighter stable hadrons, but this was neglected in the calculations. The resonant nature of these states has consequences for the spectrum calculated in lattice QCD, where the important difference with respect to experiment is the use of a finite spatial volume. 

The discrete spectrum of eigenstates in a finite periodic spatial volume can be related to infinite-volume scattering amplitudes using an approach that is commonly referred to as the L\"uscher method~\cite{Luscher:1985dn,Luscher:1986pf,Luscher:1990ux,Luscher:1991cf}, a formalism that has been extended to systems moving with respect to the lattice, hadrons with non-zero spin and any number of coupled hadron-hadron channels~\cite{Rummukainen:1995vs,He:2005ey,Kim:2005gf,Christ:2005gi,Fu:2011xz,Guo:2012hv,Hansen:2012tf,Briceno:2012yi,Gockeler:2012yj,Leskovec:2012gb,Briceno:2014oea}. Obtaining the complete spectrum of eigenstates requires a larger basis of operators than that used in the calculations referred to above~\cite{Dudek:2012xn,Wilson:2015dqa}, and it has been demonstrated that operators constructed as products of meson operators are sufficient. The coupled-channel $t$-matrix can then be obtained through the use of parameterizations which are constrained at the discrete real values of energy provided by the finite-volume spectra. The $t$-matrix is then continued into the complex energy plane and any pole singularities identified. From these the mass and width of a resonance can be determined, along with its couplings to different decay channels, in what can be argued to be the most rigorous way possible. In the past few years this approach has been used extensively in the study of elastic scattering, in cases like isospin--1 $\pi \pi$ where the $\rho$ resonance appears~\cite{Aoki:2007rd,Feng:2010es,Aoki:2011yj,Lang:2011mn,Dudek:2012xn,Pelissier:2012pi,Bali:2015gji,Wilson:2015dqa,Bulava:2016mks,Guo:2016zos,Hu:2016shf,Fu:2016itp,Alexandrou:2017mpi,Andersen:2018mau,Werner:2019hxc,Erben:2019nmx,Fischer:2020fvl}, and in several pioneering calculations of coupled-channel scattering~\cite{Dudek:2014qha,Wilson:2014cna,Wilson:2015dqa,Dudek:2016cru,Moir:2016srx,Briceno:2017qmb,Woss:2018irj,Woss:2019hse}.

\vspace{3mm}

In this paper we will report on the first calculation of an exotic $J^{PC} = 1^{-+}$ meson appearing as a resonance in coupled-channel meson-meson scattering.
By working with an exact $\SU(3)$ flavor symmetry where the $u,d$ quark masses are raised to the physical strange quark mass, we will reduce the effective number of decay channels and make three-body decays irrelevant.

The paper is structured as follows: In Section~\ref{Method} we review the techniques needed to compute finite-volume spectra in lattice QCD and to relate these to scattering amplitudes. Section~\ref{SU3F} discusses generalities of working with an exact $\SUF$ symmetry. In Section~\ref{lattice} we present calculational details and finite-volume spectra relevant for a $1^{-+}$ resonance on six lattice volumes. In Section~\ref{scattering} these spectra are used to constrain a scattering matrix of eight channels using a range of parameterizations, and in Section~\ref{poles_section} these parameterizations are analytically continued into the complex energy plane where a resonance pole singularity is found.
Section~\ref{interpret} interprets the decay couplings obtained from the residue of the resonance pole, comparing to existing models of hybrid meson decay, and attempts an extrapolation to physical kinematics. Finally, we summarize in Section~\ref{summary}. Some additional technical points are discussed in appendices, and details of the various parameterizations used can be found in Supplemental Material.

%% file: sections/method.tex
\noindent

Our approach to determining resonant physics in lattice QCD requires the computation of discrete spectra in the finite-volume defined by the lattice, and analysis of these spectra in terms of a scattering matrix using the L\"uscher method. In this section we will review our approach for doing this -- if further details are required, the field is reviewed in Ref.~\cite{Briceno:2017max}.

\subsection{Finite-volume spectra}

In order to constrain the scattering $t$-matrix over a range of energies, we are required to calculate a large number of discrete finite-volume levels sampling the energy region. An approach which has proven to be highly effective for the reliable extraction of many excited states is through the diagonalization of a large matrix of correlation functions, $C_{ij}(t) = \langle 0 | \mathcal{O}^{}_i(t) \mathcal{O}^\dag_j(0) | 0 \rangle$. This can be achieved by solving a \emph{generalized eigenvalue problem}~\cite{Michael:1985ne,Luscher:1990ck,Blossier:2009kd}, with our implementation described in Refs.~\cite{Dudek:2007wv,Dudek:2010wm}. This approach makes use of
orthogonality between energy eigenstates to distinguish contributions of even near-degenerate states, supplying their energies through the time-dependence of the eigenvalues while the eigenvectors provide linear combinations of the basis operators which serve as the optimal operator, in the variational sense, for each state.

One possible basis of operators, $\big\{ \mathcal{O}_i \big\}$, that can be used to form a matrix of meson correlation functions is built from fermion bilinears featuring gauge-covariant derivatives. A large basis can be constructed both with zero momentum~\cite{Dudek:2010wm} and non-zero momentum~\cite{Thomas:2011rh}. For the determination of stable hadrons, such a basis is typically sufficient and leads to reliable determinations of the mass (or energy with non-zero momentum) and \emph{optimized operators} which relax to the desired state more rapidly than any single operator in the basis (see for example Figure 2 of Ref.~\cite{Dudek:2012gj} or Figure 3 of Ref.~\cite{Shultz:2015pfa}).

The reduced rotational symmetry of a cubic lattice means that meson states are characterized not by integer spin values and parity, but by the irreducible representations (\emph{irreps}) of the octahedral group or the appropriate little group\footnote{the set of allowed octahedral group rotations and reflections which leave the momentum vector unchanged} for non-zero momentum, with the allowed momenta in an $L \!\times\! L \!\times\! L$ periodic volume given by $\vec{p} = \frac{2\pi}{L} \big( n_x, n_y, n_z \big)$ where $n_i$ are integers. In general, this means that examination of a particular irrep requires considering multiple $J^P$ values, but the group theory describing how spin \emph{subduces} into irreps~\cite{Johnson:1982yq,Moore:2005dw} and the construction of operators in appropriate irreps~\cite{Dudek:2010wm,Thomas:2011rh} are well understood.

When considering energies near and above meson-meson decay thresholds, a basis of only fermion bilinears is insufficient to capture the complete finite-volume spectrum, while augmenting this \emph{single-meson-like} basis with a set of \emph{meson-meson-like} constructions has proven to be highly effective~\cite{Dudek:2012xn,Wilson:2015dqa}. Such operators are built by combining optimized stable meson operators using appropriately weighted products.
For an $\mathbb{M}_1 \mathbb{M}_2$-like operator with overall momentum $\vec{P}$ in irrep $\Lambda$,
\begin{align*}
\mathcal{O}^{\Lambda \mu \dag}_{\mathbb{M}_1 \mathbb{M}_2}(\vec{P}) = 
& \sum_{\mu_1, \mu_2} \sum_{\hat{p}_1, \hat{p}_2}
\mathcal{C}( [\vec{P}] \Lambda , \mu ; [\vec{p}_1]\Lambda_1 , \mu_1 ; [\vec{p}_2]\Lambda_2 , \mu_2 ) \\
& \qquad \times 
\Omega^{\Lambda_1 \mu_1 \dag}_{\mathbb{M}_1}(\vec{p}_1) \;
\Omega^{\Lambda_2 \mu_2 \dag}_{\mathbb{M}_2}(\vec{p}_2) \, .
\end{align*}
Here the optimized stable meson operator, $\Omega^{\Lambda_i \mu_i \dag}_{\mathbb{M}_i}(\vec{p}_i)$, for meson $\mathbb{M}_i$ with momentum $\vec{p}_i$, is labelled by the irrep, $\Lambda_i$, and the row of that irrep, $\mu_i$ (analogous to the $J_z$ value for a spin-$J$ meson in an infinite-volume continuum). The sum over momentum directions related by allowed cubic rotations is subject to the constraint that $\vec{p}_1 + \vec{p}_2 = \vec{P}$.
The generalised Clebsch-Gordan coefficients, $\mathcal{C}$, are discussed in Ref.~\cite{Dudek:2012gj}.

Each meson-meson operator can be characterized by the magnitudes of meson momenta that went into its construction, $\big( |\vec{p}_1|, |\vec{p}_2| \big)$. This leads to a natural truncation of the basis of operators following from the energy we would expect if the mesons had no residual interactions,
\begin{equation*}
E^{(2)}_\mathrm{n.i.} = \sqrt{m_1^2 + |\vec{p}_1|^2} + \sqrt{m_2^2 + |\vec{p}_2|^2} \, .
\end{equation*}
Clearly, as the individual meson momenta increase, the non-interacting energy increases, and at some point becomes sufficiently far above the energy region of interest that we are justified in not including that operator, or any above it, in our basis.

Constructing operators which resemble meson-meson-meson systems, relevant in the energy region above three-meson thresholds, can be done by a recursive application of the approach described above~\cite{Woss:2019hse}. However, one subtlety that arises here is that intermediate meson-meson subsystems
may feature resonant behaviour which a single meson-meson operator alone will not efficiently capture.
In this case, one or more optimized operators can be constructed for the lowest energy eigenstates in the meson-meson subsystem by diagonalizing a matrix of correlation functions formed from a basis of single-meson-like and meson-meson-like operators. These optimized operators are then combined with the remaining optimized stable meson operator to form three-meson-like operators that efficiently interpolate the energy eigenstates. Details of this type of construction are given in Ref.~\cite{Woss:2019hse}.
 
The inclusion of multi-meson and isospin--0 single-meson operators in our bases naturally leads to Wick contractions which feature quark-antiquark annihilations; in the context of lattice QCD these appear via $t$-to-$t$ quark propagators. The \emph{distillation} approach to computing correlation functions~\cite{Peardon:2009gh} efficiently handles these, along with the required source-sink propagators, without the need to make any further approximations or to introduce any stochastic noise. The propagators, which factorize from the operator constructions, are extremely general. They can be extensively reused in other calculations which require propagation of the same flavor of quarks such that the computational cost of obtaining them is spread over many physics results.

\vspace{1cm}

\subsection{Scattering amplitudes}

Once the finite-volume spectrum has been extracted from a variational analysis of a matrix of correlation functions
it can be used as a constraint on the energy dependence of the coupled-channel $t$-matrix. The relationship is encoded in the L\"uscher quantization condition~\cite{Luscher:1985dn,Luscher:1986pf,Luscher:1990ux,Luscher:1991cf,Rummukainen:1995vs,He:2005ey,Kim:2005gf,Christ:2005gi,Fu:2011xz,Guo:2012hv,Hansen:2012tf,Briceno:2012yi,Gockeler:2012yj,Leskovec:2012gb,Briceno:2014oea},
\begin{equation}
\det \Big[ \mathbf{1} + i \bm{\rho} \, \mathbf{t} \, \big( \mathbf{1} + i \bm{\mathcal{M}} \big) \Big] = 0,
	\label{luescher}
\end{equation}
where the diagonal matrix of phase-space factors, $\bm{\rho}(E_\cm)$, and $\bm{\mathcal{M}}(E_\cm, L)$ are known functions of essentially kinematic origin -- see Ref.~\cite{Woss:2020cmp} for our conventions. The matrix space over which the determinant acts is the set of partial-waves subduced into a particular irrep, for all kinematically accessible meson-meson scattering channels. For a given $t$-matrix,\footnote{related to the scattering $S$-matrix via $\bm{S} = \bm{1} + 2 i \, \sqrt{\bm{\rho}} \, \bm{t} \, \sqrt{\bm{\rho}}$} $\bm{t}(E_\cm)$,
the discrete set of solutions of this equation, $[E_\mathsf{cm}(L)]_{\mathfrak{n}=1,2,\ldots}$, for a fixed value of $L$ is the finite-volume spectrum
in an $L \!\times\! L \!\times\! L$ periodic box. A practical approach for reliably finding solutions to this equation when there are multiple partial waves and/or hadron-hadron scattering channels, which makes use of an eigenvalue decomposition of a suitable transformation of the matrix under the determinant, was presented in Ref.~\cite{Woss:2020cmp}.

Eq.~\ref{luescher} is capable of describing any number of coupled hadron-hadron channels, but must be supplemented with further formalism once three-hadron channels are accessible. Recent progress is reviewed in Refs.~\cite{Hansen:2019nir,Rusetsky:2019gyk}.\footnote{What role the experimentally observed dominance of quasi-two-body isobars plays in these formalisms is not yet known, but it may lead to considerable simplifications in practice.}

An approach that allows computed finite-volume spectra to constrain scattering amplitudes is to propose \emph{parameterizations} of $\bm{t}(E_\cm)$, whose parameters can be varied, with the corresponding finite-volume spectra from solution of Eq.~\ref{luescher} at each iteration compared to the computed spectra~\cite{Dudek:2012gj}. In this way, a $\chi^2$ can be defined which can be minimized to find the best description of the computed lattice QCD spectra (Eq.~(9) in Ref.~\cite{Dudek:2012xn}). Use of a \mbox{$K$-matrix} in the parameterization of the $t$-matrix ensures coupled-channel unitarity, and sensitivity to the particular choice of form chosen for $\bm{K}(E_\cm)$ can be explored by varying the form~\cite{Guo:2012hv}.

This method provides coupled-channel amplitudes constrained for real values of $E_\cm$, but use of explicit functional forms in the parameterizations means that we can analytically continue into the complex-energy plane to explore the singularity content of the $t$-matrix. \emph{Poles} at complex values of $E_\cm$ can be identified with \emph{resonances}, with the real and imaginary parts of the pole position having an interpretation in terms of, respectively, the mass and width of the resonance. Factorizing the residues of elements of $\bm{t}$ at the pole position leads to decay couplings of the resonance to the various scattering channels. The statistical uncertainty originating in the finite number of Monte-Carlo samples in the lattice QCD calculation can be propagated through this process, and in addition the scatter over parameterizations can be used to estimate a systematic uncertainty from the choice of parametrization.

\vspace{0.5cm}

This approach has been applied successfully in several recent calculations of coupled-channel scattering, most notably in a series of papers computing on three lattice volumes with $m_\pi \sim 391$ MeV.
In the first calculations~\cite{Dudek:2014qha,Wilson:2014cna}, coupled $\pi K$, $\eta K$ scattering was investigated. A virtual bound state and a broad resonance were found in $J^P = 0^+$, a bound state in $1^-$, and there was evidence for a narrow resonance in $2^+$, but for all these $J^P$ the coupling to the $\eta K$ channel was found to be small in energy region studied.
In Ref.~\cite{Dudek:2016cru}, the $J^P=0^+$ coupled $\pi \eta, K\Kbar$ scattering sector was considered, where an asymmetrical peak in $\pi \eta \to \pi \eta$ at the $K\Kbar$ threshold was found to correspond to a resonance pole that could be compared to the experimental $a_0(980)$. In Ref.~\cite{Briceno:2017qmb}, the $J^P = 0^+$ and $2^+$ coupled $\pi\pi, K\Kbar, \eta \eta$ isospin-0 sectors were studied. The scalar amplitudes show a sharp dip in $\pi\pi \to \pi \pi$ at $K\Kbar$ threshold that could be associated with a resonance pole related to the experimental $f_0(980)$, while a rapid turn on of $\pi\pi$ at threshold was found to be due to a bound-state related to the $\sigma/f_0(500)$. The tensor sector was more straightforward, with clear bumps related to two resonances poles, the lighter of which was found to be dominantly coupled to $\pi\pi$ and the heavier to $K\Kbar$, in line with the experimental $f_2(1270)$ and $f'_2(1525)$. In Ref.~\cite{Woss:2019hse}, coupled $\pi \omega$, $\pi \phi$ scattering was considered, with the vector nature of the $\omega$ (which is stable at this quark mass) leading to dynamically coupled partial-waves in $J^P=1^+$. A bump was found in $\pi \omega (\threeSone) \to \pi \omega (\threeSone)$ whose origin is a $b_1$-like resonance pole.

Before computing finite-volume spectra and determining scattering amplitudes relevant for the exotic $J^{PC} = 1^{-+}$ channel, we now discuss some of the consequences of working with exact SU(3) flavor symmetry.

%% file: sections/SU3F.tex
In this paper we will present the first attempt to compute the properties of a resonance with \emph{exotic} $J^{PC}$, the lightest $\pi_1$, which is suspected to be a \emph{hybrid meson}. As indicated in the introduction, this is a challenging problem owing to the large number of possible decay channels. A significant simplification would occur if we had an \emph{exact $\mathrm{SU}(3)$ flavor symmetry}, as opposed to the approximate one present in nature, as then many of the apparently independent channels would coalesce into particular representations of $\mathrm{SU}(3)_\mathrm{F}$. In this first calculation, we opt to make this symmetry exact by working with three flavors of light quark all with a mass value tuned to approximately match the physical strange-quark mass. In this world, the lightest pseudoscalar octet, containing the pion, kaon and $\eta$-like unflavored member, has a mass around 700 MeV. This relatively large mass has the additional useful effect of pushing three-meson thresholds to higher energies such that they become irrelevant in our calculation.

With exact $\mathrm{SU}(3)$ flavor symmetry, the `conventional' mesons (having flavor quantum numbers accessible to $q\bar{q}$) lie in octet ($\bm{8}$) and singlet ($\bm{1}$) representations following from the decomposition of $\bm{3} \otimes \bar{\bm{3}}$. The lightest of these is the pseudoscalar octet, containing degenerate mesons which we can associate with the pion, the kaon and something close to the $\eta$ meson.
We choose to use
the zero-isospin, zero-strangeness member of the octet as a label to indicate the $J^{P( C)}$, e.g. $\etaOctet$ in this case of $0^{-(\!+\!)}$. There is also a light pseudoscalar singlet, $\etaSinglet$, whose sole member is close to the familiar $\eta^\prime$.\footnote{The physical $\eta$, $\eta^\prime$ eigenstates (with broken $\SUF$) are believed to be admixtures of the octet/singlet basis states with a small mixing angle as discussed in Section~\ref{interpret}.
The dependence of this mixing angle on the light-quark mass was explored using lattice QCD in Ref.~\cite{Dudek:2013yja}.}
The lightest octet of vectors, $\omegaOctet$, contains mesons we identify with the $\rho$ and the $K^\ast$, but its neutral member cannot easily be associated with either the $\omega$ or the $\phi$, as the experimental $\omega$ is believed to have approximate quark content $u\bar{u} + d\bar{d}$, while the $\phi$ is dominantly $s\bar{s}$. These correspond to significant admixtures of the octet ($u\bar{u} + d\bar{d} - 2 s\bar{s}$) and singlet ($u\bar{u} + d\bar{d} + s\bar{s}$). Clearly, when $\SUF$ is broken, the flavorless members of $\omegaOctet$ and $\omegaSinglet$ must mix to form the physical eigenstates. 

The notable difference between the pseudoscalar and vector sectors was explored in lattice QCD in terms of the $q\bar{q}$ \emph{annihilation}, or `\emph{disconnected}', contributions to two-point correlation functions in Ref.~\cite{Dudek:2013yja}. As can be seen in Figs.~4 and 5 of that paper, the vector correlators have extremely small disconnected pieces, both at and away from the $\text{SU}(3)_\text{F}$ limit, leading to a lack of hidden-light--hidden-strange mixing and the $\rho$ and $\omega$ mesons being close to degenerate. This can be compared to the same quantities in the pseudoscalar sector shown in \mbox{Figs.~2 and 3 therein}.

These observations are related to the Okubo Zweig Iizuka (OZI) rule which states that processes where there are no quark lines connecting the initial-state hadrons to the final-state hadrons are suppressed. Empirically this holds
for many $J^{PC}$, including vectors, where a famous example is the suppression of the otherwise allowed decay $\phi \to \pi \pi \pi$ which leads to the $s\bar{s}$ assignment for the $\phi$. The OZI rule does not seem to apply to the pseudoscalar sector.

A major advantage of an exact $\text{SU}(3)$ flavor symmetry comes when we consider meson-meson scattering, as channels that with broken $\text{SU}(3)_\text{F}$ were independent and had differing thresholds, like $\pi\pi$, $K\Kbar$, \ldots, are now equivalent, being a single channel, $\etaOctet \etaOctet$. Since the stable scattering hadrons lie in octets and singlets, the meson-meson products $\bm{8} \otimes \bm{8}$, $\bm{8} \otimes \bm{1}$ and $\bm{1} \otimes \bm{1}$ are of interest, with the first of these being decomposed into $\bm{1} \oplus \bm{8}_1 \oplus \bm{8}_2 \oplus \bm{10} \oplus \overline{\bm{10}} \oplus \bm{27}$. The representations $\mathbf{10}, \mathbf{\overline{10}}, \mathbf{27}$ lie outside the `conventional' sector, requiring at least $qq\bar{q}\bar{q}$, and are unlikely to be resonant~\cite{Dudek:2010ew, Dudek:2012gj, Woss:2018irj}.
The two octets, $\bm{8}_1, \bm{8}_2$, can be distinguished by their symmetries under the exchange of the flavor of the two hadrons in the product.
We follow the conventions of Ref.~\cite{deSwart:1963pdg}, where $\bm{8}_1$ is symmetric and $\bm{8}_2$ is antisymmetric, and we summarize the relevant results in Appendix~\ref{su3sym}.  
As an example, using the $\text{SU}(3)$ analogues of Clebsch-Gordan coefficients in that reference, the flavor structure of the $I=0$, $I_z=0$, zero-strangeness members of the two octets in the vector-pseudoscalar case can be expressed as,
\begin{align}\label{VPs_CGs}
\bm{8}_1 &= \tfrac{1}{\sqrt{20}} \Big( K^{*+} K^- + K^{*-} K^+ 
-K^{*0} \KbarSuper{0}
- \KbarSuper{\ast 0} K^0 \Big) \nonumber \\
&\quad- \tfrac{1}{\sqrt{5}} \Big( \rho^+ \pi^- + \rho^- \pi^+ - \rho^0 \pi^0 \Big) - \tfrac{1}{\sqrt{5}} \omega_8  \eta_8, \nonumber \\
\bm{8}_2 &= \tfrac{1}{2} \Big( K^{*+} K^- - K^{*-} K^+ 
- K^{*0}\KbarSuper{0}
+ \KbarSuper{\ast 0} K^0 \Big) ,
\end{align}
which makes manifest that $\bm{8}_1$ is symmetric under the interchange of the flavor of the two hadrons while $\bm{8}_2$ is antisymmetric.

In determining what decays are possible, it is important to pay attention to the generalization of charge-conjugation symmetry. With exact isospin symmetry it is useful to consider $G$-parity and there are natural extensions of this in the $\SUF$ case. Because we are at liberty to consider any member of the target $\mathrm{SU}(3)$ multiplet, here we focus on the neutral zero-strangeness element where charge-conjugation symmetry itself is good and so $C$-parity is the relevant quantum number to consider. The resulting selection rules apply to all members of the multiplet. Details are provided in Appendix~\ref{su3sym} and the relevant results are summarized in Table~\ref{Ctable} where the different symmetries of $\bm{8}_1$ and $\bm{8}_2$ are apparent.

\begin{table}[bt]
{\renewcommand{\arraystretch}{1.5}
	\begin{tabular}{lc|c}
	 $\bm{F_a} \otimes \bm{F_b} \to \bm{F}$ & $C$ & e.g.~($1^{-(\!-\!)} 0^{-(\!+\!)} \to 1^{+(C)}$) \\[1ex]
	\hline
	$\bm{8}_a \otimes \bm{8}_b \to \bm{8}_1$ & 	$C_a C_b$ 		& ($\omegaOctet \etaOctet \to \hOneOctet $ $C=-$) \\
	$\bm{8}_a \otimes \bm{8}_b \to \bm{8}_2$ & 	$- C_a C_b$ 	& ($ \omegaOctet \etaOctet \to \fOneOctet$ $C=+$) \\
	$\bm{8}_a \otimes \bm{8}_b \to \bm{1}$ & 	$C_a C_b$ 		& ($\omegaOctet \etaOctet \to \hOneSinglet$ $C=-$) \\
	$\bm{8}_a \otimes \bm{1}_b \to \bm{8}$ & 	$C_a C_b$ 		& ($\omegaOctet \etaSinglet \to \hOneOctet$ $C=-$) \\
	$\bm{1}_a \otimes \bm{1}_b \to \bm{1}$ & 	$C_a C_b$ & ($\omegaSinglet \etaSinglet \to \hOneSinglet$ $C=-$) \\
	\end{tabular}
	}
	\caption{$C$-parity values for the neutral zero-strangeness components of the $\text{SU}(3)$ octets and singlets from meson-meson products. $C_a$ and $C_b$ denote the $C$-parity of the neutral zero-strangeness components of the product irreps. We present an example for the $1^{-(\!-\!)} 0^{-(\!+\!)} \to 1^{+(C)}$ case to illustrate notation.}
	\label{Ctable}
\end{table}

When the two scattering mesons are in the same $\SUF$ multiplet, there is the additional constraint of \emph{Bose symmetry} which requires that the state is symmetric under the interchange of the two mesons, i.e.\ the overall symmetry under the interchange of flavor, spin and spatial position. In the pseudoscalar-pseudoscalar case, where there is no spin to be dealt with, we immediately have the restriction that $\etaOctet \etaOctet$ with \emph{even} $\ell$ appear in $\bm{8}_1$ with $J^{P(C)} = \ell^{+(\!+\!)}$, while \emph{odd} $\ell$ appear in $\bm{8}_2$ with $J^{P(C)} = \ell^{-(\!-\!)}$. It is therefore not possible to have an octet $1^{-(\!+\!)}$ resonance decay to $\etaOctet \etaOctet$. Slightly more complicated is the case of $\omegaOctet \omegaOctet$ where the spin of the two vectors can combine to total spin $S=0,1,2$ (symmetric, antisymmetric, symmetric respectively) which is then coupled to orbital angular momentum $\ell$. The spin+space symmetric options (such as $\omegaOctet \omegaOctet\{ \oneSzero \}$) appear in $\bm{8}_1$, while the spin+space antisymmetric options (such as $\omegaOctet \omegaOctet\{ \threeSone \}$) appear in $\bm{8}_2$. A more complete discussion of these constraints can be found in Appendix~\ref{bosesym}.

%%%%%%%%%%%%%%%%%%%%%%%%%%%%%%%%%%%%%%%%%%%%%%%%%%%%%%%%%%%%%%%%%%%%%%%%%%%%

\vspace{5mm}
In this study we will present the result of a calculation of the $J^{P( C)} = 1^{-(\!+\!)}$ octet, labelled $\etaOneOctet$. We will choose to focus our later interpretation on the isovector member, the $\pi_1$, even though with exact $\SUF$ symmetry the properties of the isoscalar member, the $\eta_1$, and the strange members are exactly the same. The reason for this choice is that as we move away from the $\SUF$ limit by reducing the $u,d$ quark masses, retaining an isospin symmetry, we expect that the $\eta_1$ can mix with an $\eta_1$ living in the $\SUF$ singlet, the $\etaOneSinglet$, while the kaonic states can mix with $1^{-(\!-\!)}$ kaons owing to there being no relevant \mbox{$C$-parity-like} symmetry for mesons with net strangeness. On these grounds it seems plausible that the properties of the $\pi_1$ will change least as we move away from the exact $\SUF$ limit. There may be some mixing with the corresponding states in the $\bm{10},\bm{\overline{10}},\bm{27}$ representations, but this is expected to be negligible given that there is no evidence for anything beyond rather weak non-resonant interactions in these multiplets.

The meson-meson scattering channels capable of coupling to the $1^{-(\!+\!)}$ octet include $\etaSinglet \etaOctet$, $\omegaOctet \etaOctet$, $\omegaOctet \omegaOctet$, $\omegaSinglet \omegaOctet$, $\fOneOctet \omegaOctet$, $\hOneOctet \etaOctet$, $\fOneSinglet \etaOctet \ldots$ .  How many of these are kinematically accessible in the decay of a potential lightest $1^{-(\!+\!)}$ resonance depends upon QCD dynamics which we will now explore in a lattice QCD calculation.

%% file: sections/lattice.tex
Calculations of correlation functions were performed on six anisotropic lattices with volumes $(L/a_s)^3 \!\times\! (T/a_t) = 12^3 \!\times\! 96$ and $\{14^3,16^3,18^3,20^3,24^3\} \!\times\! 128$. The spatial and temporal lattice spacings are $a_s \sim 0.12\text{ fm}$
and $a_t=a_s/\xi \sim (4.7 \text{ GeV})^{-1}$ respectively, where the anisotropy $\xi \sim 3.5$. Gauge fields were generated from a tree-level Symanzik improved gauge action and a Clover fermion action with \emph{three} degenerate flavors of dynamical quarks~\cite{Edwards:2008ja,Lin:2008pr}, tuned to approximately the value of the physical strange quark mass, such that the pion mass is \mbox{$\sim 700$ MeV}.
On all volumes, exponentially-suppressed finite-volume and thermal effects remain negligible as $m_\pi L \gtrsim 6$ and $m_\pi T \gtrsim 14$. 

Correlation functions were computed using the \emph{distillation} framework~\cite{Peardon:2009gh} and we give the rank of the distillation space, $N_{\text{vecs}}$, number of gauge configurations, $N_{\text{cfgs}}$, and time-sources, $N_{\text{tsrcs}}$, used on each volume in Table~\ref{lattice_specs}. We typically compute all the elements of the matrix of correlation functions; however, in a few cases we made use of hermiticity to infer $C_{ji}(t)$ from a computed $C_{ij}(t)$.

\begin{table}[b]
        \centering
        {\renewcommand{\arraystretch}{1.2}
        \begin{tabular}[t]{c | c c c}
                $(L/a_s)^3 \!\times\! (T/a_t)$ & $N_{\text{vecs}}$ & $N_{\text{cfgs}}$ & $N_{\text{tsrcs}}$ \\
                \hline
                $12^3 \times 96$ & 48 & 219 & 24\\
                $14^3 \times 128$ & 64 & 397 & 16\\
                $16^3 \times 128$ & 64 & 529 & 4\\
                $18^3 \times 128$ & 96 & 358 & 4\\
                $20^3 \times 128$ & 128 & 501 & 4\\
                $24^3 \times 128$ & 160 & 607 & 4\\
        \end{tabular}
        }
        \caption{Number of distillation vectors ($N_{\text{vecs}}$), gauge configurations ($N_{\text{cfgs}}$) and time-sources ($N_{\text{tsrcs}}$) used in computation of correlation functions on each lattice volume, as described in the text.}
        \label{lattice_specs}
\end{table}

\begin{figure}[t]
  	\centering
    \includegraphics[width=\columnwidth]{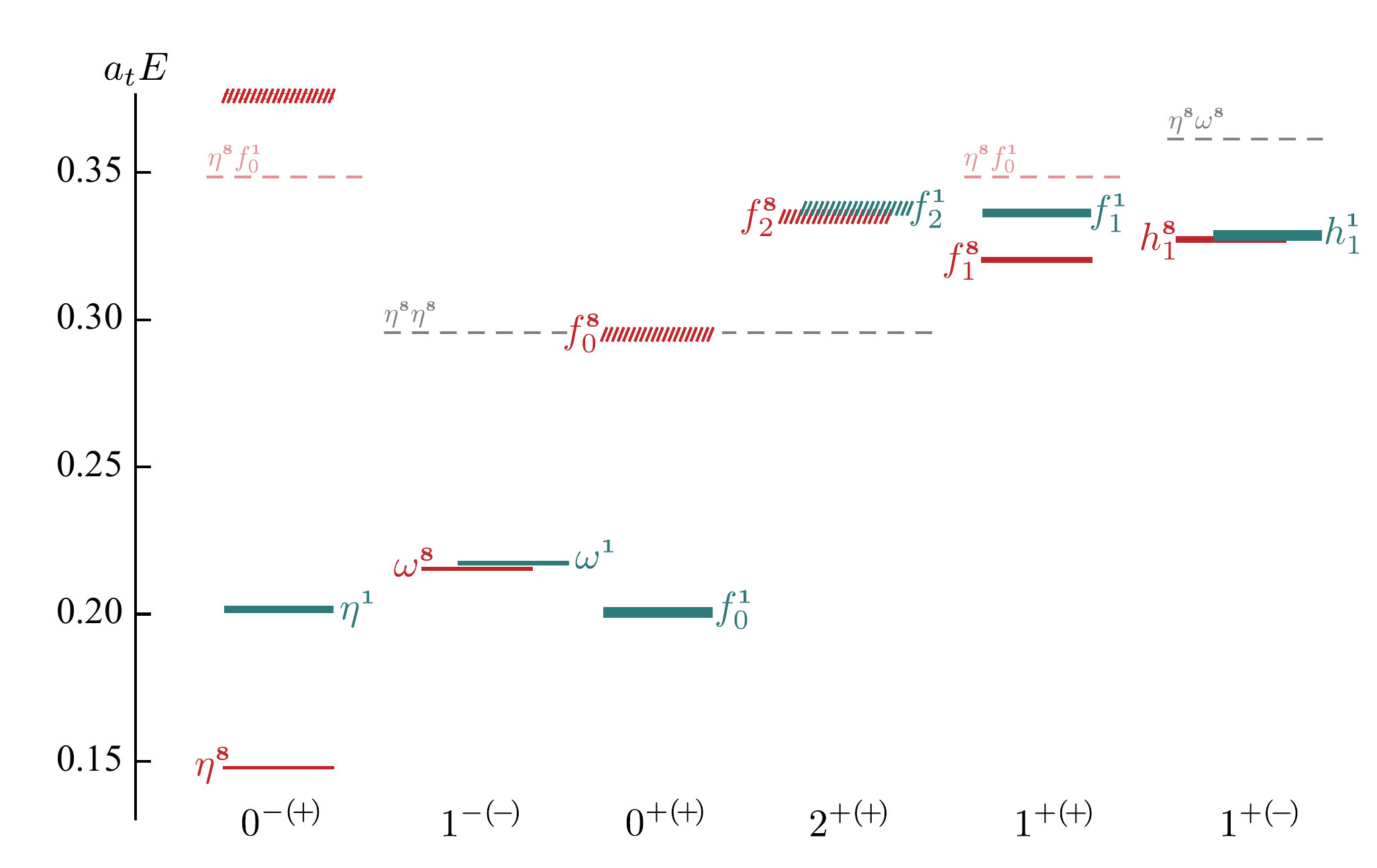}    
	\caption{Spectrum of low-lying octet (red) and singlet (cyan) mesons by $J^{P(\!C\!)}$ obtained using only single-meson operators. Solid boxes show mesons which lie below relevant meson-meson thresholds and are thus stable, while hatched boxes show mesons which lie above threshold and which will require a full finite-volume analysis to resolve their resonant nature. Dashed lines show the lowest relevant meson-meson thresholds.\\
	}
	\label{hadrons}
\end{figure}

The spectrum of low-lying mesons on these lattices is shown in Figure~\ref{hadrons}, obtained as the ground states in variational analysis of matrices of correlation functions using a basis of fermion bilinear operators in either $\SUF$ octet or singlet representations\footnote{More details of the operator construction, and decomposition in terms of connected and disconnected contributions can be found in Ref.~\cite{Dudek:2013yja}. The $16^3$ and $20^3$ volumes used in that reference are supplemented with the other volumes in Table~\ref{lattice_specs} in the current work.}. As we might expect, the pseudoscalar octet (containing the analogues of the pion, kaon and $\eta$) is lightest, with the pseudoscalar singlet (comparable to the $\eta'$) being somewhat heavier. The octet and singlet vector mesons are close to degenerate reflecting that this $J^{PC}$ has a very small disconnected contribution which distinguishes the singlet from the octet.

The singlet scalar meson ($\fZeroSinglet$) is rather light, at a similar mass to the pseudoscalar singlet. As it does not appear in the decays of the $1^{-(\!+\!)}$ resonance we are studying in this paper, we will not discuss it further here. The extracted scalar octet meson ($\fZeroOctet$) mass lies very close to the $\etaOctet \etaOctet$ threshold. This indicates that to properly understand the $\fZeroOctet$, which may be a resonance or a shallow bound state, we would have to include meson-meson operators in our basis. Levels corresponding to the tensor mesons ($\fTwoSinglet, \fTwoOctet$) are found some way above the $\etaOctet \etaOctet$ threshold, strongly suggesting that these states will be resonances capable of decaying into $\etaOctet \etaOctet$.

The axial mesons, the $J^{P( C )} = 1^{+(\!-\!)}$ $\hOneSinglet$ and $\hOneOctet$, and the $J^{P( C )} = 1^{+(\!+\!)}$ $\fOneSinglet$ and $\fOneOctet$, all lie quite far below their relevant decay thresholds, indicating that they are stable. As in the pseudoscalar-vector complex, the $C\!=\!+$ states show some octet-singlet splitting owing to a significant disconnected contribution, while the $C\!=\!-$ states are close to degenerate.

%%%%%%%%%%%%%%%%%%%%%%%%%%%%%%%%%%%%%%%%%%%%%%%%%%%%%%%%%%
%%%%%%%%%%%%%%%%%%%%%%%%%%%%%%%%%%%%%%%%%%%%%%%%%%%%%%%%%%

%%%%%%%%%%%%%%%%%%%%%%%%%%%%%%%%%%%%%%%%%%%%%%%%%%%%%%%%%%
\begin{table}[tb]
        \centering      
        {\renewcommand{\arraystretch}{1.2}
        \begin{tabular}[t]{ r l @{\hskip 3.0ex}|@{\hskip 3.0ex} r l } 
        %%%%
                        $\etaOctet$ & 0.1478(1) 	& $\etaSinglet$ & 0.2017(11) \\
                        $\omegaOctet$ & 0.2154(2) 	& $\omegaSinglet$ & 0.2174(3) \\
                        							&& $\fZeroSinglet$ & 0.2007(18) \\
                        $\fOneOctet$ & 0.3203(6)  	& $\fOneSinglet$ & 0.3364(14)\\
                        $\hOneOctet$ & 0.3272(6) 	& $\hOneSinglet$ & 0.3288(17) \\    
                        %%%%
        \end{tabular}
        }
        \caption{Relevant stable hadron masses, $a_t m$.}
        \label{masses}
\end{table} 
%%%%%%%%%%%%%%%%%%%%%%%%%%%%%%%%%%%%%%%%%%%%%%%%%%%%%%%%%%

\vspace{5mm}
As well as the computations in the rest frame from which the hadron masses in Table~\ref{masses} are obtained, matrices of correlation functions are also computed with non-zero values of allowed lattice momentum, $\vec{p}~=~\tfrac{2\pi}{L}(n_x, n_y, n_z)$, and from these the dispersion relations, $E(|\vec{p}|)$, for the stable mesons determined -- these are found
to be well described by the expected relativistic form, 
\begin{equation}\label{disp_relation}
\big( a_t E_{\vec{n}} \big)^2 = \big( a_t m \big)^2 + \frac{1}{\xi^2} \left( \frac{2\pi}{L/a_s} \right)^{\!2} |\vec{n}|^2 \, ,
\end{equation}
with the fitted values of anisotropy found for each meson being broadly compatible up to small variations due to discretization effects. An estimate of the anisotropy with an uncertainty that reflects the small variation over different mesons is $\xi = 3.486(43)$ -- see Ref.~\cite{Woss:2018irj} for further details.

\begin{figure}[h]
  	\centering
    \includegraphics[width=0.72\columnwidth]{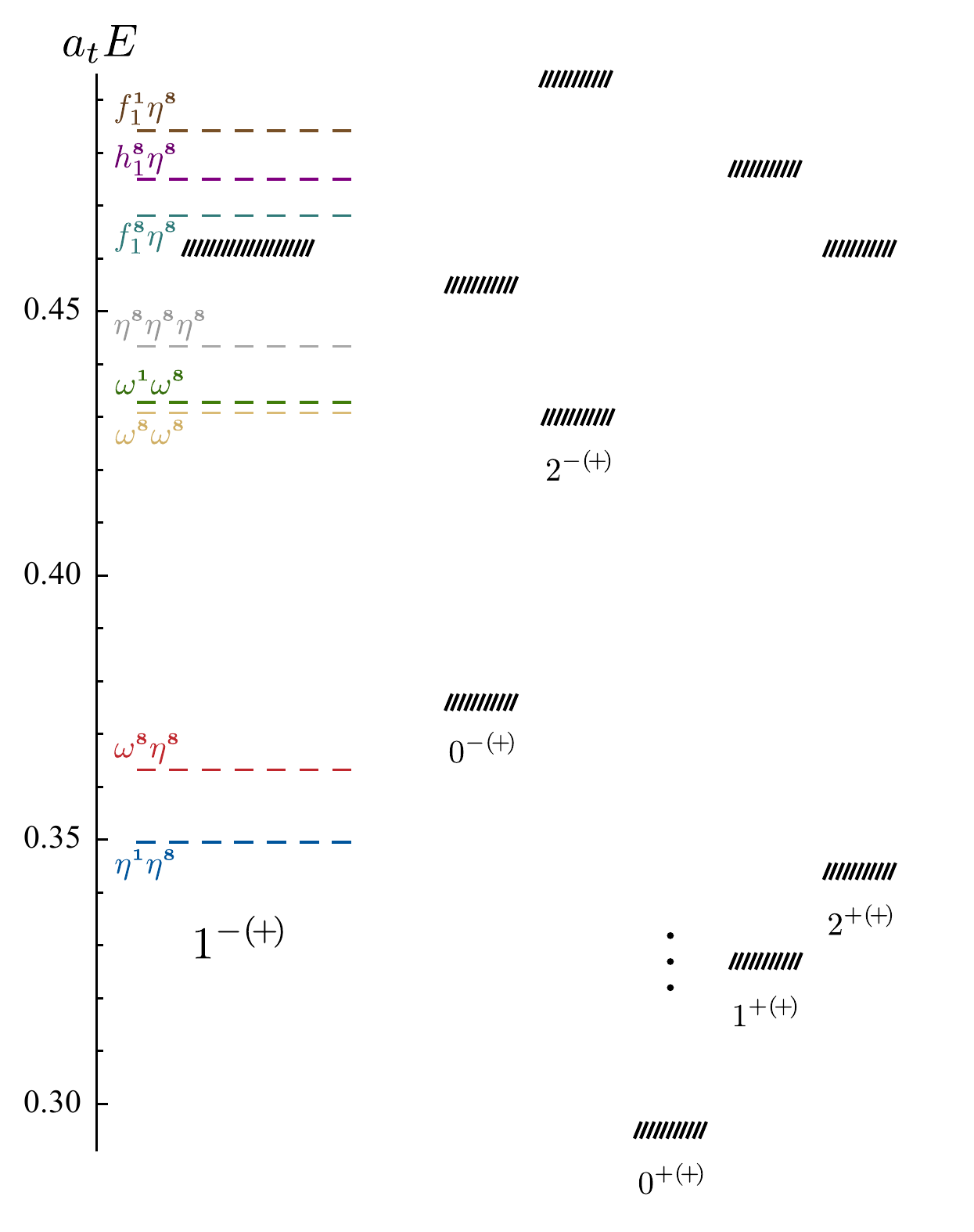}    
	\caption{Masses of $C=+$ octet mesons obtained using only single-meson operators (taken from Ref.~\cite{Dudek:2013yja}). Thresholds relevant for $J^{P(C)} = 1^{-(\!+\!)}$ are shown.}
	\vspace{5mm}
		\label{CpSpectrum}
\end{figure}

\vspace{2mm}
Figure~\ref{CpSpectrum} illustrates the position of a likely octet $1^{-(\!+\!)}$ resonance based upon variational analysis of correlation matrices using only fermion-bilinear constructions, along with the decay thresholds given in Table~\ref{thresholds} which follow from the masses in Table~\ref{masses}. Also shown are the expected octet resonance spectra with other $J^{P(\!+\!)}$ taken from Ref.~\cite{Dudek:2013yja}. These quantum numbers would contribute if spectra with non-zero overall momentum were to be considered, significantly complicating the analysis. For this reason, in this first calculation of the exotic $1^{-(\!+\!)}$ scattering system, we will restrict our attention to the spectrum in the overall rest-frame, considering the $T_1^{-(\!+\!)}$ irrep. We will consider the role played by $3^{-(\!+\!)}$, $4^{-(\!+\!)}$ scattering, which in principle contribute in this irrep, later in the manuscript.

%%%%%%%%%%%%%%%%%%%%%%%%%%%%%%%%%%%%%%%%%%%%%%%%%%%%%%%%%%
%%%%%%%%%%%%%%%%%%%%%%%%%%%%%%%%%%%%%%%%%%%%%%%%%%%%%%%%%%

%%%%%%%%%%%%%%%%%%%%%%%%%%%%%%%%%%%%%%%%%%%%%%%%%%%%%%%%%%
\begin{table}[b]
        \centering      
        {\renewcommand{\arraystretch}{1.2}
        \begin{tabular}[t]{ r l } 
                $\etaSinglet \etaOctet$ 		& 0.3495(11) \\        
                $\omegaOctet \etaOctet$ 		& 0.3632(2) \\       
                $\omegaOctet \omegaOctet$ 		& 0.4308(3) \\
                $\omegaSinglet \omegaOctet$ 	& 0.4324(7) \\
                $\etaOctet \etaOctet \etaOctet$ & 0.4434(2) \\
                $\fOneOctet \etaOctet$ 			& 0.4681(6) \\
                $\hOneOctet \etaOctet$ 			& 0.4750(6) \\
                $\fOneSinglet \etaOctet$ 		& 0.4842(14) 
        \end{tabular}
        }
        \caption{Multi-meson thresholds relevant for $J^{P(C)} = 1^{-(\!+\!)}$ shown in Fig.~\ref{CpSpectrum}. Uncertainties are determined by adding the uncertainties on the single-meson masses in quadrature.}
        \label{thresholds}
\end{table} 
%%%%%%%%%%%%%%%%%%%%%%%%%%%%%%%%%%%%%%%%%%%%%%%%%%%%%%%%%%%%%

\pagebreak
\subsection{Operator Bases}
We construct a suitable basis of operators in the $T_1^{-(\!+\!)}$ irrep from a set of single-meson-like operators and a set of meson-meson-like operators. A total of 18 fermion bilinears, $\bar{\psi}\bm{\Gamma}\psi$, are used following Ref.~\cite{Dudek:2010wm}, with a spin and spatial structure built from Dirac $\gamma$-matrices and gauge-covariant derivatives.
Gluonic degrees of freedom enter through the gauge-covariant derivatives. For example, one
simple \oneMP~ bilinear operator, constructed using the vector cross product of $\gamma_i$ and the commutator of two derivatives, is given by,
\begin{align}\label{SecIVA:Eq_hybrid_meson}
(\bar{\psi}\bm{\Gamma}\psi)_i = \underbrace{\epsilon_{ijk} ( \bar{\psi}\gamma_j\psi)\,B_k }_{1^{--}\otimes 1^{+-} \rightarrow 1^{-+}}\, ,
\end{align}
where $B_k\propto \epsilon_{kpq}[\overleftrightarrow{D_p},\, \overleftrightarrow{D_q}]$ is the chromomagnetic field. In practice, when we determine the spectra we vary the number of single-meson operators to establish insensitivity to the details of the choice of operator basis.

In Table~\ref{thresholds}, we show the relevant multi-hadron thresholds for two- and three-meson channels that appear in \oneMP~and that transform in the flavor octet. To ensure all relevant meson-meson operators are included in the operator basis, we calculate the non-interacting energies for each multi-meson system by considering all momenta combinations that sum to zero. All meson-meson operators with a corresponding non-interacting energy below $a_t E_\cm = 0.48$, a modest distance below $\fOneSinglet\etaOctet$ threshold, are included.\footnote{In addition, we include an $\fOneSinglet\etaOctet$ operator corresponding to a non-interacting level at $\fOneSinglet\etaOctet$ threshold. A small number of meson-meson operators that lie a modest distance above the $\fOneSinglet\etaOctet$ threshold were also added to explore the (very mild) sensitivity to our choice of largest energy.} These operators are presented in Table~\ref{SecVI:Tab:Ops}, listed by increasing non-interacting energy. 

The only relevant three-meson threshold, $\etaOctet \etaOctet \etaOctet$, lies slightly below the expected $1^{-(\!+\!)}$ resonance position. The lowest non-interacting three-meson energies appear at $a_tE^{(3)}_{\text{n.i.}}>0.51$.
As discussed in Sec.~\ref{Method}, resonant excitations in two-meson subsystems may be present and operators that capture these subsystem interactions need to be considered for inclusion. 
To do this we examine the `two-plus-one' non-interacting energies, $a_tE^{(2+1)}_{\text{n.i.}}$, which follow from assuming no residual interaction between the interacting two-meson subsystem and the third meson -- details are provided in Ref.~\cite{Woss:2019hse}. The lowest-energy combination of three $\etaOctet$ that appears in the $T_1^-$ irrep is

\begin{equation}\label{3mes}
\underbrace{[011]A_2}_{\etaOctet} \otimes \underbrace{[001]A_2}_{\etaOctet} \otimes  \underbrace{[001]A_2}_{\etaOctet} \rightarrow \underbrace{[000]T_1^-}_{\etaOneOctet} \oplus \, \dots .
\end{equation}
We consider all possible meson-meson subsystems here that could feature bound states or resonances. Combining the first two pseudoscalar octets appearing in Eq.~\ref{3mes} into definite momentum type $[001]$, we find the only irrep combination that yields the $T_1^-$ irrep is,
\begin{align*}
\Big(\underbrace{[011]A_2}_{\etaOctet} \otimes \underbrace{[001]A_2}_{\etaOctet}\Big)\otimes \, \underbrace{[001]A_2}_{\etaOctet}\, &\rightarrow \underbrace{[000]T_1^-}_{\etaOneOctet}\notag \\
\underbrace{[001]E_2}_{\etaOctet\etaOctet}  \otimes
\underbrace{[001]A_2}_{\etaOctet} &\rightarrow \underbrace{[000]T_1^-}_{\etaOneOctet}.
\end{align*}
The irrep $[001]E_2$ houses the $\omegaOctet$ and $\fOneOctet$, which we treat as stable scattering particles -- any excited finite-volume energy level coupling to $\etaOctet \etaOctet$ (in any flavor combination) will lie above the $\fOneOctet$ level, and hence no three-meson-like operators are needed in the basis to study a $1^{-(\!+\!)}$ resonance near $a_t E \sim 0.46$.

%%%%%%%%%%%%%%%%%%%%%%%%%%%%%%%%%%%%%%%%%%%%%%%%%%%%%
\begin{table*}
	\centering
	{\renewcommand{\arraystretch}{1.5}
		\begin{tabular}{r @{\hskip 4.0ex} r @{\hskip 4.0ex} r @{\hskip 4.0ex} r @{\hskip 4.0ex} r @{\hskip 4.0ex} r}
			\hline
			\hline
			$L/a_s=12$ & $L/a_s=14$ & $L/a_s=16$ & $L/a_s=18$ & $L/a_s=20$ & $L/a_s=24$  \\
			\hline
			${18}\times \bar{\psi}\bm{\Gamma}\psi$ &${18}\times \bar{\psi}\bm{\Gamma}\psi$ &${18}\times \bar{\psi}\bm{\Gamma}\psi$ &${18}\times \bar{\psi}\bm{\Gamma}\psi$ &${18}\times \bar{\psi}\bm{\Gamma}\psi$ &${18}\times \bar{\psi}\bm{\Gamma}\psi$ \\
			$\etaSinglet_{[001]}\etaOctet_{[001]}$ &
			$\etaSinglet_{[001]}\etaOctet_{[001]}$ &
			$\etaSinglet_{[001]}\etaOctet_{[001]}$ &
			$\etaSinglet_{[001]}\etaOctet_{[001]}$ &
			$\etaSinglet_{[001]}\etaOctet_{[001]}$ & 
			$\etaSinglet_{[001]}\etaOctet_{[001]}$  \\
			${\fOneOctet}_{[000]}\etaOctet_{[000]}$ &
			$\omegaOctet_{[001]}\etaOctet_{[001]}$ 	  & 
			$\omegaOctet_{[001]}\etaOctet_{[001]}$ 	  & 
			$\omegaOctet_{[001]}\etaOctet_{[001]}$ 	  & 
			$\omegaOctet_{[001]}\etaOctet_{[001]}$    & 
			$\omegaOctet_{[001]}\etaOctet_{[001]}$ \\
			 $\omegaOctet_{[001]}\etaOctet_{[001]}$ &
			${\fOneOctet}_{[000]}\etaOctet_{[000]}$ &
			${\fOneOctet}_{[000]}\etaOctet_{[000]}$ &
			$\etaSinglet_{[011]}\etaOctet_{[011]}$ &
			$\etaSinglet_{[011]}\etaOctet_{[011]}$ & 
			$\etaSinglet_{[011]}\etaOctet_{[011]}$ \\
			${\hOneOctet}_{[000]}\etaOctet_{[000]}$ &
			${\hOneOctet}_{[000]}\etaOctet_{[000]}$ & 
			$\etaSinglet_{[011]}\etaOctet_{[011]}$ & 
			$\{2\}\,\omegaOctet_{[011]}\etaOctet_{[011]}$ & 
			$\{2\}\,\omegaOctet_{[011]}\etaOctet_{[011]}$ & 
			$\{2\}\,\omegaOctet_{[011]}\etaOctet_{[011]}$ \\
			${\fOneSinglet}_{[000]}\etaOctet_{[000]}$ &
			${\fOneSinglet}_{[000]}\etaOctet_{[000]}$ &
			${\hOneOctet}_{[000]}\etaOctet_{[000]}$ &
			${\fOneOctet}_{[000]}\etaOctet_{[000]}$ &
			$\omegaOctet_{[001]}\omegaOctet_{[001]}$ &
			$\etaSinglet_{[111]}\etaOctet_{[111]}$ \\
			&
			$\omegaOctet_{[001]}\omegaOctet_{[001]}$  &
			${\fOneSinglet}_{[000]}\etaOctet_{[000]}$ &
			${\hOneOctet}_{[000]}\etaOctet_{[000]}$ &
	        ${\fOneOctet}_{[000]}\etaOctet_{[000]}$ &
			$\omegaOctet_{[111]}\etaOctet_{[111]}$ \\
			&
			$\{4\}\, \omegaSinglet_{[001]}\omegaOctet_{[001]}$ &
			$\{2\}\, \omegaOctet_{[011]} \etaOctet_{[011]}$ &
			$\omegaOctet_{[001]}\omegaOctet_{[001]}$  &
			$\{4\}\, \omegaSinglet_{[001]}\omegaOctet_{[001]}$ &
			$\omegaOctet_{[001]}\omegaOctet_{[001]}$ \\
			&
			$\etaSinglet_{[011]}\etaOctet_{[011]}$ &
			$\omegaOctet_{[001]}\omegaOctet_{[001]}$  &
			$\{4\}\, \omegaSinglet_{[001]} \omegaOctet_{[001]}$  &
			$\etaSinglet_{[111]} \etaOctet_{[111]}$ &
			$\{4\}\, \omegaSinglet_{[001]} \omegaOctet_{[001]}$ \\
			&
			$\{2\}\,\omegaOctet_{[011]}\etaOctet_{[011]}$ &  
			$\{4\}\, \omegaSinglet_{[001]} \omegaOctet_{[001]}$  &
			${\fOneSinglet}_{[000]} \etaOctet_{[000]}$ &
			${\hOneOctet}_{[000]} \etaOctet_{[000]}$ &
			$\etaSinglet_{[002]} \etaOctet_{[002]}$ \\
			& & &
			$\etaSinglet_{[111]} \etaOctet_{[111]}$ &
			$\omegaOctet_{[111]} \etaOctet_{[111]}$ &
			${\fOneOctet}_{[000]} \etaOctet_{[000]}$ \\   	
			& & &
			$\omegaOctet_{[111]} \etaOctet_{[111]}$ &
			${\fOneSinglet}_{[000]} \etaOctet_{[000]}$ &
			$\omegaOctet_{[002]} \etaOctet_{[002]}$ \\
			& & & &
			$\omegaOctet_{[002]} \etaOctet_{[002]}$ &
			${\hOneOctet}_{[000]} \etaOctet_{[000]}$ \\
			& & & & &
			${\fOneSinglet}_{[000]} \etaOctet_{[000]}$ \\
			& & & & &
			$\etaSinglet_{[012]} \etaOctet_{[012]}$ \\
		\end{tabular}
	}
\caption{$T_1^{-(\!+\!)}$ operator basis for each lattice volume.
Meson-meson operators are ordered by increasing $E_\text{n.i.}$ and labelled with the momentum types of the two mesons; different momentum directions are summed over as discussed in Section~\ref{Method}.
The number in braces, $\{N_\text{mult}\}$, denotes the multiplicity of linearly-independent meson-meson operators if this is larger than one.
The maximum number of single-meson operators, $N$, is denoted by $N \times \bar{\psi}\bm{\Gamma}\psi$ and various subsets of these were considered to investigate sensitivity to the details of the choice of operator basis.}
\label{SecVI:Tab:Ops}
\end{table*}
%%%%%%%%%%%%%%%%%%%%%%%%%%%%%%%%%%%%%%%%%%%%%%%%%%%%%

\vspace{5mm}
As discussed previously, the $T_1^{-(\!+\!)}$ irrep also features contributions from $J^{P(C)}=3^{-(\!+\!)}$. 
Considering the $A_2^{-(\!+\!)}$ irrep, which for $J\leq 4$ features only $J^{P(C)}=3^{-(\!+\!)}$ subductions, we can isolate the contribution from the $J=3$ partial-waves. We will use the finite-volume energy levels in this irrep to constrain the $J=3$ partial-waves and show these are small over the energy range considered here. The operator basis used in the $A_2^{-(\!+\!)}$ irrep for each lattice volume is given in Table~\ref{Tab:Ops_A2m}.

%%%%%%%%%%%%%%%%%%%%%%%%%%%%%%%%%%%%%%%%%%%%%%%%%%%%%
\begin{table}
	\centering
	{\renewcommand{\arraystretch}{1.5}
		\begin{tabular*}{0.95\linewidth}{@{\extracolsep{\fill}}rrrr}
			\hline
			\hline
			$L/a_s=16$ & $L/a_s=18$ & $L/a_s=20$ & $L/a_s=24$  \\
			\hline
			${4}\times \bar{\psi}\bm{\Gamma}\psi$ &${4}\times \bar{\psi}\bm{\Gamma}\psi$ &${4}\times \bar{\psi}\bm{\Gamma}\psi$ &${4}\times \bar{\psi}\bm{\Gamma}\psi$ \\
			$\omegaOctet_{[011]}\etaOctet_{[011]}$ &
			$\omegaOctet_{[011]}\etaOctet_{[011]}$ &
			$\omegaOctet_{[011]}\etaOctet_{[011]}$ &
			$\omegaOctet_{[011]}\etaOctet_{[011]}$ \\
			$\omegaSinglet_{[001]}\omegaOctet_{[001]}$ & 
			$\omegaSinglet_{[001]}\omegaOctet_{[001]}$ &
			$\omegaSinglet_{[001]}\omegaOctet_{[001]}$ & 
			$\etaSinglet_{[111]}\etaOctet_{[111]}$ \\
			&
			$\etaSinglet_{[111]}\etaOctet_{[111]}$ &
			$\etaSinglet_{[111]}\etaOctet_{[111]}$ & 
			$\omegaSinglet_{[001]}\omegaOctet_{[001]}$ \\
		\end{tabular*}
	}
	\caption{As Table~\ref{SecVI:Tab:Ops} but showing the $A_2^{-(\!+\!)}$ operator basis for each lattice volume, with meson-meson operators ordered by increasing $E_\text{n.i.}$. \\}
	\label{Tab:Ops_A2m}
\end{table}
%%%%%%%%%%%%%%%%%%%%%%%%%%%%%%%%%%%%%%%%%%%%%%%%%%%%%

\vspace{1cm}
\subsection{Finite-volume spectra}

%%%%%%%%%%%%%%%%%%%%%%%%%%%%%%%%%%%%%%%%%%%%%%%%%%%%%%%%%%
\begin{figure*}
  	\centering
    \includegraphics[width=1\textwidth]{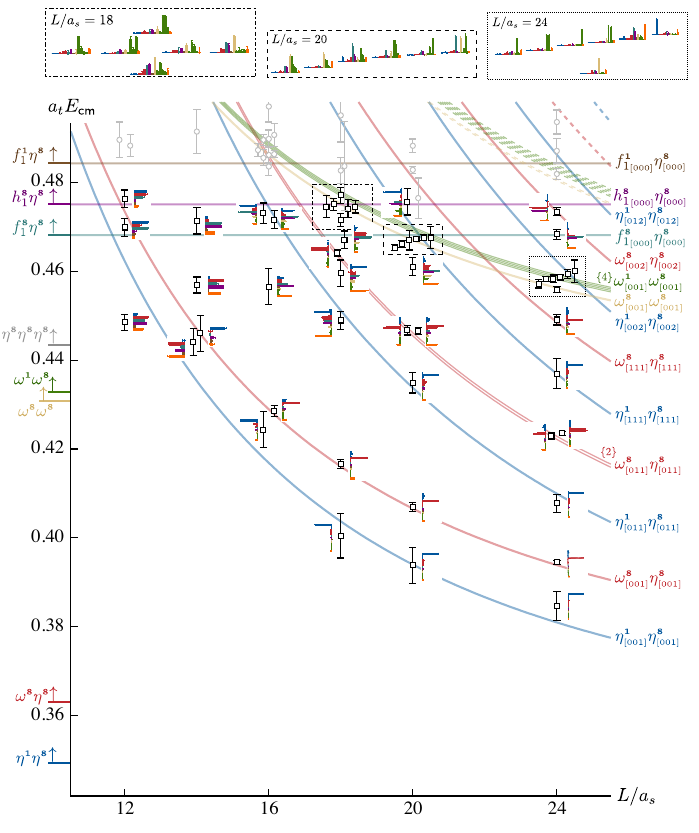}   
    \vspace{-10mm} 
	\caption{Finite-volume spectrum in the $T_1^{-(\!+\!)}$ irrep on six lattice volumes. Points show the extracted energy levels, including uncertainties, from a variational analysis using the operator bases in Table~\ref{SecVI:Tab:Ops}; black points are included in the subsequent scattering analysis and grey points are not. Some points are slightly displaced horizontally for clarity when near-degenerate energies appear. Curves show meson-meson non-interacting energies, with multiplicities greater than one labelled by $\{ n \}$ and shown as slightly split curves. Dashed curves correspond to meson-meson operators not included in the basis. Relevant thresholds transcribed from Table~\ref{thresholds} are shown on the vertical axis. Accompanying each energy level is a histogram of the operator-state overlap factors, $Z_i^\mathfrak{n}=\braket{ \mathfrak{n}|\mathcal{O}_i^\dagger(0)|0}$, for $\etaSinglet\etaOctet$~(dark blue), $\omegaOctet\etaOctet$~(red), $\omegaOctet\omegaOctet$~(sand), $\omegaSinglet\omegaOctet$~(green), $\fOneOctet\etaOctet$~(cyan), $\hOneOctet\etaOctet$~(purple) and $\fOneSinglet\etaOctet$~(brown) meson-meson operators and a sample of $\bar{\psi}\bm{\Gamma}\psi$~(orange) fermion-bilinear operators. The overlaps are normalized such that the largest value for any given operator across all energy levels is equal to one. For clarity, the histograms accompanying the cluster of levels on the $L/a_s=18,20,24$ volumes are displayed at the top of the figure.}
		\label{SpectrumHistos}
\end{figure*}
%%%%%%%%%%%%%%%%%%%%%%%%%%%%%%%%%%%%%%%%%%%%%%%%%%%%%%%%%%

Variational analysis of matrices of $T_1^{-(\!+\!)}$ correlation functions on the six volumes leads to the spectrum presented in Figure~\ref{SpectrumHistos}. Errorbars reflect the statistical uncertainty and an estimate of the systematic uncertainty from varying the details of the variational analysis (such as operator basis and fit range). For each finite-volume eigenstate that will be used to constrain scattering amplitudes, we also show a histogram illustrating the overlap strength with operators in the basis. 

We notice that below $a_t E_\cm \sim 0.44$, the energy levels lie very close to the $\etaSinglet \etaOctet$ and $\omegaOctet \etaOctet$ non-interacting energies, and each level has dominant overlap with just the operator(s) corresponding to the particular non-interacting momentum combination lying nearby (blue and red bars). This tends to suggest weak, uncoupled scattering at lower energies. The somewhat larger errorbars on levels with large overlap onto $\etaSinglet \etaOctet$ operators is a consequence of the substantial disconnected contribution to the $\etaSinglet$.

In an energy region around $a_t E_\cm \sim 0.46$ on each volume we find one more energy level than expected on the basis of the non-interacting energies, and we begin to observe levels having significant overlaps onto hybrid-like single-meson operator constructions (orange bar). This energy region is where the $1^{-(\!+\!)}$ state proposed to be a hybrid meson was observed in the analysis using only single-meson operators discussed earlier. The finite-volume eigenstates having overlap onto the hybrid-like operator are also observed to have overlap onto meson-meson constructions, notably $\etaSinglet \etaOctet$ (dark blue), $\omegaOctet \etaOctet$ (red), $\fOneOctet \etaOctet$ (cyan) and/or $\hOneOctet \etaOctet$ (purple), which might suggest a resonance coupling to these scattering channels. 

A level lying very close to the two-fold degenerate $\omegaOctet_{[011]} \etaOctet_{[011]}$ non-interacting curve is observed at each volume above $L/a_s = 16$ with a characteristic histogram that couples strongly to the two $\omegaOctet_{[011]} \etaOctet_{[011]}$ operators but is decoupled from all other operators. Such behavior would be expected if the $\omegaOctet \etaOctet \{ \threeFthree \}$ wave is weak.

On the $L/a_s = 18, 20, 24$ volumes, a cluster of states appears in the energy region of interest close to the lowest $\omegaOctet \omegaOctet$ (sand) and $\omegaSinglet \omegaOctet$ (green) non-interacting energies. The histograms for these states, presented at the top of the figure, show that in each case there are five energies which have large overlap with these vector-vector operators, but not large overlap with hybrid-like operators. This might be taken as a suggestion that a hybrid resonance (if present) may not be strongly coupled to these vector-vector scattering channels.

Finally, the only states which show any significant coupling to the $\fOneSinglet \etaOctet$ (brown) operator lie at rather high energies, suggesting that this channel is probably not relevant to any resonance near $a_t E_\cm \sim 0.46$.

%%%%%%%%%%%%%%%%%%%%%%%%%%%%%%%%%%%%%%%%%%%%%%%%%%%%%%%%%%
\begin{figure}
	\centering
	\includegraphics[width=\columnwidth]{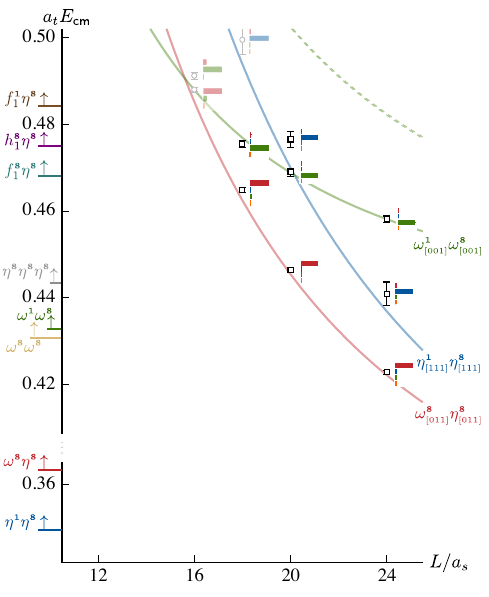}    
	\caption{Analogous to Figure~\ref{SpectrumHistos} but for the $A_2^{-(\!+\!)}$ irrep (operator lists shown in Table~\ref{Tab:Ops_A2m}). Note the vertical axis is broken to emphasise the relatively low-lying $\etaSinglet \etaOctet$ and $\omegaOctet \etaOctet$ thresholds.}
	\label{SpectrumA2}
\end{figure}
%%%%%%%%%%%%%%%%%%%%%%%%%%%%%%%%%%%%%%%%%%%%%%%%%%%%%%%%%%

Figure~\ref{SpectrumA2} shows the spectrum obtained in the $A_2^{-(\!+\!)}$ irrep. It is clear from the histograms, which are dominated in each case by a single meson-meson-like operator, and the proximity of each level to the corresponding non-interacting curves, that there are only relatively weak interactions. There is no sign of any resonant behaviour that might be associated with a low-lying $3^{-(\!+\!)}$ state.

\vspace{3mm}
While a qualitative discussion of the spectra like the one just presented can suggest possible features of the scattering system, a rigorous determination requires an analysis using the coupled-channel finite-volume formalism described in Section~\ref{Method} from which the $t$-matrix can be extracted, and from it properties of any resonance poles.

%% file: sections/scattering.tex
We wish to use the spectra computed in the $T_1^{-(\!+\!)}$ and $A_2^{-(\!+\!)}$ irreps, presented in the previous section, to determine the matrix describing scattering with $J^{P(C)} = 1^{-(\!+\!)}$. We expect $T_1^{-(\!+\!)}$ to be dominated by $1^{-(\!+\!)}$, with $3^{-(\!+\!)}$, $4^{-(\!+\!)}$, and still higher $J$ being weak at these energies -- these require higher orbital angular momentum $\ell$ and so are suppressed close to threshold in the absence of any dynamical enhancement. There is no evidence from the single-meson operator study in Ref.~\cite{Dudek:2013yja} of a low-lying $3^{-(\!+\!)}$ resonance, and while $4^{-(\!+\!)}$ is non-exotic (it can be constructed as the $q\bar{q}(\oneGfour)$ state), Ref.~\cite{Dudek:2013yja} suggests that such a state lies at $a_t E_\cm \sim 0.58$, far above our region of interest. By computing the $A_2^{-(\!+\!)}$ spectrum we are able to directly constrain the strength of scattering with $J^{P(C)} = 3^{-(\!+\!)}$ in the energy region of interest.

The first step in analysing the finite-volume spectrum is to establish the basis of relevant \emph{meson-meson partial-waves} in the considered energy region which define the matrix space in Eq.~\ref{luescher}. The set of meson-meson channels kinematically accessible was presented in the previous section and in Table~\ref{partialWaves} we show the set of partial waves we will use.

\begin{table}[t]
	\centering
	{\renewcommand{\arraystretch}{1.5}
		\begin{tabular}{r|l}
		\multirow{4}{*}{$1^{-(\!+\!)}$} 	& $ \etaSinglet \etaOctet \big\{ \!\onePone \!\big\}$ \\
											& $\omegaOctet \etaOctet \big\{\! \threePone  \!\big\}$ \\ 
											& $\omegaOctet \omegaOctet \big\{\! \threePone \!\big\}$, $\omegaSinglet \omegaOctet \big\{\! \onePone, \threePone, \fivePone \!\big\}$ \\
											& $\fOneOctet \etaOctet \big\{\! \threeSone \!\big\}$, $\hOneOctet \etaOctet \big\{\! \threeSone \!\big\}$ \\ 
		\hline
		\multirow{3}{*}{$3^{-(\!+\!)}$}		& $\etaSinglet \etaOctet \{ \oneFthree \}$	\\
											& $\omegaOctet \etaOctet \{ \threeFthree \}$ \\
											& $\omegaSinglet \omegaOctet \{ \fivePthree \}$									
		\end{tabular}
	}
	\caption{Scattering partial waves included in the description of $T_1^{-(\!+\!)}$ finite-volume spectra.}
	\label{partialWaves}
\end{table}

\pagebreak
A small number of possible partial waves have been excluded from Table~\ref{partialWaves} under the expectation that they will not contribute significantly. In the $1^{-(\!+\!)}$ sector, $\fOneOctet \etaOctet \big\{\! \threeDone \!\big\}$ and $\hOneOctet \etaOctet \big\{\! \threeDone \!\big\}$ are not included, as the thresholds for these channels are very high-lying in our energy region such that we expect a significant angular momentum suppression from the $D$-wave, relative to the leading $S$-wave, that will render them practically irrelevant. Similarly, in the vector-vector channels, we exclude $\omegaSinglet \omegaOctet \big\{\! \fiveFone \! \big\}$ on the basis of $F$-wave angular momentum suppression.\footnote{Bose symmetry forbids $\omegaOctet \omegaOctet \big\{\! \onePone, \fivePone \!\big\}$ and $\omegaOctet \omegaOctet \big\{\! \fiveFone \! \big\}$.}

In the $3^{-(\!+\!)}$ sector, $\omegaOctet \etaOctet \big\{\! \threeFthree \!\big\}$ is included despite the large angular momentum barrier. As can be seen in Table~\ref{SecVI:Tab:Ops}, there are two independent operators for $\omegaOctet_{[011]} \etaOctet_{[011]}$ and there is a corresponding two-fold degenerate non-interacting energy. In order that there be two solutions of Eq.~\ref{luescher} near this energy, as observed in our computed spectra and commented on in the previous section, higher $\omegaOctet \etaOctet$ partial-waves must be considered, so we include the $\omegaOctet \etaOctet \big\{\! \threeFthree \!\big\}$ wave along with the dominant $\omegaOctet \etaOctet \big\{\! \threePone \!\big\}$. We also include $\etaSinglet \etaOctet \big\{\! \oneFthree \!\big\}$ as the $\etaSinglet \etaOctet$ threshold is relatively low compared with the resonant region, such that the angular momentum barrier may not sufficiently suppress contributions from this higher partial-wave in the energy region of interest. Other possible $F$-waves, $\omegaOctet \omegaOctet \big\{\! \threeFthree \!\big\}$, $\omegaSinglet \omegaOctet \big\{\! \oneFthree, \threeFthree, \fiveFthree \!\big\}$ only generate additional solutions to Eq.~\ref{luescher} at somewhat higher energies and have relatively high-lying thresholds for which we expect the angular momentum suppression to be significant. In practice we will find that all the $3^{-(\!+\!)}$ partial waves we consider are modest over the energy range considered, with direct constraints coming from the computed $A_2^{-(\!+\!)}$ spectra. 

The $4^{-(\!+\!)}$ sector is populated only by partial-waves that are $F$-wave or higher, all of which we assume to be small enough as to be negligible, and none of which generate additional solutions of Eq.~\ref{luescher} in the energy region considered.

One partial wave with $1^{-(\!+\!)}$ is excluded on dynamical grounds: $\fOneSinglet \etaOctet \big\{\! \threeSone \!\big\}$ is observed to be completely decoupled from the other scattering channels when operator overlaps (as presented in Figure~\ref{SpectrumHistos}) are examined. This leads to a natural choice of energy cutoff at $a_tE_\cm = 0.48$, a modest distance below the $\fOneSinglet \etaOctet$ threshold, and we only use energies with no significant dependence on the \mbox{$\fOneSinglet \etaOctet$-like} operator. The levels to be used in constraining amplitudes are shown in black in Figs.~\ref{SpectrumHistos} and \ref{SpectrumA2}.

The contribution of the three vector-vector partial-waves, $\omegaSinglet \omegaOctet \big\{\! \onePone, \threePone, \fivePone \!\big\}$, which differ only in the total coupled intrinsic spin of the two vector mesons, to Eq.~\ref{luescher} requires some care. In the $[000]T_1^-$ irrep that we are considering, Eq.~\ref{luescher} is invariant under the interchange of any of these partial-waves, and it follows that the corresponding rows and columns of the $t$-matrix cannot be uniquely determined (see also Appendix~\ref{vvpwaves}). There is reason, from an approximate extension of Bose symmetry, to expect that only amplitudes featuring $\omegaSinglet \omegaOctet \big\{\!\threePone \!\}$ could be significant while those with $\omegaSinglet \omegaOctet \big\{\!\onePone, \fivePone \!\}$ will be very small.
The Wick contractions for diagrams featuring these channels differ only from those featuring $\omegaOctet \omegaOctet$ by the presence of the disconnected contribution to the $\omegaSinglet$, but this contribution is very small (reflected in the near degeneracy of $\omegaSinglet, \omegaOctet$). In practice we expect the $\omegaSinglet$ and $\omegaOctet$ to have almost identical spatial wavefunctions, and since $\omegaOctet \omegaOctet \big\{\!\onePone, \fivePone \!\}$ are forbidden by Bose symmetry, we anticipate that the corresponding $\omegaSinglet \omegaOctet$ amplitudes will be heavily suppressed. In fact we will observe that \emph{all} vector-vector amplitudes are found to be very small over the energy range considered. 

While the three-meson channel $\etaOctet \etaOctet \etaOctet$ becomes kinematically accessible at the upper end of the energy region we are considering, we do not include such partial waves. To couple to $J^{P(C)} = 1^{-(\!+\!)}$, this channel requires at least \emph{two} $P$-waves, and since our expected resonance lies barely above the $\etaOctet \etaOctet \etaOctet$ threshold, the angular momentum suppression implied is expected to render the partial waves irrelevant.

\vspace{0.3cm}

We now seek to use the 61 energy levels shown in black in Figures~\ref{SpectrumHistos} and~\ref{SpectrumA2} to constrain parameterizations of the $t$-matrix in the partial-wave basis presented in Table~\ref{partialWaves} by solving Eq.~\ref{luescher}. Solutions of Eq.~\ref{luescher} are only possible for $t$-matrix parameterizations which satisfy multi-channel unitarity. The simplest way to implement that constraint is to make use of the $K$-matrix, writing,
\begin{align*}
\big[t^{-1}(s)\big]_{\ell S J a, \ell' S' J b} &= 
\tfrac{1}{\left(2k_a\right)^\ell}
\big[K^{-1}(s)\big]_{\ell S J a, \ell' S' J b}
\tfrac{1}{\left(   2k_b \right)^{\ell'}} \nonumber \\
&\quad\quad\quad\quad+\delta_{\ell \ell'} \delta_{SS'}  \, I_{ab}(s),
\end{align*}
where $\mathbf{K}$ is a symmetric matrix taking real values on the real energy axis and $\mathbf{I}(s)$ is a diagonal matrix satisfying $\mathrm{Im} \,I_{ab}(s) = - \rho_a(s)$ above the threshold for channel $a$. The simplest choice is to set $\mathbf{I}(s) = -i \bm{\rho}(s)$, but other options may have better analytic properties below threshold and away from the real energy axis; for example, the Chew-Mandelstam prescription for which our implementation is described in Ref.~\cite{Dudek:2016cru}. The $K$-matrix is block-diagonal in $J$, reflecting the fact that total angular momentum is a good quantum number in infinite volume and only `mixes' in a finite volume, through the matrix $\bm{\mathcal{M}}$, due to the reduced symmetry of the lattice.

The presence in the spectrum of an additional level around $a_tE_\cm \sim 0.46$ and the lack of significant energy shifts at lower energies hints at a likely narrow resonance in the energy region around $a_tE_\cm \sim 0.46$. This is also consistent with the exotic \oneMP~octet level seen in Figure~\ref{CpSpectrum}.
The large overlap with axial-vector pseudoscalar meson-meson operators seen in Figure~\ref{SpectrumHistos} suggests significant coupling to these channels, whose thresholds lie just above the anticipated resonant region.

An efficient way to parameterize coupled-channel scattering when a narrow resonance appears is to use a \mbox{$K$-matrix} featuring an explicit pole. For the case of a single channel this form of parameterisation is closely related to the conventional Breit-Wigner and for coupled channels it is related to a multi-channel Breit-Wigner, sometimes referred to as a Flatt\'e amplitude in the two-channel case~\cite{Flatte:1976xv,Flatte:1976xu}. The $K$-matrix can also be straightforwardly augmented by the addition of a polynomial matrix in $s$, which in the simplest case can just be a constant matrix, that allows additional freedom beyond a pure resonance interpretation. This is crucial to test the robustness of scattering amplitudes and allow more flexible forms, as, for example, a pure pole parameterisation exhibits the phenomenon of ``trapped'' levels, where a single energy level is forced to appear between every pair of non-interacting energies -- see Appendix~\ref{squeezed_levels}.

In addition to varying the form of the $K$-matrix, the choice of $\mathbf{I}(s)$ may also be varied. The Chew-Mandelstam prescription improves the analytic continuation below thresholds, which is particularly useful here where, as discussed above, the axial-vector--pseudoscalar thresholds lie above the resonant region.

In this study, we will consider a variety of parameterisations, finding the best description of the finite-volume spectrum for each choice, ultimately leading to compatible results for the amplitudes and their resonant content. As we are only using rest-frame energy levels to determine the large coupled-channel scattering system (see Section~\ref{lattice}), we have less constraint than in previous calculations of simpler systems where in-flight spectra were computed~\cite{Dudek:2014qha,Wilson:2014cna,Wilson:2015dqa,Dudek:2016cru,Moir:2016srx,Briceno:2017qmb,Woss:2018irj,Woss:2019hse}.
However, the use of six volumes appears to provide enough information to isolate most of the important features.

\begin{figure}
	\centering
	\includegraphics[width=0.5\textwidth]{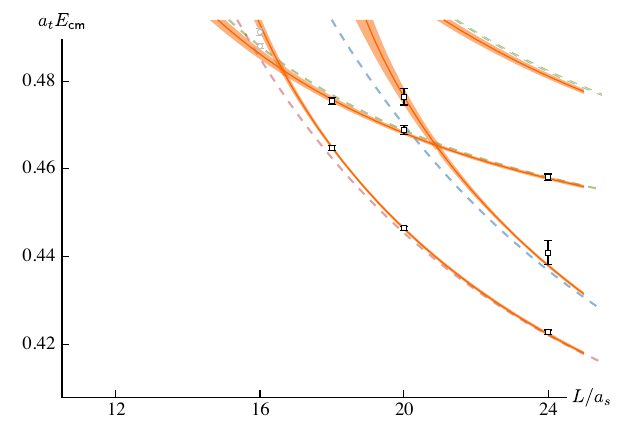}    
	\caption{As Figure~\ref{SpectrumA2} but including, as orange bands, the energy levels calculated from the amplitude in Eq.~\ref{Eq:KmatA2}. The thickness of the bands reflects the statistical uncertainty. The dashed curves show the non-interacting energy levels for $\omegaOctet \etaOctet$(red), $\etaSinglet \etaOctet$(blue) and $\omegaSinglet \omegaOctet$(green).  }
	\label{SpectrumA2orange}
\end{figure}

\vspace{1cm}
The $A_2^{-(\!+\!)}$ spectra allow us to determine the $J=3$ amplitudes which provide a `background' contribution to the $T_1^{-(\!+\!)}$ spectra. As discussed in Section~\ref{lattice}, there is no sign of any resonant behaviour associated with a $3^{-(\!+\!)}$ state in this energy region and the histograms in Figure~\ref{SpectrumA2} suggest a totally decoupled system. A reasonable form of parameterisation, capable of successfully describing the finite-volume spectra, is a diagonal constant $K$-matrix, 
\begin{align}\label{Eq:KmatA2}
\bm{K}_3(s) = \begin{bmatrix}
\gamma_{\etaSinglet \etaOctet \{ \oneFthree \}} & 0 								 & 0  \\
0							& \gamma_{\omegaOctet \etaOctet \{ \threeFthree \}} & 0 \\ 
0 							& 0 								 & \gamma_{\omegaSinglet \omegaOctet \{ \fivePthree \}}
\end{bmatrix} \, ,
\end{align}
where the Chew-Mandelstam prescription with subtraction at thresholds was used for $\mathbf{I}(s)$. 
The resulting fit describes the $A_2^{-(+)}$ finite-volume spectra with a ${\chi^2/N_{\text{dof}} = 2.53/(8-3) = 0.51}$, as shown in Figure~\ref{SpectrumA2orange}. Other parameterisations give a compatible set of amplitudes and quality of fit. The $3^{-(\!+\!)}$ amplitudes are modest over the entire energy range, with the $\etaSinglet \etaOctet$ and $\omegaOctet \etaOctet$ being mildly repulsive, and the $\omegaSinglet \omegaOctet$ being mildly attractive -- at $a_t E_\cm = 0.48$ the decoupled phase shifts reach only $-13(3)^\circ$, $-6(1)^\circ$ and $5(4)^\circ$ respectively.

%%%%%%%%%%%%%%%%%%%%%%%%%%%%%%%%%%%%%%%%%%%%%%%%%%%%%%%%%%
%%%%%%%%%%%%%%%%%%%%%%%%%%%%%%%%%%%%%%%%%%%%%%%%%%%%%%%%%%

\subsection{An illustrative $t$-matrix parameterization}

\begin{figure}[b]
	\centering
	\includegraphics[width=1.05 \columnwidth]{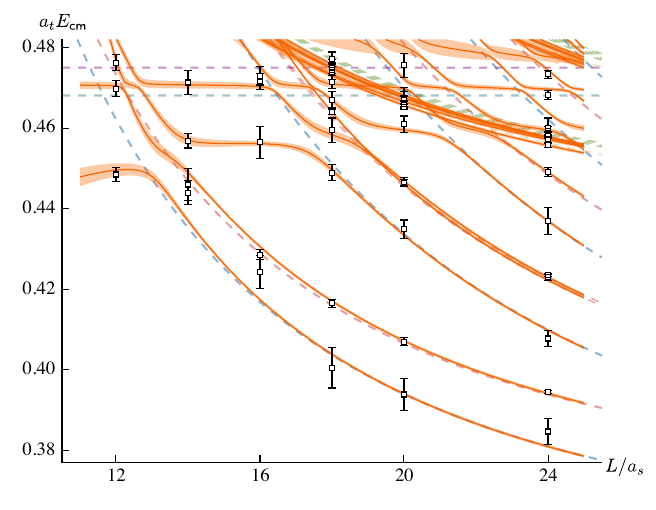}   
	\vspace{-10mm} 
	\caption{As Figure~\ref{SpectrumA2orange} but for the $T_1^{-(\!+\!)}$ irrep using the illustrative amplitude described in Eq.~\ref{Eq:Kmat}.}
	\label{SpectrumT1morange}
\end{figure}

We now consider the eight coupled-channel \oneMP~scattering system that features in $T_1^{-(\!+\!)}$. We will illustrate the scattering analysis using a single choice of amplitude parameterization, and later explore variations in that choice. The properties of the illustrative amplitude choice are motivated by the observations of the finite-volume spectra made in Section~\ref{lattice}. The four vector-vector channels appear to be decoupled for all considered energies and show no significant energy shifts, so in this parameterization we make the decoupling manifest, parameterizing the amplitudes with a diagonal $K$-matrix of constants,\footnote{In some of the parameterizations we will consider, vector-vector and non vector-vector channels are allowed to couple to each other though their coupling to the pole term.}
\begin{align*}
\bm{K}_{\scriptscriptstyle{VV}}(s) = \begin{bmatrix}
\gamma_{\omegaOctet \omegaOctet \{\!\threePone \! \} } & 0 								 & 0  & 0\\
0							& \gamma_{\omegaSinglet \omegaOctet \{\!\onePone \!\}}  & 0 & 0 \\ 
0 							& 0 								 & \gamma_{\omegaSinglet \omegaOctet \{\!\threePone \!\}}  & 0 \\
0 & 0 & 0 & \gamma_{\omegaSinglet\omegaOctet \{\!\fivePone \!\}}  
\end{bmatrix}\, .
\end{align*}
For the remaining four $1^{-(\!+\!)}$ channels, motivated by the likely presence of a narrow resonance, we parameterize the amplitudes using a `pole plus constant' form,
\begin{align*}
\bm{K}_{ \scriptscriptstyle{\cancel{VV}} }(s) = \frac{\bm{g}\, \bm{g}^T}{m^2 - s} + \begin{bmatrix}
\gamma_{\etaSinglet \etaOctet \{ \onePone \}} & 0 								 & 0  & 0\\
0							& \gamma_{\omegaOctet \etaOctet \{ \threePone \}} & 0  & 0\\ 
0&0&0&0\\
0&0&0&0
\end{bmatrix}\, ,
\end{align*}
where
\begin{align} \label{Kmatgs}
\bm{g} = \big(\, g_{\etaSinglet\etaOctet\{\onePone\}},g_{\omegaOctet\etaOctet\{\threePone\}},g_{\fOneOctet\etaOctet\{\threeSone\}},g_{\hOneOctet\etaOctet\{\threeSone\}} \,\big) \, ,
\end{align}
so that all four channels are coupled to the resonance as motivated by the histograms in Figure~\ref{SpectrumHistos}. We also add a constant term in the lowest two channels as the corresponding thresholds lie very low relative to the resonant region, and the close proximity of the energy levels with the non-interacting energies low down in the spectra suggested a region of non-resonant behavior (see the discussion in Section~\ref{lattice}).
We use the Chew-Mandelstam prescription for $\mathbf{I}(s)$ subtracting at the $K$-matrix pole mass ($s=m^2$). The eight-channel $1^{-(\!+\!)}$ $K$-matrix appears combined with the three-channel $3^{-(\!+\!)}$ $K$-matrix as given in Eq.~\ref{Eq:KmatA2},
\begin{align}   \label{Eq:Kmat}
\bm{K}(s) = \begin{bmatrix}
\bm{K}_{\scriptscriptstyle{VV}}(s) & 0  & 0  \\
0 & \bm{K}_{ \scriptscriptstyle{\cancel{VV}} }(s)& 0  \\ 
 0	& 0 &\bm{K}_3(s)
\end{bmatrix} \, ,
\end{align}
in the finite-volume spectrum condition, Eq.~\ref{luescher}. We minimise the $\chi^2$ by varying the 11 parameters in $\bm{K}_{\scriptscriptstyle{VV}}(s)$ and $\bm{K}_{ \scriptscriptstyle{\cancel{VV}} }$, with the parameters in $\bm{K}_3$ fixed according to the fit to the $A_2^{-(\!+\!)}$ lattice spectra. The resulting description of the $T_1^{-(\!+\!)}$ spectra gives a very reasonable $\chi^2/N_{\text{dof}}= 43.6 / (53 - 11) =  1.04$, shown in Figure~\ref{SpectrumT1morange}. 
%

%%%%%%%%%%%%%%%%%%%%%%%%%%%%%%%%%%%%%%%%%%%%%%%%%%%%%%%%%%
%%%%%%%%%%%%%%%%%%%%%%%%%%%%%%%%%%%%%%%%%%%%%%%%%%%%%%%%%%

\begin{figure*}
	\centering
	\includegraphics[width=0.9\textwidth]{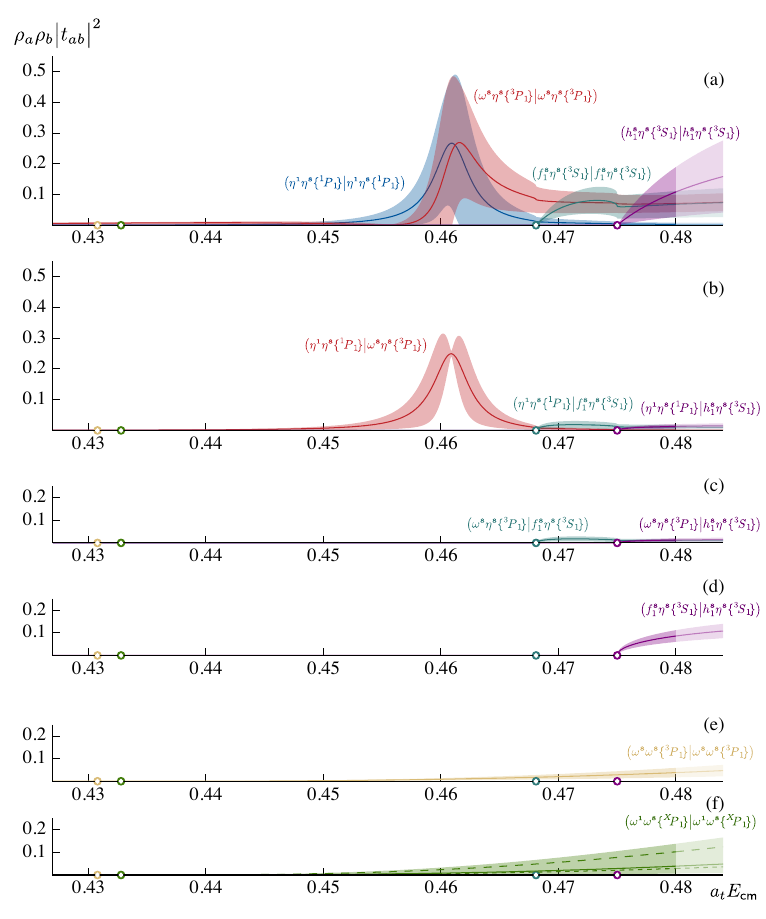}    
	\caption{\textbf{(a):} Diagonal $t$-matrix elements, plotted as $\rho_a\rho_b|t_{ab}|^2$, for the illustrative amplitude presented in Eq.~\ref{Eq:Kmat} for non vector-vector channels: $\etaSinglet \etaOctet \big\{ \!\onePone \!\big\}$, $\omegaOctet \etaOctet \big\{\! \threePone  \!\big\}$, $\fOneOctet \etaOctet \big\{\! \threeSone \!\big\}$ and $\hOneOctet \etaOctet \big\{\! \threeSone \!\big\}$. Shaded bands reflect statistical uncertainties on the scattering parameters. \textbf{(b), (c), (d):} As above, but for the off-diagonal amplitudes between the four channels displayed above. \textbf{(e), (f):} Diagonal vector-vector amplitudes: $\omegaOctet \omegaOctet \big\{\! \threePone \!\big\}$, $\omegaSinglet \omegaOctet \big\{\! \onePone, \threePone, \fivePone \!\big\}$. As discussed in the text, the $\omegaSinglet\omegaOctet$ partial-waves are indistinguishable and are combined in a single plot, labelled $\omegaSinglet\omegaOctet\{\XPone\}$, the shaded band reflecting an envelope over the statistical uncertainties for each partial-wave.}
	\label{refamps_rtsq}
\end{figure*}

\pagebreak
Plotting the resulting $t$-matrix elements as $\rho_a\rho_b|t_{ab}|^2$, shown in Figure~\ref{refamps_rtsq}, we can make a number of qualitative and quantitative observations. 

%%%%%%%%%%%%%%%%%%%%%%%%%%%%%%%%%%%%%%%%%%%%%%%%%%%%%%%%%%
%%%%%%%%%%%%%%%%%%%%%%%%%%%%%%%%%%%%%%%%%%%%%%%%%%%%%%%%%%

The diagonal amplitudes for the $\etaSinglet\etaOctet$, $\omegaOctet \etaOctet$, $\fOneOctet \etaOctet$, $\hOneOctet \etaOctet$ channels are shown in Figure~\ref{refamps_rtsq} (a) where a clear bump-like enhancement can be seen in the $\etaSinglet\etaOctet$ and $\omegaOctet\etaOctet$ channels at $a_tE_\cm \sim 0.46$, close to the mass
obtained using only single-meson operators (see Figure~\ref{CpSpectrum}). We observe a sharp turn-on of the axial-vector--pseudoscalar channels ($\fOneOctet \etaOctet$, $\hOneOctet \etaOctet$) at threshold, allowed for $S$-wave amplitudes.
The associated off-diagonal amplitudes are plotted in panels (b), (c), (d) of Figure~\ref{refamps_rtsq}. Here, we see again a bump-like enhancement in the $\etaSinglet\etaOctet \rightarrow \omegaOctet\etaOctet$ amplitude at ${a_tE_\cm \sim 0.46}$, with the other off-diagonal amplitudes being mostly small with the exception of the $\fOneOctet\etaOctet\rightarrow\hOneOctet\etaOctet$ amplitude which shows a modest rise from threshold.

The four decoupled vector-vector channels are presented in panels (e) and (f) of Figure~\ref{refamps_rtsq}. We observe that the single $\omegaOctet \omegaOctet$ amplitude, in the $\threePone$ partial-wave, is weak across the entire energy range, consistent with our observations from the finite-volume spectra in Sec~\ref{lattice}. For the $\omegaSinglet\omegaOctet$ amplitudes, we require four partial-waves, three $J^P=1^-$ $(\onePone,\,\threePone,\,\fivePone)$ and one $3^-$ $(\fivePthree)$, in order to obtain the correct number of finite-volume energies at the corresponding four-fold degenerate non-interacting energy. As discussed in Appendix~\ref{vvpwaves}, using only rest-frame energies does not uniquely constrain the three $1^-$ $\omegaSinglet\omegaOctet$ amplitudes and there is a freedom to permute these channels within the $t$-matrix. We therefore consider the envelope of these three amplitudes, as determined from the minimisation, as our best estimate for the size of the $\omegaSinglet\omegaOctet\{\XPone\}$ amplitudes. This is shown in Figure~\ref{refamps_rtsq} panel (f) where we see that they
are weak over the entire range, consistent with the observations made in Sec~\ref{lattice}. It is important to note that, as shown in Appendix~\ref{vvpwaves}, energy spectra obtained in moving-frame irreps modify the boundary conditions of the quantisation condition and \emph{do} distinguish the contributions of the $\{\onePone,\,\threePone,\,\fivePone\}$ partial-waves. 
As discussed in Sec~\ref{lattice}, we do not include moving-frame energy spectra owing to the appearance of the relatively low-lying positive-parity resonances,
as parities mix at non-zero momentum, and this would significantly complicate the analysis.

%%%%%%%%%%%%%
For this particular parameterisation, we also examine the effects of varying the stable hadron masses and anisotropy within their respective uncertainties, as given in Sec.~\ref{lattice}, to get an estimate of some of the systematic uncertainties on the amplitudes.
We adopt a conservative approach where we repeat the $\chi^2$ minimisation procedure using the extremal values $m_i \rightarrow m_i + \delta m_i$ and $\xi \rightarrow \xi - \delta \xi$, and vice-versa.
These combinations yield the largest deviations in the non-interacting energies, and therefore the largest shifts in the energy differences between the computed energy levels and the non-interacting values -- these ultimately constrain the scattering parameters. We anticipate that these combinations will therefore result in the largest changes in the scattering parameters and so yield a conservative estimate of the systematic uncertainties on the parameters from uncertainties in the hadron masses and anisotropy.

For the $J^{P}=3^-$ amplitudes, we find that varying the anisotropy yields the largest systematic uncertainties. The rather weak interactions in this system lead to small shifts in energies from their non-interacting values, as seen in Fig.~\ref{SpectrumA2orange}, which receive significant adjustment as the anisotropy is varied\footnote{This effect was observed previously in $\rho\pi$ isospin-2 scattering where the very small interactions meant that the systematic uncertainties dominated over the statistical ones~\cite{Woss:2018irj}.}. The quality of fits under these systematic variations also became rather poor: $\chi^2/N_{\text{dof}}=  2.26,$ for ${m_i \rightarrow m_i + \delta m_i}$ and ${\xi \rightarrow \xi - \delta \xi}$, and $\chi^2/N_{\text{dof}}=  4.82$ for ${m_i \rightarrow m_i - \delta m_i}$ and ${\xi \rightarrow \xi +\delta \xi}$, reflecting that even small discretization effects can be visible in weakly interacting systems where the energy levels have been determined with high statistical precision.
Nevertheless, we find all $J^{P}=3^-$ amplitudes remain small over the entire energy region considered.

Regarding the $J^P=1^-$ amplitudes, having fixed the (newly determined) $J^P=3^-$ parameters, we find the effects of varying the masses and anisotropy are much smaller relative to those for $J^P=3^-$, as expected in a more strongly interacting system. There are some modest variations in the amplitudes, but these are broadly within the statistical uncertainties and certainly within the differences we will see in the subsequent variation in the parameterisation. For example, we find the peak of the bump-like enhancements in the $\etaSinglet\etaOctet$ and $\omegaOctet\etaOctet$ amplitudes are consistent in height and only slightly displaced in energy (higher or lower  depending upon the sign of the systematic variations). This will be reflected in the position of a pole singularity of the $t$-matrix which varies at a level comparable to the statistical uncertainty.

A larger source of uncertainty arises when we consider varying the form of parameterisation, to which we now turn.

\subsection{Parameterization variations}
\begin{figure}
	\centering
	\includegraphics[width=\columnwidth]{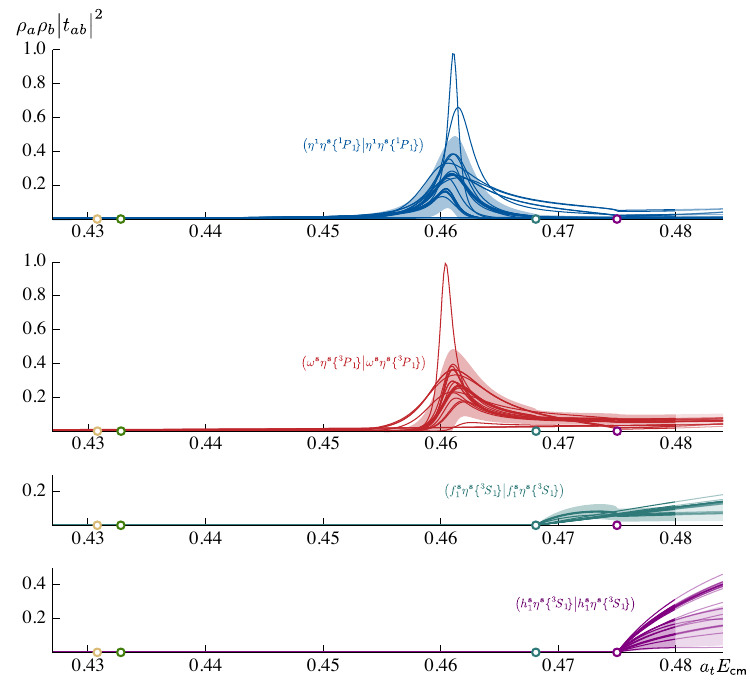}
	\vspace*{-8mm}
	\caption{Diagonal $t$-matrix elements, plotted as $\rho_a\rho_b|t_{ab}|^2$, for each parameterisation successfully describing the finite-volume spectra as discussed in the text, for non vector-vector channels: $\etaSinglet \etaOctet \big\{ \!\onePone \!\big\}$, $\omegaOctet \etaOctet \big\{\! \threePone  \!\big\}$, $\fOneOctet \etaOctet \big\{\! \threeSone \!\big\}$ and $\hOneOctet \etaOctet \big\{\! \threeSone \!\big\}$. Shaded bands reflect statistical uncertainties on the illustrative amplitude shown in Figure~\ref{refamps_rtsq}.}
			\vspace{1mm}
	\label{param_var_diag}
\end{figure}

\begin{figure}
	\centering
	\includegraphics[width=\columnwidth]{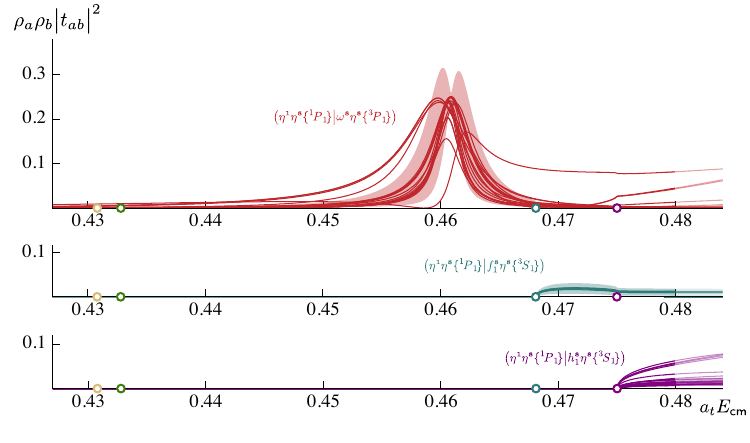}  
		\vspace*{-8mm}  
	\caption{As Figure~\ref{param_var_diag} but for off-diagonal $\etaSinglet\etaOctet \{\onePone\} \rightarrow \omegaOctet\etaOctet\{\threePone\}$, $\fOneOctet\etaOctet\{\threeSone\}$, $\hOneOctet\etaOctet\{\threeSone\}$ amplitudes.}
			\vspace{1mm}
	\label{param_var_offdiag_e8e1}
\end{figure}

\begin{figure}
	\centering
	\includegraphics[width=\columnwidth]{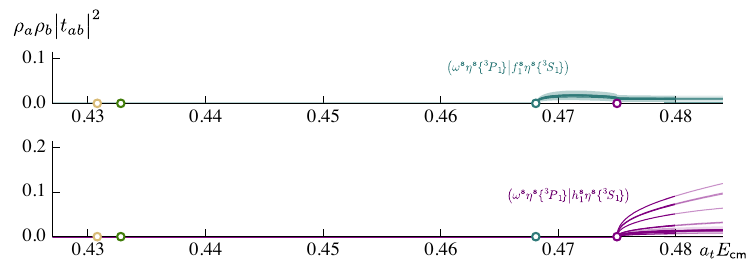}   
		\vspace*{-8mm} 
	\caption{As Figure~\ref{param_var_offdiag_e8e1} but for $\omegaOctet\etaOctet\{\threePone\}\rightarrow\fOneOctet\etaOctet\{\threeSone\}$, $\hOneOctet\etaOctet\{\threeSone\}$ amplitudes.}
		\vspace{1mm}
	\label{param_var_offdiag_e8o8}
\end{figure}

\begin{figure}
	\centering
	\includegraphics[width=\columnwidth]{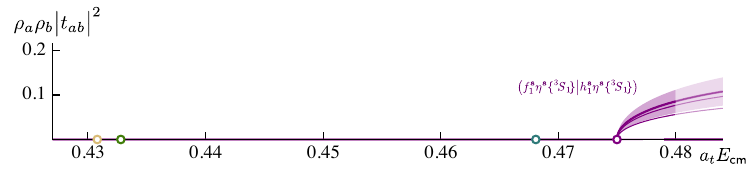}  
		\vspace*{-8mm}
	\caption{As Figure~\ref{param_var_offdiag_e8e1} but for the $\fOneOctet\etaOctet\{\threeSone\}\rightarrow\hOneOctet\etaOctet\{\threeSone\}$ amplitude.}
			\vspace{1mm}
	\label{param_var_h8e8_f8e8}
\end{figure}

\begin{figure}
	\centering
	\includegraphics[width=0.9\columnwidth]{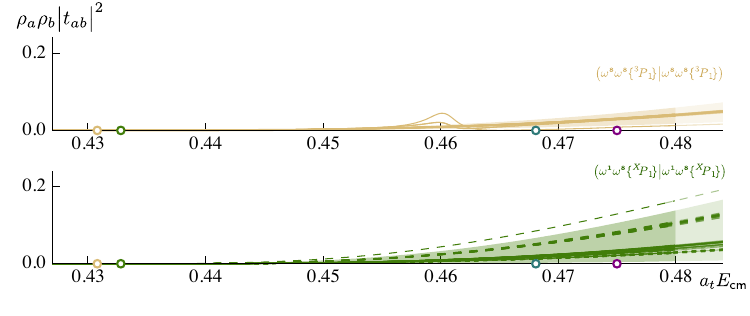}  
		\vspace*{-4mm}
	\caption{As Figure~\ref{param_var_diag} but for vector-vector channels: $\omegaOctet \omegaOctet \big\{\! \threePone \!\big\}$, $\omegaSinglet \omegaOctet \big\{\! \XPone \!\big\}$ amplitudes.}
			\vspace{1mm}
	\label{param_var_VV}
\end{figure}

In order to determine the extent to which the amplitude results presented in Figure~\ref{refamps_rtsq} are a unique description of the scattering system, we try a number of parameterizations, attempting to describe the finite-volume spectrum with each choice. Variations in the $K$-matrix include allowing energy-dependence in the numerator of the pole-term, and changes in the polynomial matrix added to the pole. The prescription used for $\bm{I}(s)$ is also adjusted, while maintaining coupled-channel unitarity in all parameterizations.
We retain 27 parameterizations\footnote{A full description of each of these parameterisations is provided in the Supplemental Material.} which are able to describe the finite-volume spectra with $\chi^2/N_{\text{dof}} \leq 1.25$, showing the resulting amplitudes in Figures~\ref{param_var_diag} --~\ref{param_var_VV}.

%%%%%%%%%%%%%%%%%%%%%%%%%%%%%%%%%%%%%%%%%%%%%%%%%%%%%%%%%
%%%%%%%%%%%%%%%%%%%%%%%%%%%%%%%%%%%%%%%%%%%%%%%%%%%%%%%%%%
%%%%%%%%%%%%%%%%%%%%%%%%%%%%%%%%%%%%%%%%%%%%%%%%%%%%%%%%%%
%%%%%%%%%%%%%%%%%%%%%%%%%%%%%%%%%%%%%%%%%%%%%%%%%%%%%%%%%%
%%%%%%%%%%%%%%%%%%%%%%%%%%%%%%%%%%%%%%%%%%%%%%%%%%%%%%%%%%
%%%%%%%%%%%%%%%%%%%%%%%%%%%%%%%%%%%%%%%%%%%%%%%%%%%%%%%%%%

For the diagonal amplitudes in the lowest two channels $\etaSinglet \etaOctet \{ \onePone \} $ and $\omegaOctet \etaOctet \{ \threePone \}  $, shown in Figure~\ref{param_var_diag}, we see a bump-like enhancement around $a_tE_\cm \sim 0.46$ for the majority of parameterisations, but we note that it is possible to describe our finite-volume spectra without seeing such a clear bump. We will revisit this observation when we examine the pole singularities of the $t$-matrix and the corresponding couplings. For the remaining two diagonal amplitudes in channels $\fOneOctet \etaOctet \{ \threeSone \}$ and $\hOneOctet\etaOctet \{ \threeSone \}  $, we observe that the relatively sharp turn-on at threshold is a quite general feature, with only the magnitude of the effect varying somewhat. That there should be some parameterization dependence here should not come as too much of a surprise given the relatively small number of finite-volume energy levels constraining the amplitudes above the axial-vector--pseudoscalar thresholds.

The off-diagonal amplitude, $\etaSinglet \etaOctet \{ \onePone \} \rightarrow \omegaOctet \etaOctet \{ \threePone \} $, shown in Figure~\ref{param_var_offdiag_e8e1}, typically features a bump-like enhancement around $a_tE_\cm\sim 0.46$, but as for the diagonal entries, it is possible to describe the spectra without such a bump and indeed without any coupling between these two channels. The remaining off-diagonal amplitudes remain modest under parameterization variation and are shown in Figures~\ref{param_var_offdiag_e8e1} --~\ref{param_var_h8e8_f8e8}.  

The vector-vector amplitudes shown in Figure~\ref{param_var_VV} have the same qualitative behavior as in the illustrative example presented previously. The small bump around $a_tE_\cm \sim 0.46$ for $\omegaOctet\omegaOctet\{\threePone\}\rightarrow\omegaOctet\omegaOctet\{\threePone\}$ on a small number of parameterisations reflects allowing freedom for this channel to couple to the $K$-matrix pole -- it is observed to be a very weak effect and is statistically compatible with zero.

Collectively, these parameterisation variations tell us that the limited number of rest-frame energy levels with which we are constraining the large number of coupled channels is not sufficient to \emph{completely uniquely} determine the $t$-matrix. Nevertheless, behavior consistent with a single resonant enhancement can typically be seen in the $\etaSinglet \etaOctet \{ \onePone \} $ and $\omegaOctet \etaOctet \{ \threePone \}  $ amplitudes. We will find that even those parameterisations that do not appear to show significant enhancement in either $\etaSinglet \etaOctet \{ \onePone \} $ or $\omegaOctet \etaOctet \{ \threePone \}  $ still feature a nearby resonance. The rapid turn-on of the axial-vector--pseudoscalar amplitudes will prove to be due to a large coupling of this resonance to one or both of these channels.

In order to demonstrate the presence of a resonance, we will now examine the amplitudes presented in this section at complex values of $s=E_\cm^2$ where a pole singularity is expected to feature.

%% file: sections/poles.tex
%%%%%%%%%%%%%%%%%%%%%%%%%%%%%%%%%%%%%%%%%%%%%%%%%%%%%%%%%%
\begin{figure*}[t]
	\centering
	\includegraphics[width=0.8\textwidth]{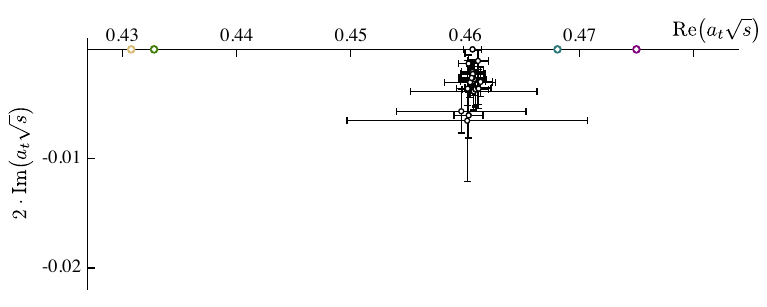}   
		\vspace*{-4mm}   
	\caption{Pole singularities on the proximal sheet for all successful parameterisations as described in the text. Error bars reflect the statistical uncertainties on the pole position for each parameterisation. \\ }
	\label{poles}
\end{figure*}
%%%%%%%%%%%%%%%%%%%%%%%%%%%%%%%%%%%%%%%%%%%%%%%%%%%%%%%%%%

At each meson-meson threshold, unitarity necessitates a branch-point singularity and the corresponding branch-cut divides the complex $s$-plane into two Riemann sheets. For the system we are considering, there are six relevant kinematic thresholds and hence a total of 64 Riemann sheets. The \emph{physical} sheet, the sheet on which physical scattering occurs just above the real energy axis, is identified by all scattering momenta having positive imaginary parts, i.e.~\smash{$\text{Im}(k^{(a)}_\cm)>0$} for all channels, $(a)$. Sheets with other sign combinations of the imaginary component of momenta are called \emph{unphysical}, and it is on these sheets where pole singularities corresponding to resonances are allowed to live as complex-conjugate pairs away from the real axis. 

In each energy region between thresholds, the unphysical sheet closest to the region of physical scattering, has \smash{$\text{Im}(k^{(a)}_\cm)<0$} for all kinematically open channels and \smash{$\text{Im}(k^{(a)}_\cm)>0$} for all kinematically closed channels. For convenience we will refer to this as the \emph{proximal} sheet, and a nearby pole singularity on the proximal sheet will have a significant impact on physical scattering. 

For brevity, sheets are labelled as an ordered list of six signs, where the order reflects increasing threshold energies ($\etaSinglet \etaOctet$, $\omegaOctet \etaOctet$, $\omegaOctet \omegaOctet$, $\omegaSinglet \omegaOctet$, $\fOneOctet \etaOctet$, $\hOneOctet \etaOctet$), and the sign reflects the imaginary component of momenta for that channel. For example~\smash{${[++++++]}$} represents the physical sheet, and~\smash{${[--++++]}$} represents the proximal sheet for scattering above the $\omegaOctet \etaOctet$ threshold, but below the $\omegaOctet \omegaOctet$ threshold.

The position of pole singularities can be related to conventional pictures of meson states. Poles on the real axis below the lowest threshold on the physical sheet correspond to stable \emph{bound states}, while poles in that location on unphysical sheets are \emph{virtual bound states} that do not appear as asymptotic particles. Poles off the real axis on unphysical sheets\footnote{Causality forbids poles on the physical sheet off the real axis, and any amplitudes featuring such singularities close enough to the real axis to have a non-negligible effect should be discarded as unphysical.} are associated with \emph{resonances}, and it is common to interpret the real and imaginary components of the pole position $s_0$ in terms of the mass $m_R$ and width $\Gamma_R$, via $\sqrt{s_0}=m_R \pm \tfrac{i}{2}\Gamma_R $. Near the pole, the $t$-matrix takes the form, 
\begin{equation*}
t_{\ell S J a, \ell' S' J b} \sim \frac{c_{\ell S J a}\,c_{\ell' S' J b}}{s_0-s}
\end{equation*}
where the factorized residues give access to $c_{\ell S J a}$, which are interpreted as complex-valued \emph{resonance couplings} for the channel $a$ in partial-wave $^{2S+1}\!\ell_J$.

The amplitudes presented in Sec.~\ref{scattering} suggest a likely resonance with real energy $a_t\sqrt{s} \sim 0.46$, in which case the proximal sheet is \smash{$[----++]$}. Indeed, for every parameterisation which successfully describes the finite-volume spectrum, we find a complex-conjugate pair of poles on the proximal sheet whose real energy is in the neighborhood of the anticipated mass and which has only a small imaginary energy\footnote{For a few parameterisations, the pole is found to lie \emph{on the real axis} below $\fOneOctet\etaOctet$ threshold. These parameterizations are those which decouple the resonance from the $\etaSinglet\etaOctet$, $\omegaOctet\etaOctet$, $\omegaOctet\omegaOctet$ and $\omegaSinglet\omegaOctet$ channels, such that the pole describes a stable bound state in a coupled $\fOneOctet\etaOctet$, $\hOneOctet\etaOctet$ system.}. For the illustrative amplitude given by Eq.~\ref{Eq:Kmat}, the poles on the proximal sheet lie at
\begin{equation}\label{ref_pole}
a_t\sqrt{s_0}_{\scriptscriptstyle{[----++]}} = 0.4609(12) \pm \tfrac{i}{2}0.0036(15) ,
\end{equation}
where the uncertainty is statistical. Based upon the variation of scattering hadron masses and anisotropy described in Sec.~\ref{scattering}, an additional conservative systematic error could be added of similar size to the statistical error.

For each of the parameterisations found to successfully describe the finite-volume spectrum, we show in Figure~\ref{poles} the proximal sheet pole position situated in the lower half-plane. In every case, the pole is found with a small imaginary component and hence is very close to the region of physical scattering, strongly influencing the amplitudes at real energies. As expected there are also `mirror poles' distributed across some of the remaining unphysical Riemann sheets, but these have a negligible effect on physical scattering by virtue of lying further away.

%%%%%%%%%%%%%%%%%%%%%%%%%%%%%%%%%%%%%%%%%%%%%%%%%%%%%%%%%%
\begin{figure*}
	\centering
	\includegraphics[width=0.8\textwidth]{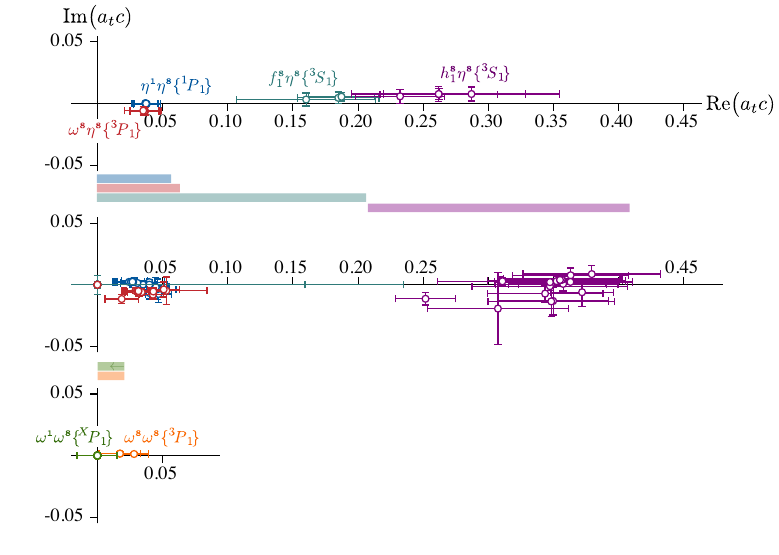}  
	\vspace*{-8mm}  
	\caption{Couplings corresponding to the pole singularities shown in Fig.~\ref{poles} as described in the text. Error bars reflect the statistical uncertainties on each coupling for each parameterisation. Shaded bars show ranges and upper limits on the couplings described in the text. \textbf{Top:} Couplings to non vector-vector channels for parameterisations where the $\fOneOctet\etaOctet\{\threeSone \}$ coupling was found to be significantly non-zero. \textbf{Middle:} As top but for parameterisations where the $\fOneOctet\etaOctet\{\threeSone \}$ coupling was found to be zero or fixed to be identically zero. \textbf{Bottom:} Couplings to vector-vector channels, $\omegaOctet \omegaOctet, \omegaSinglet \omegaOctet$.}
	\label{couplings}
\end{figure*}
%%%%%%%%%%%%%%%%%%%%%%%%%%%%%%%%%%%%%%%%%%%%%%%%%%%%%%%%%%

While it is clear that a nearby pole is required to describe the finite-volume spectra, the channel couplings which come from the factorized residues of this pole are not uniquely determined across different parameterizations. We find that the couplings of the pole to the $\etaSinglet\etaOctet\{ \onePone \}$ and $\omegaOctet\etaOctet\{ \threePone \}$ channels are small relative to a large value of the coupling to $\hOneOctet\etaOctet\{\threeSone \}$ and in some cases a large value of the coupling to $\fOneOctet\etaOctet\{\threeSone \}$.

Focusing on the axial-vector--pseudoscalar channels, we isolate two classes of results across our range of parameterization forms, one in which the coupling to $\fOneOctet\etaOctet\{\threeSone \}$ is large, of comparable size to a large coupling to $\hOneOctet\etaOctet\{\threeSone \}$, and a second in which the coupling to $\fOneOctet\etaOctet\{\threeSone \}$ is small. The couplings for these two classes are shown in the top and middle panels of Fig.~\ref{couplings}. Their sizes are governed largely by the corresponding $g$-parameters in the numerator of the pole term in the $K$-matrix, as given in Eq.~\ref{Kmatgs}. For a range of parameterisations that allow both of these $g$-parameters to freely vary, we find that the ratio of the corresponding couplings is of order one, with both found to be significantly non-zero -- these are shown in the top panel of Fig.~\ref{couplings}. 

We also find a number of parameterisations where the $\fOneOctet\etaOctet\{\threeSone\}$ coupling is negligibly small whilst the $\hOneOctet\etaOctet\{\threeSone\}$ coupling remains large, and parameterizations in which the coupling of the resonance to $\fOneOctet\etaOctet\{\threeSone\}$ is set to be exactly zero are also capable of describing the finite-volume spectra. This class of results are shown in the middle panel of Fig.~\ref{couplings}. 

Parameterizations in which the coupling of the resonance to the $\hOneOctet\etaOctet\{\threeSone\}$ channel is fixed to zero are found to be incapable of describing well the finite-volume spectra. They either have a poor $\chi^2$, or predict additional finite-volume energy levels that lie very close to our energy cutoff, levels for which there is no evidence in the lattice calculation.

The ambiguity in the relative size of the $\fOneOctet\etaOctet\{\threeSone \}$ and $\hOneOctet\etaOctet\{\threeSone \}$ couplings can be explained in terms of there being only a small gap between the relevant kinematic thresholds. These two channels both have the same partial-wave structure ($\threeSone$), so from the point of view of the finite-volume functions $\mathcal{M}$ in Eq.~\ref{luescher} they differ only in the mass difference between $\fOneOctet$ and $\hOneOctet$. If the $\fOneOctet$ and $\hOneOctet$ masses were degenerate, then the quantisation condition would be invariant under permutations of the $t$-matrix elements in these two channels, analogous to the indistinguishable vector-vector amplitudes we discuss in Appendix~\ref{vvpwaves}. It follows that we are only able to distinguish these channels by the mass splitting of the two axial-vector octets, and we explore the degree to which the finite-volume spectra are sensitive to different resonance couplings in a toy-model in Appendix~\ref{toy_model}. In this model, the scattering system is simplified to a two-channel ($\fOneOctet\etaOctet\{\threeSone \}$, $\hOneOctet\etaOctet\{\threeSone \}$) case with a bound-state pole lying below both thresholds. We find the finite-volume spectra in the rest-frame constrain very well the sum of the squared couplings, but offer relatively little constraint on the ratio of the coupling strengths. An energy level that is sensitive to the ratio lies between the two thresholds, but because the thresholds are so close together, this lever-arm is not large. 

In summary, while we can confidently state that the $\hOneOctet\etaOctet\{\threeSone \}$ coupling is large, the constraints from the finite-volume spectra can allow the $\fOneOctet\etaOctet\{\threeSone \}$ coupling to be as small as zero.

Examining the $\etaSinglet\etaOctet\{ \onePone \}$ coupling in Fig.~\ref{couplings}, we find this to be small compared with $\hOneOctet\etaOctet\{\threeSone \}$. There is a clear preference for a value close to 0.04, but there are parameterizations capable of describing the finite-volume spectra in which this coupling is set to be zero. The coupling to $\omegaOctet\etaOctet\{\threePone\}$ shows a very similar behavior.

Finally, for the vector-vector channels, we find the $\omegaOctet\omegaOctet\{\threePone\}$ coupling shows signs of being small but non-zero on some parameterisations, but again the finite-volume spectra can be equally well described with this coupling set to zero. The $\omegaSinglet\omegaOctet\{\XPone\}$ couplings are negligibly small on every parameterisation and again we find perfectly reasonable descriptions of the spectra when these are set to exactly zero.

Given this discussion, we summarize the behavior of the couplings in Figure~\ref{couplings} with the following best estimates, which we suggest are a conservative reflection of allowed ranges or limits taking into account statistical uncertainties and parameterization variations,
\begin{align}
|a_tc_{\etaSinglet\etaOctet \{ \onePone \}}| =&\, 0\rightarrow 0.055 \nonumber \\
|a_tc_{\omegaOctet\etaOctet\{ \threePone \}}| =&\, 0\rightarrow 0.060 \nonumber \\
|a_tc_{\omegaOctet\omegaOctet\{ \threePone \}}| =&\, 0\rightarrow 0.020 \nonumber \\
|a_tc_{\omegaSinglet\omegaOctet\{ \XPone \}}| \lesssim&\, 0.020 \nonumber \\
|a_tc_{\fOneOctet\etaOctet\{ \threeSone \}}| =&\, 0\rightarrow 0.21 \nonumber \\
|a_tc_{\hOneOctet\etaOctet\{ \threeSone \}}| =&\, 0.21\rightarrow 0.41\,. \label{best_couplings}
\end{align}
The upper limit for $|a_tc_{\omegaSinglet\omegaOctet\{ \XPone \}}|$ reflects the preferred zero value of this coupling, while the other couplings show evidence that they scatter around some non-zero value -- see Figure~\ref{couplings}. These ranges and upper limit are shown by the shaded bars in the figure.
Similarly, a best estimate of the pole position is given by
\begin{equation}
a_t \sqrt{s_0} = 0.4606(26) \pm \tfrac{i}{2} 0.0039(39). \label{best_pole}
\end{equation}
The small total width of the resonance, despite the large coupling to $\hOneOctet\etaOctet$, is explained by there being no phase space for this sub-threshold decay. 

The results presented in this section describe a very narrow exotic $1^{-(\!+\!)}$ resonance that appears in a version of QCD where the $u,d$ quarks are as heavy as the physical $s$ quark.
We will now discuss an interpretation of these results, aiming to provide a description of the $\pi_1$ resonance at the physical light-quark mass.

%% file: sections/interpret.tex
In this section we will discuss what can be learned from the observation of a $J^{P(C)} = 1^{-(\!+\!)}$ resonance at the $\SU(3)$ flavor point as presented above. As discussed in Section~\ref{SU3F}, we choose to focus our interpretation on the isovector member of the $\SU(3)$ octet, the $\pi_1$. We will attempt to infer possible properties of this resonance at the physical light-quark mass by performing a crude extrapolation, making use of the JPAC/COMPASS candidate state mass~\cite{Rodas:2018owy} to set the relevant decay phase-spaces. We will compare our results to existing predictions for hybrid meson decay properties made in models.

In order to present results in physical units, we must set the lattice scale using a physically measured quantity, an approach which is necessarily ambiguous, particularly given that we are far from the physical $u,d$ masses. As in previous publications, we choose to use the $\Omega$-baryon mass as a quantity which should not have a strong dependence on the $u,d$ quark masses. Calculated on the $L/a_s=16$ lattice, we find $a_tm_\Omega = 0.3593(7)$~\cite{Edwards:2012fx}, so that using the experimental mass, $1672.45(29) \text{ MeV}$~\cite{Zyla:2020zbs}, we obtain an inverse temporal lattice spacing $a_t^{-1}=4655 \text{ MeV}$. This scale setting yields stable hadron masses of
\begin{align*}
m({\etaOctet}) &= 688(1) \text{ MeV}\nonumber \\ 
m({\etaSinglet}) &= 939(5) \text{ MeV}\nonumber \\ 
m({\omegaOctet}) &=1003(1) \text{ MeV}\nonumber \\
m({\omegaSinglet}) &= 1012(1) \text{ MeV}\nonumber \\ 
m({\fOneOctet}) &= 1491(3) \text{ MeV}\nonumber \\ 
m({\hOneOctet}) &= 1523(3) \text{ MeV}.
\end{align*}
The $\etaOneOctet$ resonance pole described in the previous section when expressed in physical units has a mass, ${m_R = 2144(12)\text{ MeV}}$, and a width, ${\Gamma_R = 21(21) \text{ MeV}}$, and the couplings to meson-meson channels are
\begin{align} \label{couplings}
\big|c_{\etaSinglet\etaOctet \{ \onePone \}}\big| =& \, 0 \rightarrow 256 \text{ MeV}\nonumber \\
\big|c_{\omegaOctet\etaOctet\{ \threePone \}}\big| =& \, 0 \rightarrow 279 \text{ MeV}\nonumber \\
\big|c_{\omegaOctet\omegaOctet\{ \threePone \}}\big| =& \, 0 \rightarrow 93 \text{ MeV}\nonumber \\ 
\big|c_{\omegaSinglet\omegaOctet\{ \XPone \}}\big| \lesssim&\, 93 \text{ MeV}\nonumber \\
\big|c_{\fOneOctet\etaOctet\{ \threeSone \}}\big| =& \, 0 \rightarrow 978 \text{ MeV}\nonumber \\
\big|c_{\hOneOctet\etaOctet\{ \threeSone \}}\big| =& \, 978 \rightarrow 1909 \text{ MeV} \nonumber ,
\end{align}
where we have given an upper bound on the magnitude of $\omegaSinglet\omegaOctet\{ \XPone \}$ to acknowledge the preferred value of zero coupling to this channel.

These results can be viewed in the context of past predictions for the decays of hybrid mesons made within models. In both flux-tube breaking pictures and bag-models, decays to meson pairs in which one meson has $q\bar{q}$ in a $P$-wave and the other has $q\bar{q}$ in an $S$-wave
are enhanced over cases where both mesons have $q\bar{q}$ in an $S$-wave~\cite{Isgur:1985vy, Page:1998gz, Chanowitz:1982qj, Barnes:1982tx}.
In this particular case, that would suggest dominance of $\fOneOctet\etaOctet, \hOneOctet\etaOctet$ over $\etaSinglet\etaOctet, \omegaOctet\etaOctet,  \omegaOctet\omegaOctet, \omegaSinglet\omegaOctet$, which appears to be borne out in the couplings found in our QCD calculation.

%%%%%%%%%%%%%%%%%%%%%%%%%%%%%%%%%%%%%%%%%%%%%%%%%%%%%%%%%%
\begin{figure}[b]
	\centering
	\vspace*{-4mm}
	\includegraphics[width=0.75\columnwidth]{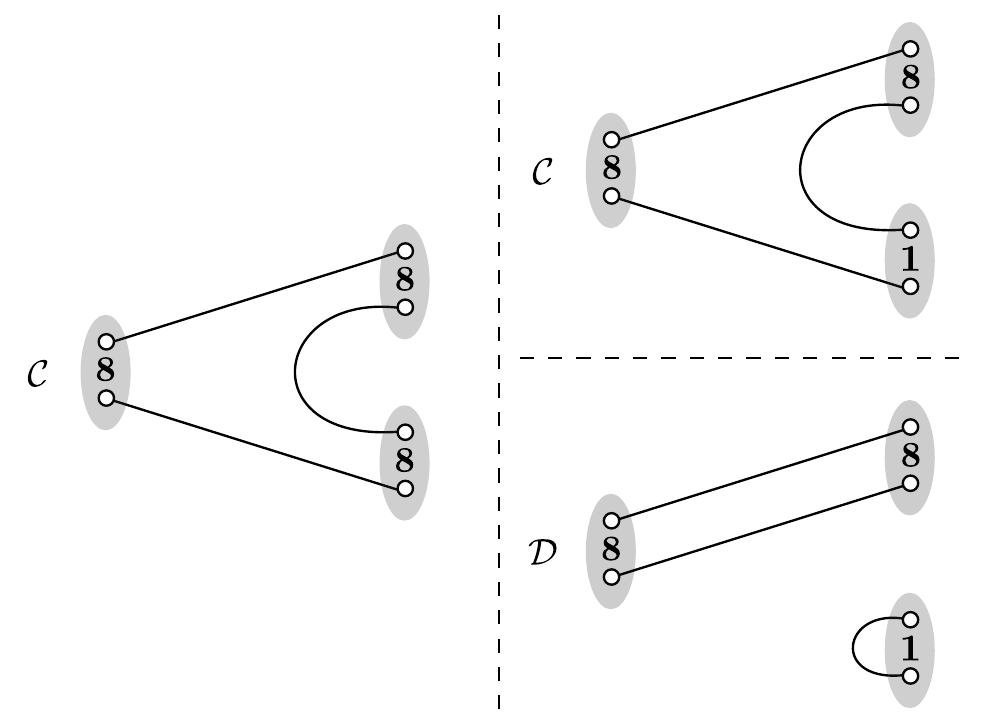}   
	\vspace*{-4mm} 
	\caption{Wick contraction topologies for $\bm{8} \rightarrow \bm{8} \otimes \bm{8} $ (left) and $\bm{8} \rightarrow \bm{8} \otimes \bm{1}$ (right).}
	\label{topologies}
\end{figure}
%%%%%%%%%%%%%%%%%%%%%%%%%%%%%%%%%%%%%%%%%%%%%%%%%%%%%%%%%%

We can explore some aspects of this observation by considering generic properties of correlation functions having a hybrid meson interpolator at the source and a meson-meson-like operator at the sink, following arguments along the lines of those given by Lipkin~\cite{Lipkin:1988um}, which were later placed in a limited field-theoretic framework by the ``Field Symmetrization Selection Rules'' (FSSR)~\cite{Page:1999ak}.
For the decay of an $\SUF$ octet into either an octet-octet pair or an octet-singlet pair, the possible Wick contractions are shown in Figure~\ref{topologies}. In the case of decays of a $1^{-(\!+\!)}$ octet to a pair of identical octet mesons, if the spin+space configuration of the meson-meson pair is antisymmetric, from Bose symmetry the flavor configuration must be antisymmetric, but this would have the wrong $C$-parity as discussed in Section~\ref{SU3F} and Appendix~\ref{bosesym}, and the correlation function is therefore zero.
Examples of such decays that are not allowed include $\etaOctet \etaOctet \{\onePone\}$ and $\omegaOctet \omegaOctet \{ \onePone, \fivePone \}$. A non-trivial implication in the $\SU(3)$ limit is for octet-singlet meson pairs. For example, in principle all of $\omegaSinglet \omegaOctet \{ \onePone, \threePone, \fivePone \}$ can have a non-zero coupling to the $\etaOneOctet$, but the fact that the disconnected contributions to the $\omegaSinglet$ are very small (see Section~\ref{SU3F}) renders the diagram $\mathcal{D}$ small, leaving only diagram $\mathcal{C}$. As the spatial $q\bar{q}$ wavefunction of the $\omegaSinglet$ is expected to be very similar to that of the $\omegaOctet$, we can anticipate that the antisymmetric combinations $\onePone$ and $\fivePone$ from diagram $\mathcal{C}$ will be small, while the symmetric combination $\threePone$ need not be suppressed. That the $\omegaSinglet \omegaOctet\{ \threePone \}$ and $\omegaOctet \omegaOctet\{ \threePone \}$ couplings prove to be small appears to be due to dynamics that go beyond simple symmetry arguments.

In the case of $\etaSinglet \etaOctet \{\onePone \}$, if the spatial $q\bar{q}$ wavefunctions of the $\etaSinglet$ and $\etaOctet$ were the same, diagram $\mathcal{C}$ would be zero owing to the antisymmetry of $\onePone$. In our calculation the optimized single-meson operators are constructed using the same fermion bilinear basis for both the octet and singlet, and we find that essentially the same optimal linear superposition is present for the $\etaSinglet$ and the $\etaOctet$, suggesting that they have similar spatial wavefunctions. However, even if diagram $\mathcal{C}$ is heavily suppressed, there remains diagram $\mathcal{D}$ which can be significant in this case owing to the large disconnected contribution to the $\etaSinglet$ (which generates the mass splitting between the $\etaOctet$ and the $\etaSinglet$).

%%%%%%%%%%%%%%%%%%%%%%%%%%%%%%%%%%%%%%%%%%%%%%%%%%%%%%%%%%%%%%%%%%%%%%%%%%%%%%%%%%%%%%%%%
\subsection{Flavor decomposition of the SU(3) amplitudes}
\label{sec:interp:decomp}

This is the first determination of the couplings of an exotic $J^{PC}$ resonance to its decay channels within a first principles approach to QCD, but of course it has been done with $u,d$ quarks that are much heavier than those in nature. In order to predict how this resonance would appear experimentally, we have to make a large extrapolation down to the physical light-quark mass. We will attempt this in a crude way, by assuming that the pole couplings are quark-mass independent except for a factor of the angular momentum barrier, $k^\ell$, evaluated at the resonance mass. To obtain this factor, and to determine the relevant phase-space, we require the mass of the $\pi_1$ at the physical light-quark mass. Given that we do not have a calculation of this, we use the experimental candidate mass, $1564$ MeV, found in the JPAC analysis of COMPASS data~\cite{Rodas:2018owy}, and we also consider a window of masses between $1500$ MeV and $1700$ MeV.

In order to extrapolate to the physical light-quark mass, we need to break the SU(3) flavor symmetry present in our calculation. We will retain isospin symmetry. Because the neutral flavorless mesons can now become admixtures of octet and singlet, we will have to introduce mixing angles, which we will take from phenomenological descriptions of experimental data. We will first break up the $\SU(3)$ octets into their component states, making use of the $\SU(3)$ Clebsch-Gordan coefficients provided in Ref.~\cite{deSwart:1963pdg}. As an example, for the decays of the $\pi_1^+$, the $I=1, I_z=+1$ member of the octet, into a vector-pseudoscalar pair we would have the combination,
\begin{equation*}
 \tfrac{1}{\sqrt{3}} \big( \pi^+ \rho^0 - \pi^0 \rho^+ \big) + \tfrac{1}{\sqrt{6}} \big( K^+ \overline{K}\vphantom{K}^{*0} - \overline{K}\vphantom{K}^0 K^{*+} \big),
\end{equation*}
such that the relevant couplings would be,
\begin{equation*}
\big| c(\pi_1 \!\to\! \pi \rho ) \big| = \sqrt{\tfrac{2}{3} } \big| c_{\omegaOctet \etaOctet} \big|, \; 
\big| c(\pi_1 \!\to\! K \overline{K}\vphantom{K}^* ) \big| = \sqrt{\tfrac{1}{3} } \big| c_{\omegaOctet \etaOctet} \big|,
\end{equation*}
where the additional factor of $\sqrt{2}$ reflects the desire to sum over all final state charge combinations when the decay rate is calculated.

It is these separated couplings which we attempt to extrapolate to the physical light-quark mass, by making the simple-minded assumption that each coupling is independent of the light-quark mass after appropriately rescaling the angular momentum barrier,
\begin{equation}\label{mom_scale}
\big|c\big|^{\text{phys}} = \Bigg|\frac{k^{\text{phys}}(m_R^{\text{phys}})}{k(m_R)}\Bigg|^{\ell} \, \big|c \big| .
\end{equation}

This approach is motivated by observations made in lattice calculations of the decays of $b_1 \to \omega \pi$ dominantly in $S$-wave~\cite{Woss:2019hse}, $\rho$ to $\pi \pi$ in $P$-wave~\cite{Wilson:2015dqa}, $K^*$ to $K\pi$ in \mbox{$P$-wave~\cite{Wilson:2019wfr}} and $f_2, f_2'$ decays to $\pi\pi$ and $K\overline{K}$ in $D$-wave~\cite{Briceno:2017qmb}, which appear to show quark-mass independence when treated this way. For example in the $b_1$ case, the coupling computed in \cite{Woss:2019hse} at \mbox{$m_\pi \sim 391$ MeV} is $|c| = 564(114)$ MeV, in good agreement with the coupling $|c|^\mathrm{phys} = 556(17)$ MeV extracted from the experimental $b_1$ decay width.
In the $P$-wave $\rho$ decay, an explicit factor of $k$ is required for the scaling to work, as presented in Ref.~\cite{Wilson:2015dqa}. In addition, as shown in Fig.~4 of Ref.~\cite{Wilson:2019wfr}, the $K^*$ coupling scaled in this way is approximately constant for four different light-quark masses corresponding to $m_\pi = 239$~MeV to 391 MeV, even when the $K^*$ is a shallow bound state, and is in agreement with the experimentally-measured coupling. Scaling the $f_2, f_2'$ \mbox{$D$-wave} couplings computed at $m_\pi \sim 391$ MeV in \cite{Briceno:2017qmb} gives, in comparison to values extracted from the Particle Data Group (PDG) review~\cite{Zyla:2020zbs},

\begin{tabular}{lcc}
& scaled & PDG \\[0.5ex]
$\big|c(f_2 \to \pi \pi)\big|$ 			& $488(28)$ & $453\substack{+9\\-4}$ \\[0.5ex]
$\big|c(f_2 \to K\overline{K})\big|$ 	& $139(27)$ & $132(7)$ \\[0.8ex]
$\big|c(f_2' \to \pi \pi )\big|$ 		& $103(32)$ & $33(4)$ \\[0.5ex]
$\big|c(f_2' \to K\overline{K})\big|$ 	& $321(50)$ & $389(12)$ \\[1.2ex]
\end{tabular}

\noindent
which is quite reasonable agreement given the large extrapolation in quark mass\footnote{An additional quark-model `form-factor' as part of the scaling is advocated by Close and Burns~\cite{Burns:2006wz}.}.

Using the couplings scaled to the physical quark mass, we can estimate partial widths for decay into kinematically open channels using the approach presented in the PDG review~\cite{Zyla:2020zbs} where the real part of the pole position is used to determine the phase-space in
\begin{equation}\label{pwid}
\Gamma(R\rightarrow i) = \frac{\big|c^{\text{phys}}_i\big|^2}{m^{\text{phys}}_R} \cdot \rho_i(m^{\text{phys}}_R).
\end{equation}
Summing up all non-zero partial widths, we can obtain an estimate for the total width\footnote{Note that doing this at the $\SU(3)$ point using the couplings in Eq.~\ref{couplings} gives a total width in the range $0 \to 45$ MeV, in reasonable agreement with our best estimate for the width from the pole position, $21(21)$ MeV.}. We will consider each constrained decay channel in turn, beginning with $\etaSinglet \etaOctet$.

%%%%%%%%%%%%%%%%%%%%%%%%%%%%%%%%%%%%%%%%%%%%%%%%%%%%%%%%%%%%%%%%%%%%%%%%%%%%%%%%%%%%%%%%%
\vspace{3mm} 
{\footnotesize $\bullet$} $\etaSinglet\etaOctet\{ \onePone\}$: For $\bm{1}\otimes\bm{8} \rightarrow \bm{8}$, we have only, trivially, $\eta_1 \pi^+$, where $\eta_1$ is the only member of the $\SU(3)$ singlet. Because $\etaOctet\etaOctet$ is forbidden in $\onePone$ by Bose symmetry, no components of the form $\eta_8 \pi^+$ can appear, where $\eta_8$ is the flavorless, neutral member of the octet. The $\eta_8$ and $\eta_1$ are related to the physical $\eta$ and $\eta^\prime$ states via a mixing angle $\theta_P$,
\begin{equation} \label{eta_etap}
\left( \begin{array}{c}
\eta_8 \\
\eta_1\end{array} \right) =
\left( \begin{array}{cc}
\cos \theta_P & \sin \theta_P \\
-\sin \theta_P & \cos \theta_P \end{array} \right) \left( \begin{array}{c}
\eta \\
\eta^\prime 
\end{array} \right),
\end{equation}
where phenomenological estimates for $\theta_P$ place it close to $-10^\circ$~\cite{Zyla:2020zbs,Feldmann:1999uf,Escribano:2007cd,Thomas:2007uy}.
The couplings of the $\pi_1$ to $\eta \pi$ and $\eta' \pi$ then follow,
\begin{align}
\big| c(\pi_1 \to \eta\pi) \big| =& \,\,\big| c_{\etaSinglet\etaOctet} \sin \theta_P \big|,\nonumber \\
\big| c(\pi_1 \to \eta^\prime\pi) \big|=& \,\, \big| c_{\etaSinglet\etaOctet} \cos \theta_P \big|\nonumber .
\end{align}
The relatively small mixing angle and the lack of coupling to Bose-forbidden $\etaOctet \etaOctet$ suggests that the $\eta' \pi$ coupling should be around six times larger than the coupling to $\eta \pi$, independent of the particular value of $c_{\etaSinglet\etaOctet}$.\footnote{Allowing a range $-10^\circ$ to $-20^\circ$ suggests an $\eta' \pi$ coupling three to six times the $\eta \pi$ coupling, in good agreement with a ratio of $3.0(3)$ suggested in the very recent analysis of COMPASS and Crystal Barrel data~\cite{Kopf:2020yoa}.}

%%%%%%%%%%%%%%%%%%%%%%%%%%%%%%%%%%%%%%%%%%%%%%%%%%%%%%%%%%%%%%%%%%%%%%%%%%%%%%%%%%%%%%%%%
\vspace{3mm}
{\footnotesize $\bullet$} $\omegaOctet\etaOctet\{ \threePone\}$: For the vector-pseudoscalar channel, the relevant flavor embedding is $\bm{8}\otimes\bm{8} \rightarrow \bm{8_2}$, and the components are
\begin{equation*}\label{o8e8_curr}
\tfrac{1}{\sqrt{3}} \big( \pi^+ \rho^0 - \pi^0 \rho^+ \big) + \tfrac{1}{\sqrt{6}} \big( K^+ \overline{K}\vphantom{K}^{*0} - \overline{K}\vphantom{K}^0 K^{*+} \big),
\end{equation*}
and the corresponding couplings, accounting for a sum over charge states to be done in the partial width calculation are
\begin{align*}
\big| c(\pi_1 \to \rho \pi ) \big| &= \sqrt{\tfrac{2}{3} } \big| c_{\omegaOctet \etaOctet} \big|, \\ 
\big| c(\pi_1 \to K^* \kbar ) \big| &= \sqrt{\tfrac{1}{3} } \big| c_{\omegaOctet \etaOctet} \big|.
\end{align*}
%

%%%%%%%%%%%%%%%%%%%%%%%%%%%%%%%%%%%%%%%%%%%%%%%%%%%%%%%%%%%%%%%%%%%%%%%%%%%%%%%%%%%%%%%%%
%
\vspace{3mm}
{\footnotesize $\bullet$} $\omegaOctet\omegaOctet\{ \threePone\}$, $\omegaSinglet\omegaOctet\{ \XPone\}$: The $\omegaOctet\omegaOctet$ and $\omegaSinglet\omegaOctet$ vector-vector channels must be considered together. Unlike the $\etaOctet\etaOctet$ channel forbidden in $\onePone$, the non-trivial spin coupling in $\omegaOctet\omegaOctet$ means that the $\threePone$ is in a totally symmetric configuration and thus not forbidden -- see Appendix~\ref{bosesym}. This means the corresponding components for $\omegaOctet\omegaOctet$ and $\omegaSinglet\omegaOctet$ in $\threePone$ both feature $\rho \omega$ and $\rho \phi$. For $\bm{8}\otimes\bm{8} \rightarrow \bm{8_1}$, the  $\omegaOctet\omegaOctet$ components are
\begin{align}\label{o8o8_curr}
&-\sqrt{\tfrac{3}{10}}  \big(K^{*+} \overline{K}\vphantom{K}^{*0} +  \overline{K}\vphantom{K}^{*0} K^{*+}\big)
+\tfrac{1}{\sqrt{5}} \big(\rho^+\omega_8 + \omega_8\rho^+ \big) \nonumber \\
=\,\, &-2\sqrt{\tfrac{3}{10}} \, K^{*+} \overline{K}\vphantom{K}^{*0} +2\sqrt{\tfrac{1}{5}} \,\omega_8\rho^+ \nonumber ,
\end{align}
and trivially the only component of $\omegaSinglet\omegaOctet$ is $\omega_1 \rho^+$. 

The $\omega_8$, $\omega_1$ mixing to give $\omega$, $\phi$ is well known to be very different to the pseudoscalar case, with the $\omega$ being dominantly $\tfrac{1}{\sqrt{2}}\big( u\bar{u} + d\bar{d}\big)$ and $\phi$ dominantly $s\bar{s}$. Using the same conventions as Eq.~\eqref{eta_etap} with $\eta \to \omega$, $\eta' \to \phi$, this `ideal' mixing would correspond to a mixing angle of ${\theta_V \approx -54.7^\circ}$. A mixing angle of $\theta_V \sim -52^\circ$, extracted from a model fit describing experimental vector to pseudoscalar radiative transitions~\cite{Escribano:2007cd}, is in good agreement with this (see also \cite{Zyla:2020zbs}).
It follows that the $\omega\rho$, $\phi\rho$ couplings for $\threePone$ are,
\begin{align}
\big| c( \pi_1 \to \omega\rho\{\threePone\}) \big| = \bigg| &2\sqrt{\tfrac{1}{5}} \big|c_{\omegaOctet\omegaOctet\{\threePone\}} \big| \cos \theta_V  \nonumber \\
&\quad\; -  \big|c_{\omegaSinglet\omegaOctet\{\threePone\}} \big| \sin \theta_V\bigg|\nonumber \\
\big| c( \pi_1 \to \phi\rho\{\threePone\}) \big| = \bigg| &2\sqrt{\tfrac{1}{5}} \big|c_{\omegaOctet\omegaOctet\{\threePone\}} \big| \sin \theta_V \nonumber \\
&\quad\; +  \big| c_{\omegaSinglet\omegaOctet\{\threePone\}} \big| \cos \theta_V\bigg| .   \nonumber 
\end{align}
These expressions are consistent with the expectations of the OZI rule: if the disconnected diagram, $\mathcal{D}$ in Fig.~\ref{topologies}, vanishes and $\omega$,$\phi$ mixing is ideal, $c( \pi_1 \to \phi\rho\{\threePone\})=0$ and ${c_{\omegaSinglet\omegaOctet\{\threePone\}} = \sqrt{\tfrac{8}{5}} c_{\omegaOctet\omegaOctet\{\threePone\}}}$.
The coupling to kaons is
\begin{align}
\big| c( \pi_1 \to K^* \kbar\vphantom{K}^{*} \{\threePone\}) \big| = 2\sqrt{\tfrac{3}{10}} \big|c_{\omegaOctet\omegaOctet\{\threePone\}} \big|. \nonumber 
\end{align}

For $\onePone$ and $\fivePone$, $\omegaOctet\omegaOctet$ is forbidden by Bose symmetry and the only contribution comes from the $\omegaSinglet\omegaOctet$. The corresponding couplings are therefore,
\begin{align}
\big| c( \pi_1 \to \omega\rho\{\onePone,\fivePone\}) \big| =& \,\, \big|  c_{\omegaSinglet\omegaOctet\{\onePone,\fivePone\}} \sin \theta_V \big| \nonumber \\
\big| c( \pi_1 \to \phi\rho\{\onePone,\fivePone\}) \big| =& \,\, \big|  c_{\omegaSinglet\omegaOctet\{\onePone,\fivePone\}} \cos \theta_V \big| . \nonumber
\end{align}
These couplings are expected to be very small because only the disconnected diagram contributes to these decays.

%%%%%%%%%%%%%%%%%%%%%%%%%%%%%%%%%%%%%%%%%%%%%%%%%%%%%%%%%%%%%%%%%%%%%%%%%%%%%%%%%%%%%%%%%%%%%
\vspace{3mm}
{\footnotesize $\bullet$} $\fOneOctet\etaOctet\{\threeSone\}$, $\hOneOctet\etaOctet\{\threeSone\}$: Similar to $\omegaOctet\omegaOctet$, $\fOneOctet\etaOctet$ embeds in $\bm{8}_1$ and decomposes into,
\begin{align}
 -\sqrt{\tfrac{3}{10}}\big(K_{1A}^{+}\,\overline{K}\vphantom{K}^0 + \overline{K} \vphantom{K}_{1A}^0\,K^+\big)+\tfrac{1}{\sqrt{5}}\big(a_1^+\eta_8 + (f_1)_8\pi^+ \big), \nonumber
\end{align}
where we see the neutral, flavorless members of the pseudoscalar and $1^{+(\!+\!)}$ octets, the $\eta_8$ and $(f_1)_8$, and the strange members of the $1^{+(\!+\!)}$ octet, $K_{1A}$.
We have not included the $\fOneSinglet\etaOctet$ channel in the scattering calculation, given that this was largely decoupled in our observations of the finite-volume spectra in Sec.~\ref{lattice}, and we therefore assume here that the $\fOneSinglet\etaOctet$ coupling is zero. 

The mixing of $(f_1)_8$ and $(f_1)_1$ to form the physical states $f_1(1285)$ and $f_1(1420)$ can be determined from the radiative decays of the $f_1(1285)$ to $\gamma\rho$ and $\gamma\phi$, which suggests a mixing angle of $\theta_A \sim -34^\circ$, following the formalism presented in~\cite{Close:1997nm}, using the PDG averages~\cite{Zyla:2020zbs}, and using the same conventions as Eq.~\eqref{eta_etap} with ${\eta \to f_1(1285)}$, ${\eta' \to f_1(1420)}$ (see also Ref.~\cite{Close:2015rza}).
The corresponding couplings in decays involving the non-strange $1^{+(\!+\!)}$ mesons are 
\begin{align*}
\big| c(\pi_1 \to a_1\eta) \big| =& \,\,\tfrac{1}{\sqrt{5}} \big|  c_{\fOneOctet\etaOctet} \cos \theta_P \big| \nonumber \\
\big| c( \pi_1 \to a_1\eta^\prime) \big| =& \,\,\tfrac{1}{\sqrt{5}} \big|  c_{\fOneOctet\etaOctet} \sin \theta_P \big| \nonumber \\
\big| c(\pi_1 \to f_1(1285)\pi) \big| =& \,\,\tfrac{1}{\sqrt{5}} \big|  c_{\fOneOctet\etaOctet} \cos \theta_A \big| \nonumber \\
\big| c( \pi_1 \to f_1(1420)\pi) \big| =& \,\,\tfrac{1}{\sqrt{5}} \big|  c_{\fOneOctet\etaOctet} \sin \theta_A \big|. \end{align*}

The other axial-vector--pseudoscalar channel, $\hOneOctet\etaOctet$, embeds in $\bm{8_2}$ and has components,
\begin{equation}\label{h8e8_curr}
\tfrac{1}{\sqrt{6}} \big(K_{1B}^{+}\,\overline{K}\vphantom{K}^0 - \overline{K}\vphantom{K}_{1B}^0\,K^+\big)+\tfrac{1}{\sqrt{3}}\big(b_1^+\pi^0 - b_1^0\,\pi^+\big), \nonumber 
\end{equation}
where $K_{1B}$ are the strange members of the $1^{+(\!-\!)}$ octet. The coupling to $b_1 \pi$ is then
\begin{equation*}
\big| c( \pi_1 \to b_1\pi) \big| = \,\,{\sqrt{\tfrac{2}{3}}}\, \big|c_{\hOneOctet\etaOctet} \big|.
\end{equation*}

The physical axial-vector kaons, the $K_1(1270)$ and $K_1(1400)$, are not eigenstates of charge-conjugation and can be considered to be admixtures of the $K_{1A}$ from the $1^{+(\!+\!)}$ octet and the $K_{1B}$ from the $1^{+(\!-\!)}$ octet. This mixing, in terms of an angle $\theta_K$, can be defined through
\begin{equation} \label{Kf_Kh2} 
\left( \begin{array}{c}
K_{1B} \\
K_{1A} \end{array} \right) =
\left( \begin{array}{cc}
\cos \theta_K & -\sin \theta_K \\
\sin \theta_K & \cos \theta_K \end{array} \right) \left( \begin{array}{c}
K_1(1270) \\
K_1(1400) 
\end{array} \right) ,
\end{equation}
which is consistent with the conventions in Ref.~\cite{Barnes:2002mu}.
There is not a clear consensus on the value of $\theta_K$, but it could be as large as $\sim 45^\circ$.
In practice there is only dependence on this mixing angle if the decay to the $K_1(1270) \kbar$ channel is open -- this requires the $\pi_1$ to have a mass above $1747$ MeV, significantly heavier than the JPAC/COMPASS candidate.

%%%%%%%%%%%%%%%%%%%%%%%%%%%%%%%%%%%%%%%%%%%%%%%%%%%%%%%%%%%%%%%%%%%%%%%%%%%%%%%%%%%%%%%%%
\subsection{Partial widths for a $\pi_1(1564)$}
\label{sec:interp:pws}

Combining the flavor decompositions in the previous section with the scaling given by Eq.~\ref{mom_scale} we obtain the couplings for a $1564$ MeV $\pi_1$ presented in Table~\ref{partialwidths}.\footnote{For the $\omega\rho\{\threePone\}$ and $\phi\rho\{\threePone\}$ momentum scaling, where there is a linear combination of two $\SU(3)$ couplings, we evaluate the momentum at the $\SU(3)$ point with $m_1=m_2=m_{\omegaOctet}$ in both cases as the mass difference between the $\omegaOctet$ and $\omegaSinglet$ is negligibly small and it simplifies the resulting algebra.} Using these couplings, we populate Table~\ref{partialwidths} with partial widths determined using Eq.~\ref{pwid}. We assume that the subsequent decays of unstable isobars (e.g.\ $\rho$, $b_1$) factorize from the initial $\pi_1$ decays given in the table.

\begin{table}[b]
	{\renewcommand{\arraystretch}{1.5}
		\begin{tabular}{r |c | c| c }
			 & thr./MeV & $\big| c^{\text{phys}}_i \big|$/MeV& $\Gamma_i$/MeV   \\[1ex]
			\hline
			$\eta \pi$ 				& 688  & $0\rightarrow 43$ & $0\rightarrow 1$  \\
			$\rho\pi$				&	910	& $0\rightarrow 203$ & $0\rightarrow 20$ \\ 
			$\eta^\prime \pi$		& 1098	& $0\rightarrow 173$ & $0\rightarrow 12$ \\
			$b_1\pi$				&	1375	&  $799\rightarrow 1559$ & $139 \rightarrow 529$ \\ 
			$K^* \kbar$		&	1386	& $0\rightarrow 87$ & $0\rightarrow 2$ \\ 
			$f_1(1285)\pi$			&	1425	& $0\rightarrow 363$ & $0\rightarrow 24 $\\
			$\rho\omega \{ \onePone\}$ 		&	1552	& $\lesssim 19$ & $\lesssim 0.03$ \\
			$\rho\omega \{ \threePone\}$	&	1552	& $\lesssim 32$ & $\lesssim 0.09$\\
			$\rho\omega \{ \fivePone\}$ 	&	1552	& $\lesssim 19$ & $\lesssim 0.03$\\
			$f_1(1420)\pi$			&	1560	& $0\rightarrow 245$ & $0\rightarrow 2$						\\
			\hline
			\multicolumn{4}{r}{$\Gamma = \sum_i \Gamma_i = 139\rightarrow 590$} \\
			\hline
		\end{tabular}
	}
	\caption{Thresholds, couplings and partial widths for each channel kinematically open at $m_R = 1564 \text{ MeV}$. Couplings are derived as discussed in the text and partial widths are determined according to the definition given in Eq.~\ref{pwid}. For both couplings and partial widths we present a range calculated from the corresponding $\SU(3)$ couplings, while those shown as an upper bound have a preferred value of zero.}
	\label{partialwidths}
\end{table}

It is clear that the dominant decay mode is $b_1\pi$, with the next largest channels, $\eta' \pi, \rho\pi$ and $f_1(1285)\pi$ being significantly smaller. Despite the larger phase space, the partial width into $\eta \pi$ is approximately ten times smaller than $\eta'\pi$, independent of the coupling and depending only on the mixing angle and phase space. Only one kaonic decay mode is kinematically accessible, $K^* \kbar$, with a very small partial width. Decays to $\rho\omega$ are negligible. Summing all partial widths we obtain an estimate for the total width in the range $139$ to $590$ MeV which includes the value $492(47)(102)$~MeV found in the JPAC/COMPASS analysis\footnote{and the somewhat smaller value $\sim 388$ MeV found in the very recent analysis of COMPASS and Crystal Barrel data~\cite{Kopf:2020yoa}.}.
If our extrapolation is accurate, it suggests that the observation of the $\pi_1$ in $\eta \pi$ and $\eta' \pi$ is through decays which are very far from being the dominant decay modes.

It is possible that this estimate of the total decay width may be missing contributions from channels which are closed at the $\SU(3)$ point, whose couplings we have not determined, but which become open at physical kinematics. Examples might include $f_2 \pi$ (although this is a $D$-wave decay with relatively little phase-space, so a large width is unlikely), or $\eta(1295) \pi$ (a $P$-wave decay with a very small phase-space). Any truly multibody decays to three or more mesons, i.e.\ those not proceeding through a resonant isobar, are also not included in this estimate, but the conventional wisdom is that such decays are not large.

\begin{figure*}
\centering
\includegraphics[width=0.8\textwidth]{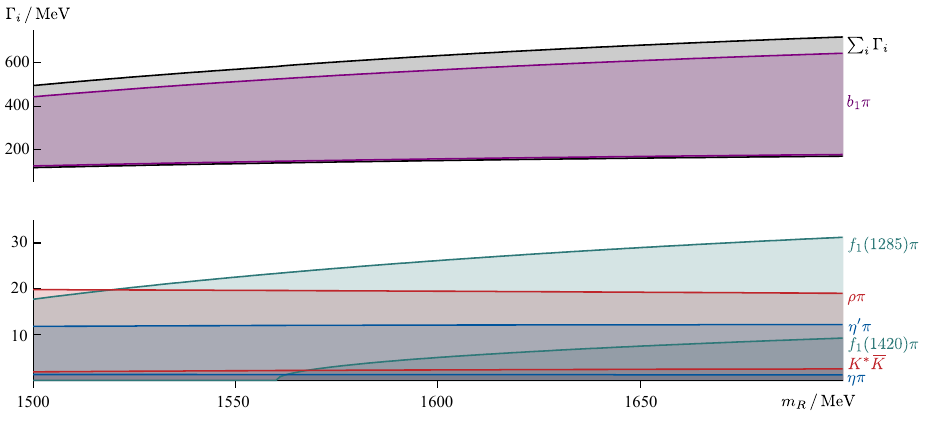}
\vspace*{-4mm}
\caption{Partial widths as a function of the $\pi_1$ pole mass. The bands reflect the coupling ranges given in Table~\ref{partialwidths}. The total width, obtained by summing the partial widths, is shown by the grey band.}
\label{mR_var}
\end{figure*}

Figure~\ref{mR_var} shows the partial widths for each channel in Table~\ref{partialwidths} as a function of the physical resonance mass, $m_R^{\text{phys}}$, allowed to vary in the range $1500-1700\text{ MeV}$. We observe only a modest dependence upon the mass of the $\pi_1$ resonance, with the exception of the $f_1(1420) \pi$ channel which becomes kinematically open in this energy range.

%%%%%%%%%%%%%%%%%%%%%%%%%%%%%%%%%%%%%%%%%%%%%%%%%%%%%%%%%%
%%%%%%%%%%%%%%%%%%%%%%%%%%%%%%%%%%%%%%%%%%%%%%%%%%%%%%%%%%

The only prior estimate of decay rates for a $\pi_1$ obtained using lattice QCD was the calculation presented in Ref.~\cite{McNeile:2006bz} which used a rather different approach to the one followed in this paper. By tuning the value of the light-quark mass in a two-flavor calculation (without strange quarks), the authors were able to make the mass of the $\pi_1$ be approximately equal to the sum of the masses of the $\pi$ and the $b_1$. They argued that the time-dependence of a single two-point function having a $1^{-(\!+\!)}$ single-meson operator at the source and a $b_1 \pi$-like operator at the sink can be used to infer a transition rate. The method makes a number of assumptions that have not yet been validated, but their result for pion masses near 500 MeV does suggest a large coupling. They also found a somewhat smaller coupling to $f_1 \pi$.

We can also compare our result extrapolated to physical kinematics with the predictions of models. Models based upon breaking of the flux-tube~\cite{Isgur:1985vy, Page:1998gz} do not allow decays to identical mesons, but these are typically prevented by Bose symmetry anyway. The ability of these models to predict decays involving the $\eta$ or $\eta'$ is somewhat questionable given that no disconnected contributions are considered. Within these models, the quark spin coupling factorizes from the spatial matrix element such that $\rho \pi$ decays are only allowed to the extent that the spatial $q\bar{q}$ wavefunctions of the $\pi$ and the $\rho$ differ. This difference is quite hard to estimate in quark models where the very light pseudo-Goldstone boson $\pi$ is typically not well described.

If this model picture of the coupling being sensitive to the difference between the $\pi$ and $\rho$ radial wavefunction is correct, our simple extrapolation of the $\rho\pi$ coupling may lead to an under estimate. We can use the charge radius as a guide to the wavefunction size, and at the $\SU(3)$ flavor symmetric point
these radii were computed in Ref.~\cite{Shultz:2015pfa}: $\langle r^2 \rangle_\pi^{1/2}$ = 0.47(6) fm, $\langle r^2\rangle_\rho^{1/2}$ = 0.55(5) fm. These sizes are not that different, as one might expect given the heaviness of the quarks, but we expect the difference to grow as the light-quark mass reduces. Our simple extrapolation of the $\rho \pi$ coupling would not capture this change, and hence our $\rho \pi$ partial width might be an under-estimate.

The flux-tube breaking models have larger couplings to axial-vector--pseudoscalar channels like $b_1 \pi$ and $ f_1 \pi$ than to, for example, $\rho \pi$, but these couplings are still much smaller than the ones we are predicting. Bag models show similar decay systematics~\cite{Chanowitz:1982qj,Barnes:1982tx}.

%% file: sections/summary.tex
Prior lattice QCD calculations which treated excited hadrons as stable particles indicated the presence of \emph{exotic hybrid mesons} in the spectrum, but until now the only theoretical information on the decay properties of these states came from models whose connection to QCD is not always clear. In this paper we presented the first  determination of the lightest $J^{P(C)}=1^{-(\!+\!)}$ \emph{resonance} within lattice QCD. The resonance was observed in a rigorous way as a pole singularity in a coupled-channel scattering amplitude obtained using constraints provided by the discrete spectrum of eigenstates of QCD in six different finite volumes. These spectra were extracted from matrices of correlation functions computed in lattice QCD using a large basis of operators.

In order to make this first calculation practical we opted to work with quark masses such that $m_u=m_d=m_s$, with the quark mass selected to approximately match the physical strange-quark mass. The resulting $\SUF$ symmetry leads to a simplified set of decay channels, and the relatively heavy quark mass means that only meson-meson decays are kinematically accessible in the energy region of interest. 

The computed lattice QCD spectra are described by an eight-channel flavor-octet $1^{-(\!+\!)}$ scattering system in which a \emph{narrow} resonance appears, lying slightly below the opening of axial-vector--pseudoscalar decay channels, but well above pseudoscalar--pseudoscalar, vector--pseudoscalar and vector--vector decay thresholds. The resonance pole shows relatively weak couplings to the open channels, hence the narrow width, but large couplings to at least one kinematically-closed axial-vector--pseudoscalar channel.

A simple-minded approach was used to predict decay properties of a $\pi_1$ resonance with physical light-quark mass from these results. We extrapolated the determined couplings, assuming their only adjustment is in the angular momentum barrier (an approach that has proven reasonably successful when applied to previous lattice QCD determinations of vector, axial-vector and tensor mesons). This suggests a potentially \emph{broad} $\pi_1$ resonance, the bulk of whose decay goes into the $b_1 \pi$ mode.

Comparing to the experimental $\pi_1(1564)$ candidate state found by the JPAC/COMPASS analysis~\cite{Rodas:2018owy},
our predicted range of total width is compatible with their width taken from the resonance pole position. We note that the $\eta \pi$, $\eta' \pi$ modes in which the resonance is observed experimentally are relatively rare decays in our picture. Although the $b_1 \pi$ decay mode is somewhat challenging experimentally, ending up in five pions through $b_1 \to \omega \pi$, these results suggest that it is a promising channel to search in.

%% file: acknow.tex
\begin{acknowledgments}
We thank our colleagues within the Hadron Spectrum Collaboration.
AJW, CET and DJW acknowledge support from the U.K. Science and Technology Facilities Council (STFC) [grant number ST/P000681/1]. 
JJD acknowledges support from the U.S. Department of Energy contract DE-SC0018416.
JJD, RGE and CET acknowledge support from the U.S. Department of Energy contract DE-AC05-06OR23177, under which Jefferson Science Associates, LLC, manages and operates Jefferson Lab. 
DJW acknowledges support from a Royal Society University Research Fellowship.
CET and JJD acknowledge support from the Munich Institute for Astro- and Particle Physics (MIAPP), which is funded by the Deutsche Forschungsgemeinschaft (DFG, German Research Foundation) under Germany's Excellence Strategy – EXC-2094 – 390783311, while attending a program. CET also acknowledges CERN TH for hospitality and support during a visit.

The software codes
{\tt Chroma}~\cite{Edwards:2004sx} and {\tt QUDA}~\cite{Clark:2009wm,Babich:2010mu,Clark:2016rdz} were used. 
The authors acknowledge support from the U.S. Department of Energy, Office of Science, Office of Advanced Scientific Computing Research and Office of Nuclear Physics, Scientific Discovery through Advanced Computing (SciDAC) program. 
Also acknowledged is support from the Exascale Computing Project (17-SC-20-SC), a collaborative effort of the U.S. Department of Energy Office of Science and the National Nuclear Security Administration.

This work was performed using the Cambridge Service for Data Driven Discovery (CSD3) operated by the University of Cambridge Research Computing Service (www.hpc.cam.ac.uk), provided by Dell EMC and Intel using Tier-2 funding from the Engineering and Physical Sciences Research Council (capital grant EP/P020259/1), and DiRAC funding from STFC (www.dirac.ac.uk). The DiRAC component of CSD3 was funded by BEIS capital funding via STFC capital grants ST/P002307/1 and ST/R002452/1 and STFC operations grant ST/R00689X/1. DiRAC is part of the National e-Infrastructure.
This work was also performed on clusters at Jefferson Lab under the USQCD Collaboration and the LQCD ARRA Project.
This research was supported in part under an ALCC award, and used resources of the Oak Ridge Leadership Computing Facility at the Oak Ridge National Laboratory, which is supported by the Office of Science of the U.S. Department of Energy under Contract No. DE-AC05-00OR22725.
This research used resources of the National Energy Research Scientific Computing Center (NERSC), a DOE Office of Science User Facility supported by the Office of Science of the U.S. Department of Energy under Contract No. DE-AC02-05CH11231.
The authors acknowledge the Texas Advanced Computing Center (TACC) at The University of Texas at Austin for providing HPC resources.
Gauge configurations were generated using resources awarded from the U.S. Department of Energy INCITE program at the Oak Ridge Leadership Computing Facility, the NERSC, the NSF Teragrid at the TACC and the Pittsburgh Supercomputer Center, as well as at the Cambridge Service for Data Driven Discovery (CSD3) and Jefferson Lab.
\end{acknowledgments}

%% file: sections/SU3F_syms.tex
Unlike in $\SU(2)$, where the product of two representations of definite isospin decomposes into a sum of isospins each of which appears only once, in $\SU(3)$ a representation can appear more than once in a product. A relevant example is $\bm{8}\otimes \bm{8} = \bm{1} \oplus \bm{8}_1 \oplus \bm{8}_2 \oplus \bm{10} \oplus \overline{\bm{10}} \oplus \bm{27}$ where we observe two octet embeddings, $\bm{8}_1$ and $\bm{8}_2$.

Following the conventions given in Ref.~\cite{deSwart:1963pdg}, the $\SU(3)$ Clebsch-Gordan coefficients, $\mathcal{C}(\dots)$, for $\bm{8}\otimes \bm{8} \rightarrow\bm{8}_1,\,\bm{8}_2$ are respectively symmetric, antisymmetric under exchanging the hadrons in the product or conjugating the hadrons in the product,
\begin{equation*}
\mathcal{C}\!\left( \begin{array}{ccc}
{\bm{8}} & {\bm{8}} & {\bm{8}_i} \\
\nu_1 & \nu_2 & \nu \end{array} \right) = \xi_1(i) \; 
\mathcal{C}\!\left( \begin{array}{ccc}
{\bm{8}} & {\bm{8}} & {\bm{8}_i} \\
\nu_2 & \nu_1 & \nu \end{array} \right)\, ,
\end{equation*}
\begin{equation*}
\mathcal{C}\!\left( \begin{array}{ccc}
{\bm{8}} & {\bm{8}} & {\bm{8}_i} \\
\nu_1 & \nu_2 & \nu \end{array} \right) = \xi_3(i) \; 
\mathcal{C}\!\left( \begin{array}{ccc}
{\bm{8}} & {\bm{8}} & {\bm{8}_i} \\
-\nu_1 & -\nu_2 & -\nu \end{array} \right)\, ,
\end{equation*}
with $\xi_1(1) = \xi_3(1) = 1$ and $\xi_1(2) = \xi_3(2) = -1$, and using $\overline{\bm{8}} = \bm{8}$. Here a particular member of the octet is labelled by its isospin $I$, hypercharge $Y$, and $z$-component of isospin $I_z$, in $\nu=(I,Y,I_z)$, and for mesons the hypercharge is simply equal to the strangeness, $Y=S$.

It is useful at this point to write out the non-zero $\SU(3)$ Clebsch-Gordan coefficients for the two embeddings explicitly. As we are at liberty to work with any member of the target octet, we choose $\nu = (0,0,0)$. We label the multiplied octets $\bm{8}_a$ and $\bm{8}_b$ in order to distinguish them. Applying the rules given in Ref.~\cite{deSwart:1963pdg}, we have for the symmetric $\bm{8}_1$ combination,
\begin{widetext}
\begin{align*}
&\Ket{\bm{8}_1;0,0,0} = \nonumber \\
&\tfrac{1}{\sqrt{20}}\Big(
\underbrace{ \Ket{\bm{8}_a; \tfrac{1}{2},        1,\tfrac{1}{2}}         }_{ K^{*+} } \,   
\underbrace{ \Ket{\bm{8}_b; \tfrac{1}{2},\text{-}1,\text{-}\tfrac{1}{2}} }_{ K^- } 
+
\underbrace{ \Ket{\bm{8}_a; \tfrac{1}{2},\text{-}1,\text{-}\tfrac{1}{2}} }_{ K^{*-} } \, 
\underbrace{ \Ket{\bm{8}_b; \tfrac{1}{2},        1,\tfrac{1}{2}}         }_{ K^+ }
- 
\underbrace{ \Ket{\bm{8}_a; \tfrac{1}{2},        1,\text{-}\tfrac{1}{2}} }_{ K^{*0} } \, 
\underbrace{ \Ket{\bm{8}_b; \tfrac{1}{2},\text{-}1,\tfrac{1}{2}}         }_{ \KbarSuper{0} }
- 
\underbrace{ \Ket{\bm{8}_a; \tfrac{1}{2},\text{-}1,\tfrac{1}{2}} }_{ \KbarSuper{*0} } \, 
\underbrace{ \Ket{\bm{8}_b; \tfrac{1}{2},        1,\text{-}\tfrac{1}{2}}         }_{ K^0 }
\Big) \nonumber \\
&-\tfrac{1}{\sqrt{5}}\Big(
\underbrace{ \Ket{\bm{8}_a; 1, 0, 1       }   }_{ \rho^+ } \,   
\underbrace{ \Ket{\bm{8}_b; 1, 0,\text{-}1}   }_{ \pi^- } 
+
\underbrace{ \Ket{\bm{8}_a; 1, 0, \text{-}1}  }_{ \rho^- } \,   
\underbrace{ \Ket{\bm{8}_b; 1, 0,         1}  }_{ \pi^+ } 
-
\underbrace{ \Ket{\bm{8}_a; 1, 0, 0}  }_{ \rho^0 } \,   
\underbrace{ \Ket{\bm{8}_b; 1, 0, 0}  }_{ \pi^0 } 
\Big)
-\tfrac{1}{\sqrt{5}}
\underbrace{ \Ket{\bm{8}_a; 0, 0, 0}  }_{ \omega_8 } \,   
\underbrace{ \Ket{\bm{8}_b; 0, 0, 0}  }_{ \eta_8 } 
\, ,
\end{align*}
while for the antisymmetric $\bm{8}_2$ combination,
\begin{align*}
& \Ket{\bm{8}_2;0,0,0}  = \nonumber \\
&\tfrac{1}{2}\Big(
\underbrace{ \Ket{\bm{8}_a; \tfrac{1}{2},        1,\tfrac{1}{2}}         }_{ K^{*+} } \,   
\underbrace{ \Ket{\bm{8}_b; \tfrac{1}{2},\text{-}1,\text{-}\tfrac{1}{2}} }_{ K^- } 
-
\underbrace{ \Ket{\bm{8}_a; \tfrac{1}{2},\text{-}1,\text{-}\tfrac{1}{2}} }_{ K^{*-} } \, 
\underbrace{ \Ket{\bm{8}_b; \tfrac{1}{2},        1,\tfrac{1}{2}}         }_{ K^+ }
- 
\underbrace{ \Ket{\bm{8}_a; \tfrac{1}{2},        1,\text{-}\tfrac{1}{2}} }_{ K^{*0} } \, 
\underbrace{ \Ket{\bm{8}_b; \tfrac{1}{2},\text{-}1,\tfrac{1}{2}}         }_{ \KbarSuper{0} }
+
\underbrace{ \Ket{\bm{8}_a; \tfrac{1}{2},\text{-}1,\tfrac{1}{2}} }_{ \KbarSuper{*0} } \, 
\underbrace{ \Ket{\bm{8}_b; \tfrac{1}{2},        1,\text{-}\tfrac{1}{2}}         }_{ K^0 }
\Big) \, ,
\end{align*}
where we have provided the PDG notation for vector and pseudoscalar mesons as an example, as was done in Eq.~\ref{VPs_CGs}.
\end{widetext}

Defining $\hat{G}$ in the usual way as $\hat{C}$ followed by a rotation by $\pi$ about the $y$-component of isospin, $\hat{R}$, it is straightforward to show~\cite{Chung:1971ri} that,
\begin{align*}
\hat{C} \, \Ket{\bm{8};I,Y,I_z} =& \,\,C \, (-1)^{Y/2\, +I_z}\, \Ket{\bm{8};I,-Y,-I_z} \nonumber \\
\hat{R} \, \Ket{\bm{8};I,Y,I_z} =& \,\, (-1)^{I-I_z}         \, \Ket{\bm{8};I,Y,-I_z} \nonumber \\
\hat{G} \, \Ket{\bm{8};I,Y,I_z} =& \,\,C \, (-1)^{Y/2 \, +I}  \, \Ket{\bm{8};I,-Y,I_z} ,
\end{align*}
where $C$ is the intrinsic charge-conjugation quantum number of the neutral element of the octet, for example, $C=+1$ for $\etaOctet$ and $C=-1$ for $\omegaOctet$. There are $\SU(3)$ analogues of $G$-parity where the rotation is between the $u,s$ or $d,s$ quarks rather than the $u,d$ quarks. When $\SU(3)$ is broken these are no longer good quantum numbers whereas $G$-parity is still good as long as there is isospin symmetry.

Acting with $\hat{C}$ or $\hat{G}$ on the decompositions above gives,
\begin{align*}
\hat{C}\Ket{\bm{8}_1;0,0,0} = \hat{G}\Ket{\bm{8}_1;0,0,0} =&+C_aC_b\ket{\bm{8}_1;0,0,0} \nonumber \\
\hat{C}\Ket{\bm{8}_2;0,0,0} = \hat{G}\Ket{\bm{8}_2;0,0,0} =& -C_aC_b\ket{\bm{8}_2;0,0,0} .
\end{align*}
where $C_{a}$ and $C_{b}$ are the intrinsic charge-conjugation quantum numbers of the neutral element of the octets $\bm{8}_a$ and $\bm{8}_b$. Therefore, $\bm{8}_1$ and $\bm{8}_2$ have isoscalar members which are eigenstates of charge-conjugation with opposite values of $C$.

In the case of $\bm{8}\otimes \bm{8}\rightarrow \bm{1}$, the $\SU(3)$ Clebsch-Gordan coefficients are symmetric under interchange -- explicitly the construction is,
\begin{widetext}
\begin{align*}
& \Ket{\bm{1};0,0,0}  = \nonumber \\
&\tfrac{1}{2\sqrt{2}}\Big(
\underbrace{ \Ket{\bm{8}_a; \tfrac{1}{2},        1,\tfrac{1}{2}}         }_{ K^{*+} } \,   
\underbrace{ \Ket{\bm{8}_b; \tfrac{1}{2},\text{-}1,\text{-}\tfrac{1}{2}} }_{ K^- } 
+
\underbrace{ \Ket{\bm{8}_a; \tfrac{1}{2},\text{-}1,\text{-}\tfrac{1}{2}} }_{ K^{*-} } \, 
\underbrace{ \Ket{\bm{8}_b; \tfrac{1}{2},        1,\tfrac{1}{2}}         }_{ K^+ }
- 
\underbrace{ \Ket{\bm{8}_a; \tfrac{1}{2},        1,\text{-}\tfrac{1}{2}} }_{ K^{*0} } \, 
\underbrace{ \Ket{\bm{8}_b; \tfrac{1}{2},\text{-}1,\tfrac{1}{2}}         }_{ \KbarSuper{0} }
- 
\underbrace{ \Ket{\bm{8}_a; \tfrac{1}{2},\text{-}1,\tfrac{1}{2}} }_{ \KbarSuper{*0} } \, 
\underbrace{ \Ket{\bm{8}_b; \tfrac{1}{2},        1,\text{-}\tfrac{1}{2}}         }_{ K^0 }
\Big) \nonumber \\
&+\tfrac{1}{2\sqrt{2}}\Big(
\underbrace{ \Ket{\bm{8}_a; 1, 0, 1       }   }_{ \rho^+ } \,   
\underbrace{ \Ket{\bm{8}_b; 1, 0,\text{-}1}   }_{ \pi^- } 
+
\underbrace{ \Ket{\bm{8}_a; 1, 0, \text{-}1}  }_{ \rho^- } \,   
\underbrace{ \Ket{\bm{8}_b; 1, 0,         1}  }_{ \pi^+ } 
-
\underbrace{ \Ket{\bm{8}_a; 1, 0, 0}  }_{ \rho^0 } \,   
\underbrace{ \Ket{\bm{8}_b; 1, 0, 0}  }_{ \pi^0 } 
\Big)
-\tfrac{1}{2\sqrt{2}}
\underbrace{ \Ket{\bm{8}_a; 0, 0, 0}  }_{ \omega_8 } \,   
\underbrace{ \Ket{\bm{8}_b; 0, 0, 0}  }_{ \eta_8 } 
\, ,
\end{align*}
and $\hat{C}\Ket{\bm{1};0,0,0} = \hat{G}\Ket{\bm{1};0,0,0} = C_a C_b \,\Ket{\bm{1};0,0,0}$.
\end{widetext}

For the cases of $\bm{8}\otimes \bm{1}\rightarrow \bm{8}$ and $\bm{1}\otimes \bm{1}\rightarrow \bm{1}$, the Clebsch-Gordan coefficients are trivial,
\begin{align*}
\Ket{\bm{8};0,0,0} =& \Ket{\bm{8}_a;0,0,0}\, \Ket{\bm{1}_b;0,0,0}  \\
\Ket{\bm{1};0,0,0} =& \Ket{\bm{1}_a;0,0,0}\, \Ket{\bm{1}_b;0,0,0}, 
\end{align*}
and obviously $C = C_a C_b$.

%% file: sections/Bose_syms.tex
 
A practical consequence of Bose symmetry is the elimination of certain partial-wave configurations in the scattering of identical mesons. A familiar example assuming only isospin symmetry is that $\pi\pi$ scattering with isospin=1 is only in \emph{odd} partial waves, while isospin=0,2 are only in \emph{even} partial waves. 
The $SU(3)$ Clebsch-Gordan coefficients discussed in Appendix~\ref{su3sym} have definite symmetry under the exchange of the two scattering hadrons, and this makes the application of Bose symmetry straightforward when we need to combine two identical meson multiplets.

Consider first identical pseudoscalar meson octets -- the total spin $S$ is zero and the spin wavefunction is trivially symmetric. To ensure overall symmetry under exchange we require the product of flavor and spatial wavefunctions to be overall symmetric, meaning they are either both symmetric or both antisymmetric. In Appendix~\ref{su3sym} we showed that $\bm{8}_1$ and $\bm{8}_2$ are symmetric and antisymmetric in flavor respectively, so we deduce that only partial waves of even $\ell$ are permitted in $\bm{8}_1$ and odd $\ell$ in $\bm{8}_2$. 
It follows that, for example, $\etaOctet\etaOctet$ appears with even $\ell$ in $\bm{8_1}$ with $J^{P(C)}=\ell^{+(\!+\!)}$ and odd $\ell$ in $\bm{8_2}$ with $J^{P(C)}=\ell^{- (\!-\!)}$. A consequence is that $\etaOctet\etaOctet$ is forbidden in decays of an $J^{P(C)}=1^{-(+)}$ octet resonance.

For identical vector meson octets, the symmetry of the spin wavefunction depends on the total spin $S$: symmetric for $S=0,2$ and antisymmetric for $S=1$. It follows that for $S=0,2$, the product of flavor and spatial wavefunctions must be totally symmetric, so either they are both symmetric or both antisymmetric, similar to the case above -- only even $\ell$ partial waves are permitted in $\bm{8}_1$, while only odd $\ell$ appear for $\bm{8}_2$. In the case of $S=1$, by an analogous argument, only partial waves of odd $\ell$ are permitted in $\bm{8}_1$ and even $\ell$ in $\bm{8}_2$. Hence $\omegaOctet\omegaOctet$ is forbidden in $\onePone$ and $\fivePone$ decays of an $J^{P(C)}=1^{-(+)}$ octet resonance, while it is allowed in $\threePone$. Table~\ref{Tab:Bose} summarises the Bose-allowed partial-wave content of $\bm{8}_1$ and $\bm{8}_2$ for identical vector meson octets.
\begin{table}[tb]
\centering
{\renewcommand{\arraystretch}{1.7}
\setlength{\tabcolsep}{10pt}
\begin{tabular}{ l | l }
$\bm{8_1}$ $(C=+)$ & $\bm{8_2}$ $(C=-)$ \\
\hline
$\prescript{1\!}{}{S}_0,\,\prescript{1\!}{}{D}_2,\,\prescript{1}{}{G}_4,\,\dots$
 & $\prescript{1\!}{}{P}_1,\,\prescript{1\!}{}{F}_3,\,\dots$ \\
$\prescript{3\!}{}{P}_{0,1,2},\,\prescript{3\!}{}{F}_{2,3,4},\,\dots$
 & $\prescript{3\!}{}{S}_1,\,\prescript{3\!}{}{D}_{1,2,3},\,\prescript{3}{}{G}_{3,4,5},\,\dots$ \\
$\prescript{5\!}{}{S}_2,\,\prescript{5\!}{}{D}_{0 \ldots 4},\,\prescript{5}{}{G}_{2 \ldots 6},\,\dots$
 & $\prescript{5\!}{}{P}_{1,2,3},\,\prescript{5\!}{}{F}_{1\ldots5},\,\dots$ \\
\end{tabular}
}
\caption{Bose-allowed partial-wave content of multiplets $\bm{8}_1$ and $\bm{8}_2$ from a product of two identical vector meson octets, $\bm{8}_a \otimes \bm{8}_a$, for $\ell \leq 4$.}
\label{Tab:Bose}
\end{table}

%% file: sections/VV_pwaves.tex
In this appendix we show that the quantization condition, Eq.~\ref{luescher}, when subduced into the $T_1^-$ irrep at rest cannot uniquely constrain the $\omegaSinglet \omegaOctet \big\{\! \onePone, \threePone, \fivePone \!\big\}$ amplitudes owing to a residual $S_3$ permutation symmetry on these channels, i.e.~the corresponding scattering parameters in the $t$-matrix can be freely interchanged while leaving the determinant invariant. We also show that the same permutation symmetry is not present for systems with overall non-zero momentum, so including energy levels obtained in such irreps would provide a unique constraint for each of these partial waves. 

Recalling the form of the quantization condition,
\begin{equation*}
\det_{\ell S J m a}\big[ \mathbf{1} + i \, \bm{\rho} \, \bm{t} \, \big( \mathbf{1} + i \boldsymbol{\mathcal{M}} \big) \big] = 0,
\end{equation*}
we note that the finite-volume nature of the problem resides in the matrix $\boldsymbol{\mathcal{M}}$ whose components are defined explicitly in App.~A of Ref.~\cite{Woss:2020cmp}.
$\boldsymbol{\mathcal{M}}$ is trivially diagonal in hadron channel and intrinsic spin, leading to it being diagonal in $\omegaSinglet \omegaOctet \big\{\! \onePone, \threePone, \fivePone \!\big\}$ channels.\footnote{It is also diagonal in the $\omegaSinglet \omegaOctet \big\{\! \fivePone,\fivePthree \!\big\}$ subspace at rest.} The reason that these channels cannot be distinguished at overall zero momentum is that the diagonal entries of $\boldsymbol{\mathcal{M}}$ in each of $\omegaSinglet \omegaOctet \big\{\! \onePone, \threePone, \fivePone \!\big\}$ are \emph{equal}.

From the product of spherical harmonics in Eq.~(A1) of Ref.~\cite{Woss:2020cmp}, $ \int\!\! d\Omega \; Y_{1 m_\ell}^*  Y_{\bar{\ell} \overline{m}_\ell}^* Y^{\,}_{1 m_\ell'}$, it is clear that only $\bar{\ell} \le 2$ contribute, and from the symmetries of the L\"{u}scher zeta-functions at zero momentum,
\begin{equation*}
Z^{\vec{0}}_{\ell \notin 2\mathbb{Z}, m \notin 4\mathbb{Z}}=0, \quad\quad Z^{\vec{0}}_{20} = 0,
\end{equation*} 
only $\bar{\ell} =0, \overline{m}_\ell=0$ survives. The elements of $\boldsymbol{\mathcal{M}}$ thus reduce to the rather simple form,
\begin{align}
\mathcal{M}
\big( \prescript{2S+1\!}{}{P}_1,m ; \prescript{2S'+1\!}{}{P}_1 ,m'\big)
= \delta_{S,S'} \delta_{m,m'}\, \tfrac{4\pi}{k} \, c^{\vec{0}}_{0,0}(k^2; L), \nonumber
\end{align} 
and it follows that the rest-frame $\boldsymbol{\mathcal{M}}$ does not distinguish between the $\omegaSinglet \omegaOctet \big\{\! \onePone, \threePone, \fivePone\!\big\}$ channels. The result of this is that permutations of the $\omegaSinglet \omegaOctet \big\{\! \onePone, \threePone, \fivePone \!\big\}$ channels will leave the determinant in Eq.~\ref{luescher} invariant.

These partial waves become distinguishable if we consider the system at overall non-zero momentum. Following a similar derivation to the zero momentum case, owing to $Z^{\vec{P}}_{20}$ being non-zero in general, we find that elements of $\bm{\mathcal{M}}$ are spin dependent. For example, in the case that $\vec{P}=[00n]$, the $m=+1$, $m'=+1$ elements are given by,
\begin{align*}
\mathcal{M}\big(\!\onePone,+1;\onePone,+1\big)&
\!= \tfrac{4\pi}{k}\, c^{\vec{P}}_{0,0} (k^2;\! L) - \tfrac{1}{\sqrt{5}} \tfrac{4\pi}{k^3} c^{\vec{P}}_{2,0}(k^2;\! L) \nonumber \\
\mathcal{M}\big(\! \threePone,+1;\threePone,+1\big)&
\!= \tfrac{4\pi}{k}\, c^{\vec{P}}_{0,0}(k^2;\! L) +  \tfrac{1}{2\sqrt{5}} \tfrac{4\pi}{k^3} c^{\vec{P}}_{2,0}(k^2;\! L) \nonumber \\
\mathcal{M}\big(\! \fivePone,+1;\fivePone,+1\big)&
\!= \tfrac{4\pi}{k} \, c^{\vec{P}}_{0,0}(k^2;\! L) -  \tfrac{1}{10\sqrt{5}} \tfrac{4\pi}{k^3} c^{\vec{P}}_{2,0}(k^2;\! L),
\end{align*}
where we observe that the coefficients of the $c^{\vec{P}}_{2,0}$ term distinguishes the different spin configurations.

%% file: sections/squeezed_levels.tex
A parameterization in common use to describe a single coupled-channel resonance with angular momentum $J$ assumes a factorized pole in the $K$-matrix and the simple phase space ($I_a(s)=-i\rho_a(s)$) in the construction of the $t$-matrix,
\begin{align*}
\mathbf{t} &= \widetilde{\bm{K}} \, \big( 1 - i \bm{\rho} \, \widetilde{\bm{K}}  \big)^{-1}, \\
\big[\widetilde{K}(s)\big]_{\ell S J a, \ell' S' J b} &= (2k_a)^\ell\frac{g_{\ell S J a}\, g_{\ell' S' J b}}{m^2-s}  (2k_b)^{\ell'}.
\end{align*} 

Here we will show that this particular form can lead to the phenomenon of ``trapped'' levels in finite-volume spectra, a situation where there is guaranteed to be exactly one finite-volume energy level lying between every neighboring non-interacting energy. In particular, we will present a proof of how trapped levels emerge in coupled meson-meson scattering in $\threeSone$ and $\{\onePone,\threePone,\fivePone\}$-wave in the rest frame irreps, as relevant for this study. This effect is not a general feature of the finite-volume method -- for example, upon adding a matrix of polynomials in $s$ to the $K$-matrix above (as we commonly do) the guarantee is removed.

The L\"{u}scher quantisation condition, Eq.~\ref{luescher}, can be rewritten in terms of the $K$-matrix defined above yielding the convenient form,
\begin{equation*}
\det [\bm{1} - \bm{\rho} \,\widetilde{\bm{K}} \, \bm{\mathcal{M}}] = 0,
\end{equation*}
where the determinant is taken over the $N$-dimensional space of hadron-hadron channels and partial waves.

When $\widetilde{\bm{K}}$ is factorized as above, the matrix $\bm{\rho} \, \widetilde{\bm{K}} \, \bm{\mathcal{M}}$ is of the form $\bm{a}\,  \bm{b}^T$ for all energies, where $\bm{a}(s)$ and $\bm{b}(s)$ are (energy dependent) vectors, and hence of rank one. It has one non-zero eigenvalue, $\mu_0(s) =  \bm{b}^T \! \, \bm{a}$, with eigenvector, $\bm{v}_0 = \bm{a}$, and $N-1$ zero eigenvalues, $\mu_i(s)=0$ for ${i=1,\dots, N-1}$, whose eigenvectors span the hyperplane orthogonal to $\bm{a}$.
It immediately follows that $\bm{1} - \bm{\rho} \, \widetilde{\bm{K}} \, \bm{\mathcal{M}}$ has exactly one eigenvalue capable of taking a zero value, $\lambda_0(s)=1-\bm{b}^T \! \, \bm{a}$ -- all other eigenvalues $\lambda_i(s)=1$ for $i=1$, $\dots$, $N-1$. The finite-volume spectrum is therefore given by the solutions to $\lambda_0(s)=0$.

For ease of illustration, consider the case of several coupled meson-meson channels, each in a single partial-wave. The nontrivial eigenvalue $\lambda_0(s)$ takes the form,
\begin{equation}\label{eigenvalue}
\lambda_0(s) = 1-\tfrac{2}{\sqrt{s}(m^2-s)} \sum_a (2k_a)^{2\ell} \, g_a^2 \, k_a\,  \mathcal{M}_a \, ,
\end{equation}
where $\mathcal{M}_a$ are the elements of the \emph{diagonal in channel-space} $\bm{\mathcal{M}}$. Recalling the definition of these presented in Ref.~\cite{Woss:2020cmp}, for $S$- and $P$-waves in the rest-frame,
\begin{equation*}
\mathcal{M}_a(s) = \tfrac{2}{\sqrt{\pi}} \tfrac{1}{k_a L}\, Z^{\vec{0}}_{00}\Big[1;\left(\tfrac{k_a L}{2\pi}\right)^{2} \Big] \, , \label{Mi}
\end{equation*}
independent of the intrinsic spin of the system. The only differences between the objects $\mathcal{M}_a(s)$ for different channels come from the momenta $k_a$. It is therefore instructive to examine the functional form of 
\begin{equation} \label{XX}
-(2 k_a(s) )^{2\ell}\, k_a(s)\, \mathcal{M}_a(s) \, ,
\end{equation}
that appears for each channel in Eq.~\ref{eigenvalue}. We now investigate the consequences of this for $S$-wave scattering before considering $P$-wave scattering.

%%%%%%%%%%%%%%%%%%%%%%%%%%%%%%%%%%%%%%%%%%%%%%%%%%%%%%%%%%%%%%%%%%%%%%%%%%%%%%%%%%%%%%%%%%
\subsection{$S$-wave scattering}

%%%%%%%%%%%%%%%%%%%%%%%%%%%%%%%%%%%%%%%%%%%%%%%%%%%%%%%%%%
\begin{figure*}
	\centering
	\includegraphics[width=1\textwidth]{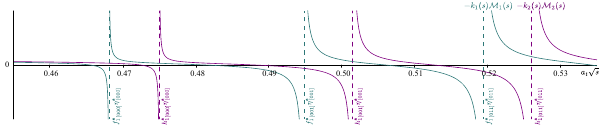}    
	\caption{Energy dependence of $-k_1(s) \mathcal{M}_1(s)$ and $-k_2(s) \mathcal{M}_2(s)$ for a lattice of spatial extent $L/a_s=24$. Both functions are purely real across this energy region. Dashed vertical lines indicate the location of non-interacting energies. }
	\label{kM}
\end{figure*}
%%%%%%%%%%%%%%%%%%%%%%%%%%%%%%%%%%%%%%%%%%%%%%%%%%%%%%%%%%

%%%%%%%%%%%%%%%%%%%%%%%%%%%%%%%%%%%%%%%%%%%%%%%%%%%%%%%%%%
\begin{figure*}
	\centering
	\vspace*{3mm}
	\includegraphics[width=1\textwidth]{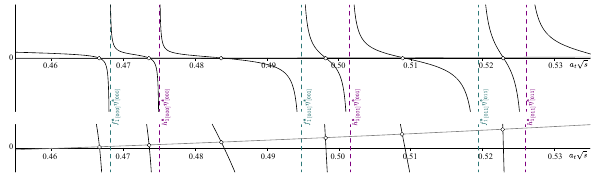}    
	\caption{\textbf{Top:} The function $-g_1^2 \,k_1(s) \mathcal{M}_1(s) - g_2^2\, k_2(s) \mathcal{M}_2(s)$, for $L/a_s=24$ and $g_1=g_2=1$, plotted in black, and the function $\tfrac{1}{2}\sqrt{s}(s-m^2)$ with $m = 0.461057$ plotted in grey. The points of intersection (circles) correspond to the finite-volume energy levels. Dashed vertical lines indicate the location of non-interacting energies. \textbf{Bottom:} Same as the top plot but with a zoomed in vertical scale.}
	\label{trapped}
\end{figure*}
%%%%%%%%%%%%%%%%%%%%%%%%%%%%%%%%%%%%%%%%%%%%%%%%%%%%%%%%%%

In Fig.~\ref{kM} we plot Eq.~\ref{XX} for the $\fOneOctet\etaOctet$ and $\hOneOctet\etaOctet$ $\threeSone$ channels. These functions are real above threshold, and show monotonic decrease between divergences at each non-interacting energy. The finite-volume spectrum in this case is given by the solutions of,
\begin{equation} \label{qmlike}
\tfrac{1}{2} \sqrt{s}(s-m^2) = -g_1^2 \,k_1(s) \mathcal{M}_1(s) - g_2^2 \, k_2(s) \mathcal{M}_2(s),
\end{equation}
where the RHS of this expression is just a weighted sum of the expressions plotted in Fig.~\ref{kM}. The effect of changing the values of $g_1$ and $g_2$ simply moves the point of inflection of the RHS in each region between neighboring non-interacting energies. As the LHS is a monotonically increasing function for $\sqrt{s} > \tfrac{1}{\sqrt{3}}m$, this will intersect the RHS \emph{exactly once} in each energy region between non-interacting energies. This results in what we refer to as ``trapped'' levels. We see exactly this in Fig.~\ref{trapped}, where above $\fOneOctet\etaOctet$ threshold we see a single solution in each region as described.

%%%%%%%%%%%%%%%%%%%%%%%%%%%%%%%%%%%%%%%%%%%%%%%%%%%%%%%%%%%%%%%%%%%%%%%%%%%%%%%%
\subsection{$P$-wave scattering}
For $P$-wave scattering, Eq.~\ref{XX} has an extra factor of a smooth real function, $4k_a^2(s)$, compared to the $S$-wave case. This is positive above threshold, negative below and has a zero exactly at threshold, and this zero is the reason for there being no non-interacting level at threshold in $P$-wave. The argument that led to ``trapped'' levels in $S$-wave applies here too.

It is interesting to revisit the indistinguishability of $\{\onePone,\threePone,\fivePone\}$ in vector-vector scattering in the context of a factorized pole $K$-matrix. If we consider this system which has only a single open channel but three partial-waves, then the single non-trivial eigenvalue which has zeros at the finite-volume energy levels is,
\begin{align*}
\lambda_0(s) = 1-\frac{8 k^3 \mathcal{M} }{\sqrt{s}(m^2-s)} \big( g_1^2+g_2^2+g_3^2 \big),
\end{align*}
as the momenta $k(s)$ and the function $\mathcal{M}(s)$ are identical in each of these partial-waves. 

Naively, we would expect to find only a single root between neighboring non-interacting energies; however, this would overlook the fact that the multiplicity of each of these non-interacting energies is in fact \emph{three}, and so we should find \emph{three} roots associated with each non-interacting energy (these roots are \emph{not} necessarily triply degenerate as we will see).

This can be seen most easily by treating each partial wave as an independent hadron-hadron scattering channel by perturbing the scattering vector meson mass slightly in each partial wave. In $\onePone$ we take $m_{\omegaOctet} \rightarrow m_{\omegaOctet} -\epsilon$ and in $\fivePone$ $m_{\omegaOctet} \rightarrow m_{\omegaOctet} +\epsilon$, so that the perturbed finite-volume energy levels are roots of,
\begin{align*}
&\widetilde{\lambda}_0(s) = \nonumber \\
 &\,\,\,\,1-\frac{8}{\sqrt{s}(m^2-s)}\big(g_1^2 \, k_{-\epsilon}^3 \mathcal{M}_{-\epsilon} 
 + g_2^2 \,k^3 \mathcal{M} + g_3^2 \, k_{+\epsilon}^3 \mathcal{M}_{+\epsilon}\big) ,
\end{align*}
where the subscript $\pm \epsilon$ means that the vector meson mass has been perturbed by $\pm \epsilon$. 
The previously triply degenerate non-interacting energies are now split by order $\epsilon$. However, there are trapped roots between these perturbed non-interacting energies which forces at least \emph{two} of the roots to lie within $\epsilon$ of the unperturbed non-interacting energy. In the limit $\epsilon \rightarrow 0$, we find $\widetilde{\lambda}_0(s) \rightarrow \lambda_0(s)$ with at least two roots positioned exactly at the non-interacting energy. The third root is free to vary in position between these two roots and the next non-interacting energy,
its location depending on the value of $g_1^2+g_2^2+g_3^2$; it is exactly at the non-interacting energy if and only if $g_1=g_2=g_3=0$, in which case the roots are triply degenerate.

%% file: sections/toy_model.tex
In this appendix, we will examine the sensitivity of finite-volume spectra to the relative size of the $\fOneOctet\etaOctet\{\threeSone\}$ and $\hOneOctet\etaOctet\{\threeSone\}$ couplings. In Sec.~\ref{poles_section}, we found the ratio of these couplings to be poorly determined, while the sum of the squared couplings was well determined. We will investigate this effect using a simplified two-channel toy model where the $t$-matrix is given by,
\begin{equation*}
t_{ab}(s) = \frac{g_a \,g_b}{m^2 - s+g_1^2 \,I_1(s)+g_2^2 \,I_2(s)} .
\end{equation*}
The mass parameter, $m = 0.46$, is chosen to be below the $\fOneOctet\etaOctet\{\threeSone\}$ threshold, a value which is comparable to the pole mass found in Sec.~\ref{poles_section}. By choosing the Chew-Mandelstam phase-space, with real part such that \mbox{$I_a(s=m^2)=0$}, we ensure that when varying the couplings the bound-state pole remains at $\sqrt{s} = m$. This enables us to test the dependence of the finite-volume spectra on the magnitude and ratio of the couplings for a fixed pole position.

\begin{figure*}[tb]
	\centering
	\includegraphics[width=0.8\textwidth]{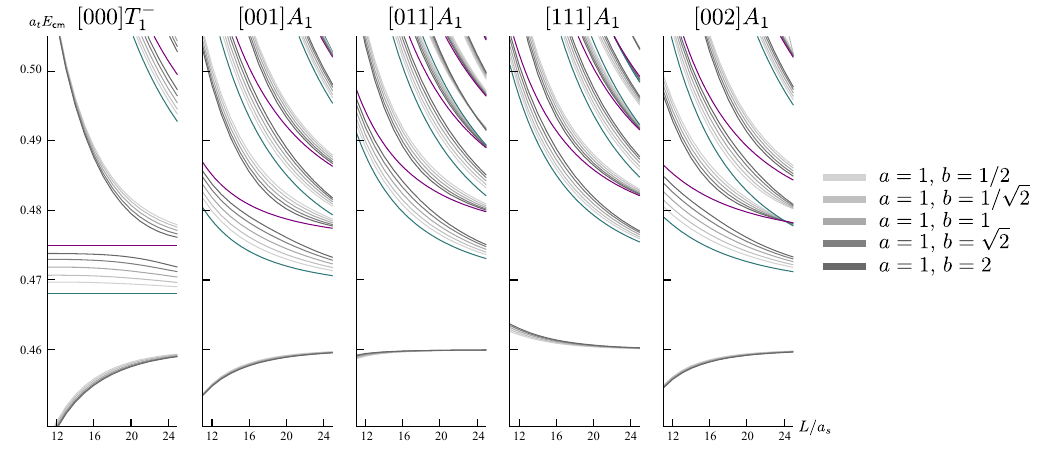}    
	\caption{Finite-volume spectra for the toy model as described in the text for the $T_1^-$ irrep at rest and the $A_1$ irreps in flight. $a^2 \equiv g_1^2+g_2^2 = 1$ is kept fixed, while the ratio, $b \equiv g_1/g_2$, is varied.}
	\vspace{1cm}
	\label{fixed_mag_vary_rat}
\end{figure*}

In Figure~\ref{fixed_mag_vary_rat}, we present finite-volume spectra obtained by solving Eq.~\ref{luescher} for several values of the ratio $b \equiv g_1/g_2$ for a fixed magnitude, $a \equiv g_1^2+g_2^2=1$. Shown are the rest-frame $T_1^-$ irrep, considered in this paper, and also the moving-frame $A_1$ irreps. It is clear that the sub-threshold level, while volume-dependent, is quite insensitive to the coupling ratio in all irreps. The level lying between the thresholds in the $T_1^-$ irrep is sensitive to the ratio, but that is of limited use because the thresholds are rather close together, split only by the mass difference between the $\fOneOctet$ and $\hOneOctet$.

\begin{figure*}[tb]
	\centering
	\includegraphics[width=0.8\textwidth]{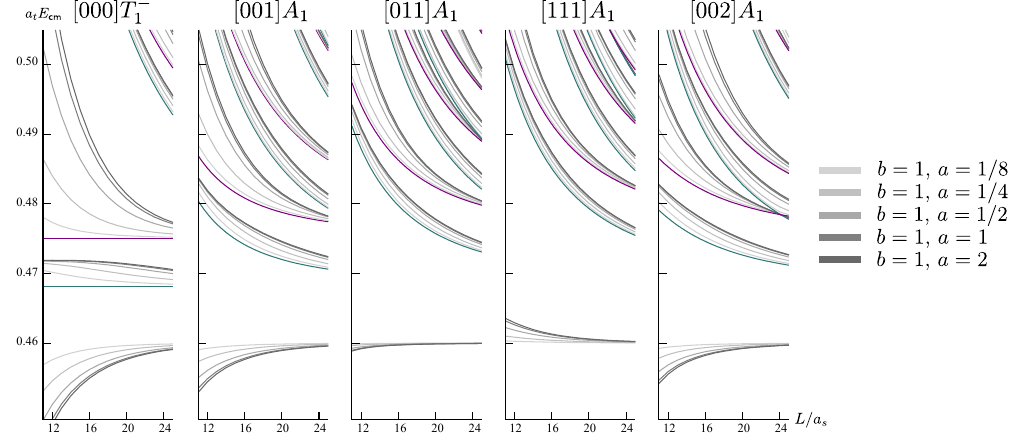}    
	\caption{As Figure~\ref{fixed_mag_vary_rat} but for a fixed ratio, $b=1$, and varying magnitude $a$.}
	\label{fixed_rat_vary_mag}
\end{figure*}

On the contrary, for a fixed ratio $g_1/g_2=1$, it is clear from Figure~\ref{fixed_rat_vary_mag} that the sub-threshold level is rather sensitive to the sum of the squared couplings, $a$, with a smaller value leading to an energy level much closer to the value of $m$ and with less volume dependence. The level between the thresholds in the $T_1^-$ irrep is somewhat less sensitive to $a$.

It appears that to have a well-determined ratio of couplings in this case, we need greater statistical precision on the finite-volume energies and/or additional constraint from several energy levels in moving frames.

%%%%%%%%%%%%%%%%%%%%%%%%%%%%%%%%%%%%%%%%%%%%%%%%%%%%%%%%%%
%%%%%%%%%%%%%%%%%%%%%%%%%%%%%%%%%%%%%%%%%%%%%%%%%%%%%%%%%%

%%%%%%%%%%%%%%%%%%%%%%%%%%%%%%%%%%%%%%%%%%%%%%%%%%%%%%%%%%
%%%%%%%%%%%%%%%%%%%%%%%%%%%%%%%%%%%%%%%%%%%%%%%%%%%%%%%%%%

%% file: paper.bbl
%merlin.mbs apsrev4-1.bst 2010-07-25 4.21a (PWD, AO, DPC) hacked
%Control: key (0)
%Control: author (72) initials jnrlst
%Control: editor formatted (1) identically to author
%Control: production of article title (-1) disabled
%Control: page (0) single
%Control: year (1) truncated
%Control: production of eprint (0) enabled
\begin{thebibliography}{105}%
\makeatletter
\providecommand \@ifxundefined [1]{%
 \@ifx{#1\undefined}
}%
\providecommand \@ifnum [1]{%
 \ifnum #1\expandafter \@firstoftwo
 \else \expandafter \@secondoftwo
 \fi
}%
\providecommand \@ifx [1]{%
 \ifx #1\expandafter \@firstoftwo
 \else \expandafter \@secondoftwo
 \fi
}%
\providecommand \natexlab [1]{#1}%
\providecommand \enquote  [1]{``#1''}%
\providecommand \bibnamefont  [1]{#1}%
\providecommand \bibfnamefont [1]{#1}%
\providecommand \citenamefont [1]{#1}%
\providecommand \href@noop [0]{\@secondoftwo}%
\providecommand \href [0]{\begingroup \@sanitize@url \@href}%
\providecommand \@href[1]{\@@startlink{#1}\@@href}%
\providecommand \@@href[1]{\endgroup#1\@@endlink}%
\providecommand \@sanitize@url [0]{\catcode `\\12\catcode `\$12\catcode
  `\&12\catcode `\#12\catcode `\^12\catcode `\_12\catcode `\%12\relax}%
\providecommand \@@startlink[1]{}%
\providecommand \@@endlink[0]{}%
\providecommand \url  [0]{\begingroup\@sanitize@url \@url }%
\providecommand \@url [1]{\endgroup\@href {#1}{\urlprefix }}%
\providecommand \urlprefix  [0]{URL }%
\providecommand \Eprint [0]{\href }%
\providecommand \doibase [0]{http://dx.doi.org/}%
\providecommand \selectlanguage [0]{\@gobble}%
\providecommand \bibinfo  [0]{\@secondoftwo}%
\providecommand \bibfield  [0]{\@secondoftwo}%
\providecommand \translation [1]{[#1]}%
\providecommand \BibitemOpen [0]{}%
\providecommand \bibitemStop [0]{}%
\providecommand \bibitemNoStop [0]{.\EOS\space}%
\providecommand \EOS [0]{\spacefactor3000\relax}%
\providecommand \BibitemShut  [1]{\csname bibitem#1\endcsname}%
\let\auto@bib@innerbib\@empty
%</preamble>
\bibitem [{\citenamefont {Horn}\ and\ \citenamefont
  {Mandula}(1978)}]{Horn:1977rq}%
  \BibitemOpen
  \bibfield  {author} {\bibinfo {author} {\bibfnamefont {D.}~\bibnamefont
  {Horn}}\ and\ \bibinfo {author} {\bibfnamefont {J.}~\bibnamefont {Mandula}},\
  }\href {\doibase 10.1103/PhysRevD.17.898} {\bibfield  {journal} {\bibinfo
  {journal} {Phys. Rev.}\ }\textbf {\bibinfo {volume} {D17}},\ \bibinfo {pages}
  {898} (\bibinfo {year} {1978})}\BibitemShut {NoStop}%
%%CITATION = PHRVA,D17,898;%%
\bibitem [{\citenamefont {Barnes}\ \emph {et~al.}(1983)\citenamefont {Barnes},
  \citenamefont {Close},\ and\ \citenamefont {de~Viron}}]{Barnes:1982tx}%
  \BibitemOpen
  \bibfield  {author} {\bibinfo {author} {\bibfnamefont {T.}~\bibnamefont
  {Barnes}}, \bibinfo {author} {\bibfnamefont {F.~E.}\ \bibnamefont {Close}}, \
  and\ \bibinfo {author} {\bibfnamefont {F.}~\bibnamefont {de~Viron}},\ }\href
  {\doibase 10.1016/0550-3213(83)90004-4} {\bibfield  {journal} {\bibinfo
  {journal} {Nucl. Phys.}\ }\textbf {\bibinfo {volume} {B224}},\ \bibinfo
  {pages} {241} (\bibinfo {year} {1983})}\BibitemShut {NoStop}%
%%CITATION = NUPHA,B224,241;%%
\bibitem [{\citenamefont {Chanowitz}\ and\ \citenamefont
  {Sharpe}(1983)}]{Chanowitz:1982qj}%
  \BibitemOpen
  \bibfield  {author} {\bibinfo {author} {\bibfnamefont {M.~S.}\ \bibnamefont
  {Chanowitz}}\ and\ \bibinfo {author} {\bibfnamefont {S.~R.}\ \bibnamefont
  {Sharpe}},\ }\href {\doibase 10.1016/0550-3213(83)90561-8,
  10.1016/0550-3213(83)90635-1} {\bibfield  {journal} {\bibinfo  {journal}
  {Nucl. Phys.}\ }\textbf {\bibinfo {volume} {B222}},\ \bibinfo {pages} {211}
  (\bibinfo {year} {1983})},\ \bibinfo {note} {[Erratum: Nucl.
  Phys.B228,588(1983)]}\BibitemShut {NoStop}%
%%CITATION = NUPHA,B222,211;%%
\bibitem [{\citenamefont {Isgur}\ \emph {et~al.}(1985)\citenamefont {Isgur},
  \citenamefont {Kokoski},\ and\ \citenamefont {Paton}}]{Isgur:1985vy}%
  \BibitemOpen
  \bibfield  {author} {\bibinfo {author} {\bibfnamefont {N.}~\bibnamefont
  {Isgur}}, \bibinfo {author} {\bibfnamefont {R.}~\bibnamefont {Kokoski}}, \
  and\ \bibinfo {author} {\bibfnamefont {J.}~\bibnamefont {Paton}},\ }\bibfield
   {booktitle} {\emph {\bibinfo {booktitle} {{Proceedings, International
  Conference on Hadron Spectroscopy: College Park, Maryland, April 20-22,
  1985}}},\ }\href {\doibase 10.1103/PhysRevLett.54.869, 10.1063/1.35357}
  {\bibfield  {journal} {\bibinfo  {journal} {Phys. Rev. Lett.}\ }\textbf
  {\bibinfo {volume} {54}},\ \bibinfo {pages} {869} (\bibinfo {year} {1985})},\
  \bibinfo {note} {[AIP Conf. Proc.132,242(1985)]}\BibitemShut {NoStop}%
%%CITATION = PRLTA,54,869;%%
\bibitem [{\citenamefont {Jaffe}\ \emph {et~al.}(1986)\citenamefont {Jaffe},
  \citenamefont {Johnson},\ and\ \citenamefont {Ryzak}}]{Jaffe:1985qp}%
  \BibitemOpen
  \bibfield  {author} {\bibinfo {author} {\bibfnamefont {R.~L.}\ \bibnamefont
  {Jaffe}}, \bibinfo {author} {\bibfnamefont {K.}~\bibnamefont {Johnson}}, \
  and\ \bibinfo {author} {\bibfnamefont {Z.}~\bibnamefont {Ryzak}},\ }\href
  {\doibase 10.1016/0003-4916(86)90035-7} {\bibfield  {journal} {\bibinfo
  {journal} {Annals Phys.}\ }\textbf {\bibinfo {volume} {168}},\ \bibinfo
  {pages} {344} (\bibinfo {year} {1986})}\BibitemShut {NoStop}%
%%CITATION = APNYA,168,344;%%
\bibitem [{\citenamefont {General}\ \emph {et~al.}(2007)\citenamefont
  {General}, \citenamefont {Cotanch},\ and\ \citenamefont
  {Llanes-Estrada}}]{General:2006ed}%
  \BibitemOpen
  \bibfield  {author} {\bibinfo {author} {\bibfnamefont {I.~J.}\ \bibnamefont
  {General}}, \bibinfo {author} {\bibfnamefont {S.~R.}\ \bibnamefont
  {Cotanch}}, \ and\ \bibinfo {author} {\bibfnamefont {F.~J.}\ \bibnamefont
  {Llanes-Estrada}},\ }\href {\doibase 10.1140/epjc/s10052-007-0298-3}
  {\bibfield  {journal} {\bibinfo  {journal} {Eur. Phys. J.}\ }\textbf
  {\bibinfo {volume} {C51}},\ \bibinfo {pages} {347} (\bibinfo {year}
  {2007})},\ \Eprint {http://arxiv.org/abs/hep-ph/0609115}
  {arXiv:hep-ph/0609115 [hep-ph]} \BibitemShut {NoStop}%
%%CITATION = HEP-PH/0609115;%%
\bibitem [{\citenamefont {Guo}\ \emph {et~al.}(2008)\citenamefont {Guo},
  \citenamefont {Szczepaniak}, \citenamefont {Galata}, \citenamefont
  {Vassallo},\ and\ \citenamefont {Santopinto}}]{Guo:2008yz}%
  \BibitemOpen
  \bibfield  {author} {\bibinfo {author} {\bibfnamefont {P.}~\bibnamefont
  {Guo}}, \bibinfo {author} {\bibfnamefont {A.~P.}\ \bibnamefont
  {Szczepaniak}}, \bibinfo {author} {\bibfnamefont {G.}~\bibnamefont {Galata}},
  \bibinfo {author} {\bibfnamefont {A.}~\bibnamefont {Vassallo}}, \ and\
  \bibinfo {author} {\bibfnamefont {E.}~\bibnamefont {Santopinto}},\ }\href
  {\doibase 10.1103/PhysRevD.78.056003} {\bibfield  {journal} {\bibinfo
  {journal} {Phys. Rev.}\ }\textbf {\bibinfo {volume} {D78}},\ \bibinfo {pages}
  {056003} (\bibinfo {year} {2008})},\ \Eprint {http://arxiv.org/abs/0807.2721}
  {arXiv:0807.2721 [hep-ph]} \BibitemShut {NoStop}%
%%CITATION = ARXIV:0807.2721;%%
\bibitem [{\citenamefont {Tanimoto}(1982)}]{Tanimoto:1982eh}%
  \BibitemOpen
  \bibfield  {author} {\bibinfo {author} {\bibfnamefont {M.}~\bibnamefont
  {Tanimoto}},\ }\href {\doibase 10.1016/0370-2693(82)91008-5} {\bibfield
  {journal} {\bibinfo  {journal} {Phys. Lett.}\ }\textbf {\bibinfo {volume}
  {116B}},\ \bibinfo {pages} {198} (\bibinfo {year} {1982})}\BibitemShut
  {NoStop}%
%%CITATION = PHLTA,116B,198;%%
\bibitem [{\citenamefont {Iddir}\ \emph {et~al.}(1988)\citenamefont {Iddir},
  \citenamefont {Le~Yaouanc}, \citenamefont {Oliver}, \citenamefont {Pene},\
  and\ \citenamefont {Raynal}}]{Iddir:1988jc}%
  \BibitemOpen
  \bibfield  {author} {\bibinfo {author} {\bibfnamefont {F.}~\bibnamefont
  {Iddir}}, \bibinfo {author} {\bibfnamefont {A.}~\bibnamefont {Le~Yaouanc}},
  \bibinfo {author} {\bibfnamefont {L.}~\bibnamefont {Oliver}}, \bibinfo
  {author} {\bibfnamefont {O.}~\bibnamefont {Pene}}, \ and\ \bibinfo {author}
  {\bibfnamefont {J.~C.}\ \bibnamefont {Raynal}},\ }\href {\doibase
  10.1016/0370-2693(88)90584-9} {\bibfield  {journal} {\bibinfo  {journal}
  {Phys. Lett.}\ }\textbf {\bibinfo {volume} {B207}},\ \bibinfo {pages} {325}
  (\bibinfo {year} {1988})}\BibitemShut {NoStop}%
%%CITATION = PHLTA,B207,325;%%
\bibitem [{\citenamefont {Ishida}\ \emph {et~al.}(1993)\citenamefont {Ishida},
  \citenamefont {Sawazaki}, \citenamefont {Oda},\ and\ \citenamefont
  {Yamada}}]{Ishida:1991mx}%
  \BibitemOpen
  \bibfield  {author} {\bibinfo {author} {\bibfnamefont {S.}~\bibnamefont
  {Ishida}}, \bibinfo {author} {\bibfnamefont {H.}~\bibnamefont {Sawazaki}},
  \bibinfo {author} {\bibfnamefont {M.}~\bibnamefont {Oda}}, \ and\ \bibinfo
  {author} {\bibfnamefont {K.}~\bibnamefont {Yamada}},\ }\href {\doibase
  10.1103/PhysRevD.47.179} {\bibfield  {journal} {\bibinfo  {journal} {Phys.
  Rev.}\ }\textbf {\bibinfo {volume} {D47}},\ \bibinfo {pages} {179} (\bibinfo
  {year} {1993})}\BibitemShut {NoStop}%
%%CITATION = PHRVA,D47,179;%%
\bibitem [{\citenamefont {Close}\ and\ \citenamefont
  {Page}(1995)}]{Close:1994hc}%
  \BibitemOpen
  \bibfield  {author} {\bibinfo {author} {\bibfnamefont {F.~E.}\ \bibnamefont
  {Close}}\ and\ \bibinfo {author} {\bibfnamefont {P.~R.}\ \bibnamefont
  {Page}},\ }\href {\doibase 10.1016/0550-3213(95)00085-7} {\bibfield
  {journal} {\bibinfo  {journal} {Nucl. Phys.}\ }\textbf {\bibinfo {volume}
  {B443}},\ \bibinfo {pages} {233} (\bibinfo {year} {1995})},\ \Eprint
  {http://arxiv.org/abs/hep-ph/9411301} {arXiv:hep-ph/9411301 [hep-ph]}
  \BibitemShut {NoStop}%
%%CITATION = HEP-PH/9411301;%%
\bibitem [{\citenamefont {Swanson}\ and\ \citenamefont
  {Szczepaniak}(1997)}]{Swanson:1997wy}%
  \BibitemOpen
  \bibfield  {author} {\bibinfo {author} {\bibfnamefont {E.~S.}\ \bibnamefont
  {Swanson}}\ and\ \bibinfo {author} {\bibfnamefont {A.~P.}\ \bibnamefont
  {Szczepaniak}},\ }\href {\doibase 10.1103/PhysRevD.56.5692} {\bibfield
  {journal} {\bibinfo  {journal} {Phys. Rev.}\ }\textbf {\bibinfo {volume}
  {D56}},\ \bibinfo {pages} {5692} (\bibinfo {year} {1997})},\ \Eprint
  {http://arxiv.org/abs/hep-ph/9704434} {arXiv:hep-ph/9704434 [hep-ph]}
  \BibitemShut {NoStop}%
%%CITATION = HEP-PH/9704434;%%
\bibitem [{\citenamefont {Page}\ \emph {et~al.}(1999)\citenamefont {Page},
  \citenamefont {Swanson},\ and\ \citenamefont {Szczepaniak}}]{Page:1998gz}%
  \BibitemOpen
  \bibfield  {author} {\bibinfo {author} {\bibfnamefont {P.~R.}\ \bibnamefont
  {Page}}, \bibinfo {author} {\bibfnamefont {E.~S.}\ \bibnamefont {Swanson}}, \
  and\ \bibinfo {author} {\bibfnamefont {A.~P.}\ \bibnamefont {Szczepaniak}},\
  }\href {\doibase 10.1103/PhysRevD.59.034016} {\bibfield  {journal} {\bibinfo
  {journal} {Phys. Rev.}\ }\textbf {\bibinfo {volume} {D59}},\ \bibinfo {pages}
  {034016} (\bibinfo {year} {1999})},\ \Eprint
  {http://arxiv.org/abs/hep-ph/9808346} {arXiv:hep-ph/9808346 [hep-ph]}
  \BibitemShut {NoStop}%
%%CITATION = HEP-PH/9808346;%%
\bibitem [{\citenamefont {Dudek}(2011)}]{Dudek:2011bn}%
  \BibitemOpen
  \bibfield  {author} {\bibinfo {author} {\bibfnamefont {J.~J.}\ \bibnamefont
  {Dudek}},\ }\href {\doibase 10.1103/PhysRevD.84.074023} {\bibfield  {journal}
  {\bibinfo  {journal} {Phys. Rev.}\ }\textbf {\bibinfo {volume} {D84}},\
  \bibinfo {pages} {074023} (\bibinfo {year} {2011})},\ \Eprint
  {http://arxiv.org/abs/1106.5515} {arXiv:1106.5515 [hep-ph]} \BibitemShut
  {NoStop}%
%%CITATION = ARXIV:1106.5515;%%
\bibitem [{\citenamefont {Meyer}\ and\ \citenamefont
  {Van~Haarlem}(2010)}]{Meyer:2010ku}%
  \BibitemOpen
  \bibfield  {author} {\bibinfo {author} {\bibfnamefont {C.~A.}\ \bibnamefont
  {Meyer}}\ and\ \bibinfo {author} {\bibfnamefont {Y.}~\bibnamefont
  {Van~Haarlem}},\ }\href {\doibase 10.1103/PhysRevC.82.025208} {\bibfield
  {journal} {\bibinfo  {journal} {Phys. Rev.}\ }\textbf {\bibinfo {volume}
  {C82}},\ \bibinfo {pages} {025208} (\bibinfo {year} {2010})},\ \Eprint
  {http://arxiv.org/abs/1004.5516} {arXiv:1004.5516 [nucl-ex]} \BibitemShut
  {NoStop}%
%%CITATION = ARXIV:1004.5516;%%
\bibitem [{\citenamefont {Meyer}\ and\ \citenamefont
  {Swanson}(2015)}]{Meyer:2015eta}%
  \BibitemOpen
  \bibfield  {author} {\bibinfo {author} {\bibfnamefont {C.~A.}\ \bibnamefont
  {Meyer}}\ and\ \bibinfo {author} {\bibfnamefont {E.~S.}\ \bibnamefont
  {Swanson}},\ }\href {\doibase 10.1016/j.ppnp.2015.03.001} {\bibfield
  {journal} {\bibinfo  {journal} {Prog. Part. Nucl. Phys.}\ }\textbf {\bibinfo
  {volume} {82}},\ \bibinfo {pages} {21} (\bibinfo {year} {2015})},\ \Eprint
  {http://arxiv.org/abs/1502.07276} {arXiv:1502.07276 [hep-ph]} \BibitemShut
  {NoStop}%
%%CITATION = ARXIV:1502.07276;%%
\bibitem [{\citenamefont {Adolph}\ \emph {et~al.}(2015)\citenamefont {Adolph}
  \emph {et~al.}}]{Adolph:2014rpp}%
  \BibitemOpen
  \bibfield  {author} {\bibinfo {author} {\bibfnamefont {C.}~\bibnamefont
  {Adolph}} \emph {et~al.} (\bibinfo {collaboration} {COMPASS}),\ }\href
  {\doibase 10.1016/j.physletb.2014.11.058} {\bibfield  {journal} {\bibinfo
  {journal} {Phys. Lett.}\ }\textbf {\bibinfo {volume} {B740}},\ \bibinfo
  {pages} {303} (\bibinfo {year} {2015})},\ \Eprint
  {http://arxiv.org/abs/1408.4286} {arXiv:1408.4286 [hep-ex]} \BibitemShut
  {NoStop}%
%%CITATION = ARXIV:1408.4286;%%
\bibitem [{\citenamefont {Rodas}\ \emph {et~al.}(2019)\citenamefont {Rodas}
  \emph {et~al.}}]{Rodas:2018owy}%
  \BibitemOpen
  \bibfield  {author} {\bibinfo {author} {\bibfnamefont {A.}~\bibnamefont
  {Rodas}} \emph {et~al.} (\bibinfo {collaboration} {JPAC}),\ }\href {\doibase
  10.1103/PhysRevLett.122.042002} {\bibfield  {journal} {\bibinfo  {journal}
  {Phys. Rev. Lett.}\ }\textbf {\bibinfo {volume} {122}},\ \bibinfo {pages}
  {042002} (\bibinfo {year} {2019})},\ \Eprint
  {http://arxiv.org/abs/1810.04171} {arXiv:1810.04171 [hep-ph]} \BibitemShut
  {NoStop}%
%%CITATION = ARXIV:1810.04171;%%
\bibitem [{\citenamefont {Kopf}\ \emph {et~al.}(2020)\citenamefont {Kopf},
  \citenamefont {Albrecht}, \citenamefont {Koch}, \citenamefont {Pychy},
  \citenamefont {Qin},\ and\ \citenamefont {Wiedner}}]{Kopf:2020yoa}%
  \BibitemOpen
  \bibfield  {author} {\bibinfo {author} {\bibfnamefont {B.}~\bibnamefont
  {Kopf}}, \bibinfo {author} {\bibfnamefont {M.}~\bibnamefont {Albrecht}},
  \bibinfo {author} {\bibfnamefont {H.}~\bibnamefont {Koch}}, \bibinfo {author}
  {\bibfnamefont {J.}~\bibnamefont {Pychy}}, \bibinfo {author} {\bibfnamefont
  {X.}~\bibnamefont {Qin}}, \ and\ \bibinfo {author} {\bibfnamefont
  {U.}~\bibnamefont {Wiedner}},\ }\href@noop {} {\  (\bibinfo {year} {2020})},\
  \Eprint {http://arxiv.org/abs/2008.11566} {arXiv:2008.11566 [hep-ph]}
  \BibitemShut {NoStop}%
\bibitem [{\citenamefont {AlekSejevs}\ \emph {et~al.}(2013)\citenamefont
  {AlekSejevs} \emph {et~al.}}]{AlekSejevs:2013mkl}%
  \BibitemOpen
  \bibfield  {author} {\bibinfo {author} {\bibfnamefont {A.}~\bibnamefont
  {AlekSejevs}} \emph {et~al.} (\bibinfo {collaboration} {GlueX}),\ }\href@noop
  {} {\  (\bibinfo {year} {2013})},\ \Eprint {http://arxiv.org/abs/1305.1523}
  {arXiv:1305.1523 [nucl-ex]} \BibitemShut {NoStop}%
\bibitem [{\citenamefont {Dobbs}(2020)}]{Dobbs:2019sgr}%
  \BibitemOpen
  \bibfield  {author} {\bibinfo {author} {\bibfnamefont {S.}~\bibnamefont
  {Dobbs}} (\bibinfo {collaboration} {GlueX}),\ }\href {\doibase
  10.1063/5.0008562} {\bibfield  {journal} {\bibinfo  {journal} {AIP Conf.
  Proc.}\ }\textbf {\bibinfo {volume} {2249}},\ \bibinfo {pages} {020001}
  (\bibinfo {year} {2020})},\ \Eprint {http://arxiv.org/abs/1908.09711}
  {arXiv:1908.09711 [nucl-ex]} \BibitemShut {NoStop}%
\bibitem [{\citenamefont {Dudek}\ \emph {et~al.}(2009)\citenamefont {Dudek},
  \citenamefont {Edwards}, \citenamefont {Peardon}, \citenamefont {Richards},\
  and\ \citenamefont {Thomas}}]{Dudek:2009qf}%
  \BibitemOpen
  \bibfield  {author} {\bibinfo {author} {\bibfnamefont {J.~J.}\ \bibnamefont
  {Dudek}}, \bibinfo {author} {\bibfnamefont {R.~G.}\ \bibnamefont {Edwards}},
  \bibinfo {author} {\bibfnamefont {M.~J.}\ \bibnamefont {Peardon}}, \bibinfo
  {author} {\bibfnamefont {D.~G.}\ \bibnamefont {Richards}}, \ and\ \bibinfo
  {author} {\bibfnamefont {C.~E.}\ \bibnamefont {Thomas}},\ }\href {\doibase
  10.1103/PhysRevLett.103.262001} {\bibfield  {journal} {\bibinfo  {journal}
  {Phys. Rev. Lett.}\ }\textbf {\bibinfo {volume} {103}},\ \bibinfo {pages}
  {262001} (\bibinfo {year} {2009})},\ \Eprint {http://arxiv.org/abs/0909.0200}
  {arXiv:0909.0200 [hep-ph]} \BibitemShut {NoStop}%
%%CITATION = ARXIV:0909.0200;%%
\bibitem [{\citenamefont {Dudek}\ \emph {et~al.}(2010)\citenamefont {Dudek},
  \citenamefont {Edwards}, \citenamefont {Peardon}, \citenamefont {Richards},\
  and\ \citenamefont {Thomas}}]{Dudek:2010wm}%
  \BibitemOpen
  \bibfield  {author} {\bibinfo {author} {\bibfnamefont {J.~J.}\ \bibnamefont
  {Dudek}}, \bibinfo {author} {\bibfnamefont {R.~G.}\ \bibnamefont {Edwards}},
  \bibinfo {author} {\bibfnamefont {M.~J.}\ \bibnamefont {Peardon}}, \bibinfo
  {author} {\bibfnamefont {D.~G.}\ \bibnamefont {Richards}}, \ and\ \bibinfo
  {author} {\bibfnamefont {C.~E.}\ \bibnamefont {Thomas}},\ }\href {\doibase
  10.1103/PhysRevD.82.034508} {\bibfield  {journal} {\bibinfo  {journal} {Phys.
  Rev.}\ }\textbf {\bibinfo {volume} {D82}},\ \bibinfo {pages} {034508}
  (\bibinfo {year} {2010})},\ \Eprint {http://arxiv.org/abs/1004.4930}
  {arXiv:1004.4930 [hep-ph]} \BibitemShut {NoStop}%
%%CITATION = ARXIV:1004.4930;%%
\bibitem [{\citenamefont {Dudek}\ \emph
  {et~al.}(2011{\natexlab{a}})\citenamefont {Dudek}, \citenamefont {Edwards},
  \citenamefont {Joo}, \citenamefont {Peardon}, \citenamefont {Richards},\ and\
  \citenamefont {Thomas}}]{Dudek:2011tt}%
  \BibitemOpen
  \bibfield  {author} {\bibinfo {author} {\bibfnamefont {J.~J.}\ \bibnamefont
  {Dudek}}, \bibinfo {author} {\bibfnamefont {R.~G.}\ \bibnamefont {Edwards}},
  \bibinfo {author} {\bibfnamefont {B.}~\bibnamefont {Joo}}, \bibinfo {author}
  {\bibfnamefont {M.~J.}\ \bibnamefont {Peardon}}, \bibinfo {author}
  {\bibfnamefont {D.~G.}\ \bibnamefont {Richards}}, \ and\ \bibinfo {author}
  {\bibfnamefont {C.~E.}\ \bibnamefont {Thomas}},\ }\href {\doibase
  10.1103/PhysRevD.83.111502} {\bibfield  {journal} {\bibinfo  {journal} {Phys.
  Rev.}\ }\textbf {\bibinfo {volume} {D83}},\ \bibinfo {pages} {111502}
  (\bibinfo {year} {2011}{\natexlab{a}})},\ \Eprint
  {http://arxiv.org/abs/1102.4299} {arXiv:1102.4299 [hep-lat]} \BibitemShut
  {NoStop}%
%%CITATION = ARXIV:1102.4299;%%
\bibitem [{\citenamefont {Thomas}\ \emph {et~al.}(2012)\citenamefont {Thomas},
  \citenamefont {Edwards},\ and\ \citenamefont {Dudek}}]{Thomas:2011rh}%
  \BibitemOpen
  \bibfield  {author} {\bibinfo {author} {\bibfnamefont {C.~E.}\ \bibnamefont
  {Thomas}}, \bibinfo {author} {\bibfnamefont {R.~G.}\ \bibnamefont {Edwards}},
  \ and\ \bibinfo {author} {\bibfnamefont {J.~J.}\ \bibnamefont {Dudek}},\
  }\href {\doibase 10.1103/PhysRevD.85.014507, 10.1103/PhysRevD.85.039901}
  {\bibfield  {journal} {\bibinfo  {journal} {Phys. Rev.}\ }\textbf {\bibinfo
  {volume} {D85}},\ \bibinfo {pages} {014507} (\bibinfo {year} {2012})},\
  \Eprint {http://arxiv.org/abs/1107.1930} {arXiv:1107.1930 [hep-lat]}
  \BibitemShut {NoStop}%
%%CITATION = ARXIV:1107.1930;%%
\bibitem [{\citenamefont {Dudek}\ \emph
  {et~al.}(2013{\natexlab{a}})\citenamefont {Dudek}, \citenamefont {Edwards},
  \citenamefont {Guo},\ and\ \citenamefont {Thomas}}]{Dudek:2013yja}%
  \BibitemOpen
  \bibfield  {author} {\bibinfo {author} {\bibfnamefont {J.~J.}\ \bibnamefont
  {Dudek}}, \bibinfo {author} {\bibfnamefont {R.~G.}\ \bibnamefont {Edwards}},
  \bibinfo {author} {\bibfnamefont {P.}~\bibnamefont {Guo}}, \ and\ \bibinfo
  {author} {\bibfnamefont {C.~E.}\ \bibnamefont {Thomas}} (\bibinfo
  {collaboration} {Hadron Spectrum}),\ }\href {\doibase
  10.1103/PhysRevD.88.094505} {\bibfield  {journal} {\bibinfo  {journal} {Phys.
  Rev.}\ }\textbf {\bibinfo {volume} {D88}},\ \bibinfo {pages} {094505}
  (\bibinfo {year} {2013}{\natexlab{a}})},\ \Eprint
  {http://arxiv.org/abs/1309.2608} {arXiv:1309.2608 [hep-lat]} \BibitemShut
  {NoStop}%
%%CITATION = ARXIV:1309.2608;%%
\bibitem [{\citenamefont {Liu}\ \emph {et~al.}(2012)\citenamefont {Liu},
  \citenamefont {Moir}, \citenamefont {Peardon}, \citenamefont {Ryan},
  \citenamefont {Thomas}, \citenamefont {Vilaseca}, \citenamefont {Dudek},
  \citenamefont {Edwards}, \citenamefont {Joo},\ and\ \citenamefont
  {Richards}}]{Liu:2012ze}%
  \BibitemOpen
  \bibfield  {author} {\bibinfo {author} {\bibfnamefont {L.}~\bibnamefont
  {Liu}}, \bibinfo {author} {\bibfnamefont {G.}~\bibnamefont {Moir}}, \bibinfo
  {author} {\bibfnamefont {M.}~\bibnamefont {Peardon}}, \bibinfo {author}
  {\bibfnamefont {S.~M.}\ \bibnamefont {Ryan}}, \bibinfo {author}
  {\bibfnamefont {C.~E.}\ \bibnamefont {Thomas}}, \bibinfo {author}
  {\bibfnamefont {P.}~\bibnamefont {Vilaseca}}, \bibinfo {author}
  {\bibfnamefont {J.~J.}\ \bibnamefont {Dudek}}, \bibinfo {author}
  {\bibfnamefont {R.~G.}\ \bibnamefont {Edwards}}, \bibinfo {author}
  {\bibfnamefont {B.}~\bibnamefont {Joo}}, \ and\ \bibinfo {author}
  {\bibfnamefont {D.~G.}\ \bibnamefont {Richards}} (\bibinfo {collaboration}
  {Hadron Spectrum}),\ }\href {\doibase 10.1007/JHEP07(2012)126} {\bibfield
  {journal} {\bibinfo  {journal} {JHEP}\ }\textbf {\bibinfo {volume} {07}},\
  \bibinfo {pages} {126} (\bibinfo {year} {2012})},\ \Eprint
  {http://arxiv.org/abs/1204.5425} {arXiv:1204.5425 [hep-ph]} \BibitemShut
  {NoStop}%
%%CITATION = ARXIV:1204.5425;%%
\bibitem [{\citenamefont {Cheung}\ \emph {et~al.}(2016)\citenamefont {Cheung},
  \citenamefont {O'Hara}, \citenamefont {Moir}, \citenamefont {Peardon},
  \citenamefont {Ryan}, \citenamefont {Thomas},\ and\ \citenamefont
  {Tims}}]{Cheung:2016bym}%
  \BibitemOpen
  \bibfield  {author} {\bibinfo {author} {\bibfnamefont {G.~K.~C.}\
  \bibnamefont {Cheung}}, \bibinfo {author} {\bibfnamefont {C.}~\bibnamefont
  {O'Hara}}, \bibinfo {author} {\bibfnamefont {G.}~\bibnamefont {Moir}},
  \bibinfo {author} {\bibfnamefont {M.}~\bibnamefont {Peardon}}, \bibinfo
  {author} {\bibfnamefont {S.~M.}\ \bibnamefont {Ryan}}, \bibinfo {author}
  {\bibfnamefont {C.~E.}\ \bibnamefont {Thomas}}, \ and\ \bibinfo {author}
  {\bibfnamefont {D.}~\bibnamefont {Tims}} (\bibinfo {collaboration} {Hadron
  Spectrum}),\ }\href {\doibase 10.1007/JHEP12(2016)089} {\bibfield  {journal}
  {\bibinfo  {journal} {JHEP}\ }\textbf {\bibinfo {volume} {12}},\ \bibinfo
  {pages} {089} (\bibinfo {year} {2016})},\ \Eprint
  {http://arxiv.org/abs/1610.01073} {arXiv:1610.01073 [hep-lat]} \BibitemShut
  {NoStop}%
%%CITATION = ARXIV:1610.01073;%%
\bibitem [{\citenamefont {Dudek}\ and\ \citenamefont
  {Edwards}(2012)}]{Dudek:2012ag}%
  \BibitemOpen
  \bibfield  {author} {\bibinfo {author} {\bibfnamefont {J.~J.}\ \bibnamefont
  {Dudek}}\ and\ \bibinfo {author} {\bibfnamefont {R.~G.}\ \bibnamefont
  {Edwards}},\ }\href {\doibase 10.1103/PhysRevD.85.054016} {\bibfield
  {journal} {\bibinfo  {journal} {Phys. Rev.}\ }\textbf {\bibinfo {volume}
  {D85}},\ \bibinfo {pages} {054016} (\bibinfo {year} {2012})},\ \Eprint
  {http://arxiv.org/abs/1201.2349} {arXiv:1201.2349 [hep-ph]} \BibitemShut
  {NoStop}%
%%CITATION = ARXIV:1201.2349;%%
\bibitem [{\citenamefont {Luscher}(1986{\natexlab{a}})}]{Luscher:1985dn}%
  \BibitemOpen
  \bibfield  {author} {\bibinfo {author} {\bibfnamefont {M.}~\bibnamefont
  {Luscher}},\ }\href {\doibase 10.1007/BF01211589} {\bibfield  {journal}
  {\bibinfo  {journal} {Commun. Math. Phys.}\ }\textbf {\bibinfo {volume}
  {104}},\ \bibinfo {pages} {177} (\bibinfo {year}
  {1986}{\natexlab{a}})}\BibitemShut {NoStop}%
%%CITATION = CMPHA,104,177;%%
\bibitem [{\citenamefont {Luscher}(1986{\natexlab{b}})}]{Luscher:1986pf}%
  \BibitemOpen
  \bibfield  {author} {\bibinfo {author} {\bibfnamefont {M.}~\bibnamefont
  {Luscher}},\ }\href {\doibase 10.1007/BF01211097} {\bibfield  {journal}
  {\bibinfo  {journal} {Commun. Math. Phys.}\ }\textbf {\bibinfo {volume}
  {105}},\ \bibinfo {pages} {153} (\bibinfo {year}
  {1986}{\natexlab{b}})}\BibitemShut {NoStop}%
%%CITATION = CMPHA,105,153;%%
\bibitem [{\citenamefont {Luscher}(1991{\natexlab{a}})}]{Luscher:1990ux}%
  \BibitemOpen
  \bibfield  {author} {\bibinfo {author} {\bibfnamefont {M.}~\bibnamefont
  {Luscher}},\ }\href {\doibase 10.1016/0550-3213(91)90366-6} {\bibfield
  {journal} {\bibinfo  {journal} {Nucl. Phys.}\ }\textbf {\bibinfo {volume}
  {B354}},\ \bibinfo {pages} {531} (\bibinfo {year}
  {1991}{\natexlab{a}})}\BibitemShut {NoStop}%
%%CITATION = NUPHA,B354,531;%%
\bibitem [{\citenamefont {Luscher}(1991{\natexlab{b}})}]{Luscher:1991cf}%
  \BibitemOpen
  \bibfield  {author} {\bibinfo {author} {\bibfnamefont {M.}~\bibnamefont
  {Luscher}},\ }\href {\doibase 10.1016/0550-3213(91)90584-K} {\bibfield
  {journal} {\bibinfo  {journal} {Nucl. Phys.}\ }\textbf {\bibinfo {volume}
  {B364}},\ \bibinfo {pages} {237} (\bibinfo {year}
  {1991}{\natexlab{b}})}\BibitemShut {NoStop}%
%%CITATION = NUPHA,B364,237;%%
\bibitem [{\citenamefont {Rummukainen}\ and\ \citenamefont
  {Gottlieb}(1995)}]{Rummukainen:1995vs}%
  \BibitemOpen
  \bibfield  {author} {\bibinfo {author} {\bibfnamefont {K.}~\bibnamefont
  {Rummukainen}}\ and\ \bibinfo {author} {\bibfnamefont {S.~A.}\ \bibnamefont
  {Gottlieb}},\ }\href {\doibase 10.1016/0550-3213(95)00313-H} {\bibfield
  {journal} {\bibinfo  {journal} {Nucl. Phys.}\ }\textbf {\bibinfo {volume}
  {B450}},\ \bibinfo {pages} {397} (\bibinfo {year} {1995})},\ \Eprint
  {http://arxiv.org/abs/hep-lat/9503028} {arXiv:hep-lat/9503028 [hep-lat]}
  \BibitemShut {NoStop}%
%%CITATION = HEP-LAT/9503028;%%
\bibitem [{\citenamefont {He}\ \emph {et~al.}(2005)\citenamefont {He},
  \citenamefont {Feng},\ and\ \citenamefont {Liu}}]{He:2005ey}%
  \BibitemOpen
  \bibfield  {author} {\bibinfo {author} {\bibfnamefont {S.}~\bibnamefont
  {He}}, \bibinfo {author} {\bibfnamefont {X.}~\bibnamefont {Feng}}, \ and\
  \bibinfo {author} {\bibfnamefont {C.}~\bibnamefont {Liu}},\ }\href {\doibase
  10.1088/1126-6708/2005/07/011} {\bibfield  {journal} {\bibinfo  {journal}
  {JHEP}\ }\textbf {\bibinfo {volume} {07}},\ \bibinfo {pages} {011} (\bibinfo
  {year} {2005})},\ \Eprint {http://arxiv.org/abs/hep-lat/0504019}
  {arXiv:hep-lat/0504019 [hep-lat]} \BibitemShut {NoStop}%
%%CITATION = HEP-LAT/0504019;%%
\bibitem [{\citenamefont {Kim}\ \emph {et~al.}(2005)\citenamefont {Kim},
  \citenamefont {Sachrajda},\ and\ \citenamefont {Sharpe}}]{Kim:2005gf}%
  \BibitemOpen
  \bibfield  {author} {\bibinfo {author} {\bibfnamefont {C.~H.}\ \bibnamefont
  {Kim}}, \bibinfo {author} {\bibfnamefont {C.~T.}\ \bibnamefont {Sachrajda}},
  \ and\ \bibinfo {author} {\bibfnamefont {S.~R.}\ \bibnamefont {Sharpe}},\
  }\href {\doibase 10.1016/j.nuclphysb.2005.08.029} {\bibfield  {journal}
  {\bibinfo  {journal} {Nucl. Phys.}\ }\textbf {\bibinfo {volume} {B727}},\
  \bibinfo {pages} {218} (\bibinfo {year} {2005})},\ \Eprint
  {http://arxiv.org/abs/hep-lat/0507006} {arXiv:hep-lat/0507006 [hep-lat]}
  \BibitemShut {NoStop}%
%%CITATION = HEP-LAT/0507006;%%
\bibitem [{\citenamefont {Christ}\ \emph {et~al.}(2005)\citenamefont {Christ},
  \citenamefont {Kim},\ and\ \citenamefont {Yamazaki}}]{Christ:2005gi}%
  \BibitemOpen
  \bibfield  {author} {\bibinfo {author} {\bibfnamefont {N.~H.}\ \bibnamefont
  {Christ}}, \bibinfo {author} {\bibfnamefont {C.}~\bibnamefont {Kim}}, \ and\
  \bibinfo {author} {\bibfnamefont {T.}~\bibnamefont {Yamazaki}},\ }\href
  {\doibase 10.1103/PhysRevD.72.114506} {\bibfield  {journal} {\bibinfo
  {journal} {Phys. Rev.}\ }\textbf {\bibinfo {volume} {D72}},\ \bibinfo {pages}
  {114506} (\bibinfo {year} {2005})},\ \Eprint
  {http://arxiv.org/abs/hep-lat/0507009} {arXiv:hep-lat/0507009 [hep-lat]}
  \BibitemShut {NoStop}%
%%CITATION = HEP-LAT/0507009;%%
\bibitem [{\citenamefont {Fu}(2012)}]{Fu:2011xz}%
  \BibitemOpen
  \bibfield  {author} {\bibinfo {author} {\bibfnamefont {Z.}~\bibnamefont
  {Fu}},\ }\href {\doibase 10.1103/PhysRevD.85.014506} {\bibfield  {journal}
  {\bibinfo  {journal} {Phys. Rev.}\ }\textbf {\bibinfo {volume} {D85}},\
  \bibinfo {pages} {014506} (\bibinfo {year} {2012})},\ \Eprint
  {http://arxiv.org/abs/1110.0319} {arXiv:1110.0319 [hep-lat]} \BibitemShut
  {NoStop}%
%%CITATION = ARXIV:1110.0319;%%
\bibitem [{\citenamefont {Guo}\ \emph {et~al.}(2013)\citenamefont {Guo},
  \citenamefont {Dudek}, \citenamefont {Edwards},\ and\ \citenamefont
  {Szczepaniak}}]{Guo:2012hv}%
  \BibitemOpen
  \bibfield  {author} {\bibinfo {author} {\bibfnamefont {P.}~\bibnamefont
  {Guo}}, \bibinfo {author} {\bibfnamefont {J.}~\bibnamefont {Dudek}}, \bibinfo
  {author} {\bibfnamefont {R.}~\bibnamefont {Edwards}}, \ and\ \bibinfo
  {author} {\bibfnamefont {A.~P.}\ \bibnamefont {Szczepaniak}},\ }\href
  {\doibase 10.1103/PhysRevD.88.014501} {\bibfield  {journal} {\bibinfo
  {journal} {Phys. Rev.}\ }\textbf {\bibinfo {volume} {D88}},\ \bibinfo {pages}
  {014501} (\bibinfo {year} {2013})},\ \Eprint {http://arxiv.org/abs/1211.0929}
  {arXiv:1211.0929 [hep-lat]} \BibitemShut {NoStop}%
%%CITATION = ARXIV:1211.0929;%%
\bibitem [{\citenamefont {Hansen}\ and\ \citenamefont
  {Sharpe}(2012)}]{Hansen:2012tf}%
  \BibitemOpen
  \bibfield  {author} {\bibinfo {author} {\bibfnamefont {M.~T.}\ \bibnamefont
  {Hansen}}\ and\ \bibinfo {author} {\bibfnamefont {S.~R.}\ \bibnamefont
  {Sharpe}},\ }\href {\doibase 10.1103/PhysRevD.86.016007} {\bibfield
  {journal} {\bibinfo  {journal} {Phys. Rev.}\ }\textbf {\bibinfo {volume}
  {D86}},\ \bibinfo {pages} {016007} (\bibinfo {year} {2012})},\ \Eprint
  {http://arxiv.org/abs/1204.0826} {arXiv:1204.0826 [hep-lat]} \BibitemShut
  {NoStop}%
%%CITATION = ARXIV:1204.0826;%%
\bibitem [{\citenamefont {Briceño}\ and\ \citenamefont
  {Davoudi}(2013)}]{Briceno:2012yi}%
  \BibitemOpen
  \bibfield  {author} {\bibinfo {author} {\bibfnamefont {R.~A.}\ \bibnamefont
  {Briceño}}\ and\ \bibinfo {author} {\bibfnamefont {Z.}~\bibnamefont
  {Davoudi}},\ }\href {\doibase 10.1103/PhysRevD.88.094507} {\bibfield
  {journal} {\bibinfo  {journal} {Phys. Rev.}\ }\textbf {\bibinfo {volume}
  {D88}},\ \bibinfo {pages} {094507} (\bibinfo {year} {2013})},\ \Eprint
  {http://arxiv.org/abs/1204.1110} {arXiv:1204.1110 [hep-lat]} \BibitemShut
  {NoStop}%
%%CITATION = ARXIV:1204.1110;%%
\bibitem [{\citenamefont {Gockeler}\ \emph {et~al.}(2012)\citenamefont
  {Gockeler}, \citenamefont {Horsley}, \citenamefont {Lage}, \citenamefont
  {Meissner}, \citenamefont {Rakow}, \citenamefont {Rusetsky}, \citenamefont
  {Schierholz},\ and\ \citenamefont {Zanotti}}]{Gockeler:2012yj}%
  \BibitemOpen
  \bibfield  {author} {\bibinfo {author} {\bibfnamefont {M.}~\bibnamefont
  {Gockeler}}, \bibinfo {author} {\bibfnamefont {R.}~\bibnamefont {Horsley}},
  \bibinfo {author} {\bibfnamefont {M.}~\bibnamefont {Lage}}, \bibinfo {author}
  {\bibfnamefont {U.~G.}\ \bibnamefont {Meissner}}, \bibinfo {author}
  {\bibfnamefont {P.~E.~L.}\ \bibnamefont {Rakow}}, \bibinfo {author}
  {\bibfnamefont {A.}~\bibnamefont {Rusetsky}}, \bibinfo {author}
  {\bibfnamefont {G.}~\bibnamefont {Schierholz}}, \ and\ \bibinfo {author}
  {\bibfnamefont {J.~M.}\ \bibnamefont {Zanotti}},\ }\href {\doibase
  10.1103/PhysRevD.86.094513} {\bibfield  {journal} {\bibinfo  {journal} {Phys.
  Rev.}\ }\textbf {\bibinfo {volume} {D86}},\ \bibinfo {pages} {094513}
  (\bibinfo {year} {2012})},\ \Eprint {http://arxiv.org/abs/1206.4141}
  {arXiv:1206.4141 [hep-lat]} \BibitemShut {NoStop}%
%%CITATION = ARXIV:1206.4141;%%
\bibitem [{\citenamefont {Leskovec}\ and\ \citenamefont
  {Prelovsek}(2012)}]{Leskovec:2012gb}%
  \BibitemOpen
  \bibfield  {author} {\bibinfo {author} {\bibfnamefont {L.}~\bibnamefont
  {Leskovec}}\ and\ \bibinfo {author} {\bibfnamefont {S.}~\bibnamefont
  {Prelovsek}},\ }\href {\doibase 10.1103/PhysRevD.85.114507} {\bibfield
  {journal} {\bibinfo  {journal} {Phys. Rev.}\ }\textbf {\bibinfo {volume}
  {D85}},\ \bibinfo {pages} {114507} (\bibinfo {year} {2012})},\ \Eprint
  {http://arxiv.org/abs/1202.2145} {arXiv:1202.2145 [hep-lat]} \BibitemShut
  {NoStop}%
%%CITATION = ARXIV:1202.2145;%%
\bibitem [{\citenamefont {Brice\~no}(2014)}]{Briceno:2014oea}%
  \BibitemOpen
  \bibfield  {author} {\bibinfo {author} {\bibfnamefont {R.~A.}\ \bibnamefont
  {Brice\~no}},\ }\href {\doibase 10.1103/PhysRevD.89.074507} {\bibfield
  {journal} {\bibinfo  {journal} {Phys. Rev.}\ }\textbf {\bibinfo {volume}
  {D89}},\ \bibinfo {pages} {074507} (\bibinfo {year} {2014})},\ \Eprint
  {http://arxiv.org/abs/1401.3312} {arXiv:1401.3312 [hep-lat]} \BibitemShut
  {NoStop}%
%%CITATION = ARXIV:1401.3312;%%
\bibitem [{\citenamefont {Dudek}\ \emph
  {et~al.}(2013{\natexlab{b}})\citenamefont {Dudek}, \citenamefont {Edwards},\
  and\ \citenamefont {Thomas}}]{Dudek:2012xn}%
  \BibitemOpen
  \bibfield  {author} {\bibinfo {author} {\bibfnamefont {J.~J.}\ \bibnamefont
  {Dudek}}, \bibinfo {author} {\bibfnamefont {R.~G.}\ \bibnamefont {Edwards}},
  \ and\ \bibinfo {author} {\bibfnamefont {C.~E.}\ \bibnamefont {Thomas}}
  (\bibinfo {collaboration} {Hadron Spectrum}),\ }\href {\doibase
  10.1103/PhysRevD.87.034505, 10.1103/PhysRevD.90.099902} {\bibfield  {journal}
  {\bibinfo  {journal} {Phys. Rev.}\ }\textbf {\bibinfo {volume} {D87}},\
  \bibinfo {pages} {034505} (\bibinfo {year} {2013}{\natexlab{b}})},\ \bibinfo
  {note} {[Erratum: Phys. Rev.D90,no.9,099902(2014)]},\ \Eprint
  {http://arxiv.org/abs/1212.0830} {arXiv:1212.0830 [hep-ph]} \BibitemShut
  {NoStop}%
%%CITATION = ARXIV:1212.0830;%%
\bibitem [{\citenamefont {Wilson}\ \emph
  {et~al.}(2015{\natexlab{a}})\citenamefont {Wilson}, \citenamefont {Briceño},
  \citenamefont {Dudek}, \citenamefont {Edwards},\ and\ \citenamefont
  {Thomas}}]{Wilson:2015dqa}%
  \BibitemOpen
  \bibfield  {author} {\bibinfo {author} {\bibfnamefont {D.~J.}\ \bibnamefont
  {Wilson}}, \bibinfo {author} {\bibfnamefont {R.~A.}\ \bibnamefont
  {Briceño}}, \bibinfo {author} {\bibfnamefont {J.~J.}\ \bibnamefont {Dudek}},
  \bibinfo {author} {\bibfnamefont {R.~G.}\ \bibnamefont {Edwards}}, \ and\
  \bibinfo {author} {\bibfnamefont {C.~E.}\ \bibnamefont {Thomas}},\ }\href
  {\doibase 10.1103/PhysRevD.92.094502} {\bibfield  {journal} {\bibinfo
  {journal} {Phys. Rev.}\ }\textbf {\bibinfo {volume} {D92}},\ \bibinfo {pages}
  {094502} (\bibinfo {year} {2015}{\natexlab{a}})},\ \Eprint
  {http://arxiv.org/abs/1507.02599} {arXiv:1507.02599 [hep-ph]} \BibitemShut
  {NoStop}%
%%CITATION = ARXIV:1507.02599;%%
\bibitem [{\citenamefont {Aoki}\ \emph {et~al.}(2007)\citenamefont {Aoki} \emph
  {et~al.}}]{Aoki:2007rd}%
  \BibitemOpen
  \bibfield  {author} {\bibinfo {author} {\bibfnamefont {S.}~\bibnamefont
  {Aoki}} \emph {et~al.} (\bibinfo {collaboration} {CP-PACS}),\ }\href
  {\doibase 10.1103/PhysRevD.76.094506} {\bibfield  {journal} {\bibinfo
  {journal} {Phys. Rev.}\ }\textbf {\bibinfo {volume} {D76}},\ \bibinfo {pages}
  {094506} (\bibinfo {year} {2007})},\ \Eprint {http://arxiv.org/abs/0708.3705}
  {arXiv:0708.3705 [hep-lat]} \BibitemShut {NoStop}%
%%CITATION = ARXIV:0708.3705;%%
\bibitem [{\citenamefont {Feng}\ \emph {et~al.}(2011)\citenamefont {Feng},
  \citenamefont {Jansen},\ and\ \citenamefont {Renner}}]{Feng:2010es}%
  \BibitemOpen
  \bibfield  {author} {\bibinfo {author} {\bibfnamefont {X.}~\bibnamefont
  {Feng}}, \bibinfo {author} {\bibfnamefont {K.}~\bibnamefont {Jansen}}, \ and\
  \bibinfo {author} {\bibfnamefont {D.~B.}\ \bibnamefont {Renner}},\ }\href
  {\doibase 10.1103/PhysRevD.83.094505} {\bibfield  {journal} {\bibinfo
  {journal} {Phys. Rev.}\ }\textbf {\bibinfo {volume} {D83}},\ \bibinfo {pages}
  {094505} (\bibinfo {year} {2011})},\ \Eprint {http://arxiv.org/abs/1011.5288}
  {arXiv:1011.5288 [hep-lat]} \BibitemShut {NoStop}%
%%CITATION = ARXIV:1011.5288;%%
\bibitem [{\citenamefont {Aoki}\ \emph {et~al.}(2011)\citenamefont {Aoki} \emph
  {et~al.}}]{Aoki:2011yj}%
  \BibitemOpen
  \bibfield  {author} {\bibinfo {author} {\bibfnamefont {S.}~\bibnamefont
  {Aoki}} \emph {et~al.} (\bibinfo {collaboration} {CS}),\ }\href {\doibase
  10.1103/PhysRevD.84.094505} {\bibfield  {journal} {\bibinfo  {journal} {Phys.
  Rev.}\ }\textbf {\bibinfo {volume} {D84}},\ \bibinfo {pages} {094505}
  (\bibinfo {year} {2011})},\ \Eprint {http://arxiv.org/abs/1106.5365}
  {arXiv:1106.5365 [hep-lat]} \BibitemShut {NoStop}%
%%CITATION = ARXIV:1106.5365;%%
\bibitem [{\citenamefont {Lang}\ \emph {et~al.}(2011)\citenamefont {Lang},
  \citenamefont {Mohler}, \citenamefont {Prelovsek},\ and\ \citenamefont
  {Vidmar}}]{Lang:2011mn}%
  \BibitemOpen
  \bibfield  {author} {\bibinfo {author} {\bibfnamefont {C.~B.}\ \bibnamefont
  {Lang}}, \bibinfo {author} {\bibfnamefont {D.}~\bibnamefont {Mohler}},
  \bibinfo {author} {\bibfnamefont {S.}~\bibnamefont {Prelovsek}}, \ and\
  \bibinfo {author} {\bibfnamefont {M.}~\bibnamefont {Vidmar}},\ }\href
  {\doibase 10.1103/PhysRevD.89.059903, 10.1103/PhysRevD.84.054503} {\bibfield
  {journal} {\bibinfo  {journal} {Phys. Rev.}\ }\textbf {\bibinfo {volume}
  {D84}},\ \bibinfo {pages} {054503} (\bibinfo {year} {2011})},\ \bibinfo
  {note} {[Erratum: Phys. Rev.D89,no.5,059903(2014)]},\ \Eprint
  {http://arxiv.org/abs/1105.5636} {arXiv:1105.5636 [hep-lat]} \BibitemShut
  {NoStop}%
%%CITATION = ARXIV:1105.5636;%%
\bibitem [{\citenamefont {Pelissier}\ and\ \citenamefont
  {Alexandru}(2013)}]{Pelissier:2012pi}%
  \BibitemOpen
  \bibfield  {author} {\bibinfo {author} {\bibfnamefont {C.}~\bibnamefont
  {Pelissier}}\ and\ \bibinfo {author} {\bibfnamefont {A.}~\bibnamefont
  {Alexandru}},\ }\href {\doibase 10.1103/PhysRevD.87.014503} {\bibfield
  {journal} {\bibinfo  {journal} {Phys. Rev.}\ }\textbf {\bibinfo {volume}
  {D87}},\ \bibinfo {pages} {014503} (\bibinfo {year} {2013})},\ \Eprint
  {http://arxiv.org/abs/1211.0092} {arXiv:1211.0092 [hep-lat]} \BibitemShut
  {NoStop}%
%%CITATION = ARXIV:1211.0092;%%
\bibitem [{\citenamefont {Bali}\ \emph {et~al.}(2016)\citenamefont {Bali},
  \citenamefont {Collins}, \citenamefont {Cox}, \citenamefont {Donald},
  \citenamefont {Göckeler}, \citenamefont {Lang},\ and\ \citenamefont
  {Schäfer}}]{Bali:2015gji}%
  \BibitemOpen
  \bibfield  {author} {\bibinfo {author} {\bibfnamefont {G.~S.}\ \bibnamefont
  {Bali}}, \bibinfo {author} {\bibfnamefont {S.}~\bibnamefont {Collins}},
  \bibinfo {author} {\bibfnamefont {A.}~\bibnamefont {Cox}}, \bibinfo {author}
  {\bibfnamefont {G.}~\bibnamefont {Donald}}, \bibinfo {author} {\bibfnamefont
  {M.}~\bibnamefont {Göckeler}}, \bibinfo {author} {\bibfnamefont {C.~B.}\
  \bibnamefont {Lang}}, \ and\ \bibinfo {author} {\bibfnamefont
  {A.}~\bibnamefont {Schäfer}} (\bibinfo {collaboration} {RQCD}),\ }\href
  {\doibase 10.1103/PhysRevD.93.054509} {\bibfield  {journal} {\bibinfo
  {journal} {Phys. Rev.}\ }\textbf {\bibinfo {volume} {D93}},\ \bibinfo {pages}
  {054509} (\bibinfo {year} {2016})},\ \Eprint
  {http://arxiv.org/abs/1512.08678} {arXiv:1512.08678 [hep-lat]} \BibitemShut
  {NoStop}%
%%CITATION = ARXIV:1512.08678;%%
\bibitem [{\citenamefont {Bulava}\ \emph {et~al.}(2016)\citenamefont {Bulava},
  \citenamefont {Fahy}, \citenamefont {Hörz}, \citenamefont {Juge},
  \citenamefont {Morningstar},\ and\ \citenamefont {Wong}}]{Bulava:2016mks}%
  \BibitemOpen
  \bibfield  {author} {\bibinfo {author} {\bibfnamefont {J.}~\bibnamefont
  {Bulava}}, \bibinfo {author} {\bibfnamefont {B.}~\bibnamefont {Fahy}},
  \bibinfo {author} {\bibfnamefont {B.}~\bibnamefont {Hörz}}, \bibinfo
  {author} {\bibfnamefont {K.~J.}\ \bibnamefont {Juge}}, \bibinfo {author}
  {\bibfnamefont {C.}~\bibnamefont {Morningstar}}, \ and\ \bibinfo {author}
  {\bibfnamefont {C.~H.}\ \bibnamefont {Wong}},\ }\href {\doibase
  10.1016/j.nuclphysb.2016.07.024} {\bibfield  {journal} {\bibinfo  {journal}
  {Nucl. Phys.}\ }\textbf {\bibinfo {volume} {B910}},\ \bibinfo {pages} {842}
  (\bibinfo {year} {2016})},\ \Eprint {http://arxiv.org/abs/1604.05593}
  {arXiv:1604.05593 [hep-lat]} \BibitemShut {NoStop}%
%%CITATION = ARXIV:1604.05593;%%
\bibitem [{\citenamefont {Guo}\ \emph {et~al.}(2016)\citenamefont {Guo},
  \citenamefont {Alexandru}, \citenamefont {Molina},\ and\ \citenamefont
  {Döring}}]{Guo:2016zos}%
  \BibitemOpen
  \bibfield  {author} {\bibinfo {author} {\bibfnamefont {D.}~\bibnamefont
  {Guo}}, \bibinfo {author} {\bibfnamefont {A.}~\bibnamefont {Alexandru}},
  \bibinfo {author} {\bibfnamefont {R.}~\bibnamefont {Molina}}, \ and\ \bibinfo
  {author} {\bibfnamefont {M.}~\bibnamefont {Döring}},\ }\href {\doibase
  10.1103/PhysRevD.94.034501} {\bibfield  {journal} {\bibinfo  {journal} {Phys.
  Rev.}\ }\textbf {\bibinfo {volume} {D94}},\ \bibinfo {pages} {034501}
  (\bibinfo {year} {2016})},\ \Eprint {http://arxiv.org/abs/1605.03993}
  {arXiv:1605.03993 [hep-lat]} \BibitemShut {NoStop}%
%%CITATION = ARXIV:1605.03993;%%
\bibitem [{\citenamefont {Hu}\ \emph {et~al.}(2016)\citenamefont {Hu},
  \citenamefont {Molina}, \citenamefont {D\"oring},\ and\ \citenamefont
  {Alexandru}}]{Hu:2016shf}%
  \BibitemOpen
  \bibfield  {author} {\bibinfo {author} {\bibfnamefont {B.}~\bibnamefont
  {Hu}}, \bibinfo {author} {\bibfnamefont {R.}~\bibnamefont {Molina}}, \bibinfo
  {author} {\bibfnamefont {M.}~\bibnamefont {D\"oring}}, \ and\ \bibinfo
  {author} {\bibfnamefont {A.}~\bibnamefont {Alexandru}},\ }\href {\doibase
  10.1103/PhysRevLett.117.122001} {\bibfield  {journal} {\bibinfo  {journal}
  {Phys. Rev. Lett.}\ }\textbf {\bibinfo {volume} {117}},\ \bibinfo {pages}
  {122001} (\bibinfo {year} {2016})},\ \Eprint
  {http://arxiv.org/abs/1605.04823} {arXiv:1605.04823 [hep-lat]} \BibitemShut
  {NoStop}%
\bibitem [{\citenamefont {Fu}\ and\ \citenamefont {Wang}(2016)}]{Fu:2016itp}%
  \BibitemOpen
  \bibfield  {author} {\bibinfo {author} {\bibfnamefont {Z.}~\bibnamefont
  {Fu}}\ and\ \bibinfo {author} {\bibfnamefont {L.}~\bibnamefont {Wang}},\
  }\href {\doibase 10.1103/PhysRevD.94.034505} {\bibfield  {journal} {\bibinfo
  {journal} {Phys. Rev.}\ }\textbf {\bibinfo {volume} {D94}},\ \bibinfo {pages}
  {034505} (\bibinfo {year} {2016})},\ \Eprint
  {http://arxiv.org/abs/1608.07478} {arXiv:1608.07478 [hep-lat]} \BibitemShut
  {NoStop}%
%%CITATION = ARXIV:1608.07478;%%
\bibitem [{\citenamefont {Alexandrou}\ \emph {et~al.}(2017)\citenamefont
  {Alexandrou}, \citenamefont {Leskovec}, \citenamefont {Meinel}, \citenamefont
  {Negele}, \citenamefont {Paul}, \citenamefont {Petschlies}, \citenamefont
  {Pochinsky}, \citenamefont {Rendon},\ and\ \citenamefont
  {Syritsyn}}]{Alexandrou:2017mpi}%
  \BibitemOpen
  \bibfield  {author} {\bibinfo {author} {\bibfnamefont {C.}~\bibnamefont
  {Alexandrou}}, \bibinfo {author} {\bibfnamefont {L.}~\bibnamefont
  {Leskovec}}, \bibinfo {author} {\bibfnamefont {S.}~\bibnamefont {Meinel}},
  \bibinfo {author} {\bibfnamefont {J.}~\bibnamefont {Negele}}, \bibinfo
  {author} {\bibfnamefont {S.}~\bibnamefont {Paul}}, \bibinfo {author}
  {\bibfnamefont {M.}~\bibnamefont {Petschlies}}, \bibinfo {author}
  {\bibfnamefont {A.}~\bibnamefont {Pochinsky}}, \bibinfo {author}
  {\bibfnamefont {G.}~\bibnamefont {Rendon}}, \ and\ \bibinfo {author}
  {\bibfnamefont {S.}~\bibnamefont {Syritsyn}},\ }\href {\doibase
  10.1103/PhysRevD.96.034525} {\bibfield  {journal} {\bibinfo  {journal} {Phys.
  Rev.}\ }\textbf {\bibinfo {volume} {D96}},\ \bibinfo {pages} {034525}
  (\bibinfo {year} {2017})},\ \Eprint {http://arxiv.org/abs/1704.05439}
  {arXiv:1704.05439 [hep-lat]} \BibitemShut {NoStop}%
%%CITATION = ARXIV:1704.05439;%%
\bibitem [{\citenamefont {Andersen}\ \emph {et~al.}(2019)\citenamefont
  {Andersen}, \citenamefont {Bulava}, \citenamefont {Hörz},\ and\
  \citenamefont {Morningstar}}]{Andersen:2018mau}%
  \BibitemOpen
  \bibfield  {author} {\bibinfo {author} {\bibfnamefont {C.}~\bibnamefont
  {Andersen}}, \bibinfo {author} {\bibfnamefont {J.}~\bibnamefont {Bulava}},
  \bibinfo {author} {\bibfnamefont {B.}~\bibnamefont {Hörz}}, \ and\ \bibinfo
  {author} {\bibfnamefont {C.}~\bibnamefont {Morningstar}},\ }\href {\doibase
  10.1016/j.nuclphysb.2018.12.018} {\bibfield  {journal} {\bibinfo  {journal}
  {Nucl. Phys.}\ }\textbf {\bibinfo {volume} {B939}},\ \bibinfo {pages} {145}
  (\bibinfo {year} {2019})},\ \Eprint {http://arxiv.org/abs/1808.05007}
  {arXiv:1808.05007 [hep-lat]} \BibitemShut {NoStop}%
%%CITATION = ARXIV:1808.05007;%%
\bibitem [{\citenamefont {Werner}\ \emph {et~al.}(2020)\citenamefont {Werner}
  \emph {et~al.}}]{Werner:2019hxc}%
  \BibitemOpen
  \bibfield  {author} {\bibinfo {author} {\bibfnamefont {M.}~\bibnamefont
  {Werner}} \emph {et~al.} (\bibinfo {collaboration} {Extended Twisted Mass}),\
  }\href {\doibase 10.1140/epja/s10050-020-00057-4} {\bibfield  {journal}
  {\bibinfo  {journal} {Eur. Phys. J. A}\ }\textbf {\bibinfo {volume} {56}},\
  \bibinfo {pages} {61} (\bibinfo {year} {2020})},\ \Eprint
  {http://arxiv.org/abs/1907.01237} {arXiv:1907.01237 [hep-lat]} \BibitemShut
  {NoStop}%
\bibitem [{\citenamefont {Erben}\ \emph {et~al.}(2020)\citenamefont {Erben},
  \citenamefont {Green}, \citenamefont {Mohler},\ and\ \citenamefont
  {Wittig}}]{Erben:2019nmx}%
  \BibitemOpen
  \bibfield  {author} {\bibinfo {author} {\bibfnamefont {F.}~\bibnamefont
  {Erben}}, \bibinfo {author} {\bibfnamefont {J.~R.}\ \bibnamefont {Green}},
  \bibinfo {author} {\bibfnamefont {D.}~\bibnamefont {Mohler}}, \ and\ \bibinfo
  {author} {\bibfnamefont {H.}~\bibnamefont {Wittig}},\ }\href {\doibase
  10.1103/PhysRevD.101.054504} {\bibfield  {journal} {\bibinfo  {journal}
  {Phys. Rev. D}\ }\textbf {\bibinfo {volume} {101}},\ \bibinfo {pages}
  {054504} (\bibinfo {year} {2020})},\ \Eprint
  {http://arxiv.org/abs/1910.01083} {arXiv:1910.01083 [hep-lat]} \BibitemShut
  {NoStop}%
\bibitem [{\citenamefont {Fischer}\ \emph {et~al.}(2020)\citenamefont
  {Fischer}, \citenamefont {Kostrzewa}, \citenamefont {Mai}, \citenamefont
  {Petschlies}, \citenamefont {Pittler}, \citenamefont {Ueding}, \citenamefont
  {Urbach},\ and\ \citenamefont {Werner}}]{Fischer:2020fvl}%
  \BibitemOpen
  \bibfield  {author} {\bibinfo {author} {\bibfnamefont {M.}~\bibnamefont
  {Fischer}}, \bibinfo {author} {\bibfnamefont {B.}~\bibnamefont {Kostrzewa}},
  \bibinfo {author} {\bibfnamefont {M.}~\bibnamefont {Mai}}, \bibinfo {author}
  {\bibfnamefont {M.}~\bibnamefont {Petschlies}}, \bibinfo {author}
  {\bibfnamefont {F.}~\bibnamefont {Pittler}}, \bibinfo {author} {\bibfnamefont
  {M.}~\bibnamefont {Ueding}}, \bibinfo {author} {\bibfnamefont
  {C.}~\bibnamefont {Urbach}}, \ and\ \bibinfo {author} {\bibfnamefont
  {M.}~\bibnamefont {Werner}} (\bibinfo {collaboration} {ETM}),\ }\href@noop {}
  {\  (\bibinfo {year} {2020})},\ \Eprint {http://arxiv.org/abs/2006.13805}
  {arXiv:2006.13805 [hep-lat]} \BibitemShut {NoStop}%
\bibitem [{\citenamefont {Dudek}\ \emph {et~al.}(2014)\citenamefont {Dudek},
  \citenamefont {Edwards}, \citenamefont {Thomas},\ and\ \citenamefont
  {Wilson}}]{Dudek:2014qha}%
  \BibitemOpen
  \bibfield  {author} {\bibinfo {author} {\bibfnamefont {J.~J.}\ \bibnamefont
  {Dudek}}, \bibinfo {author} {\bibfnamefont {R.~G.}\ \bibnamefont {Edwards}},
  \bibinfo {author} {\bibfnamefont {C.~E.}\ \bibnamefont {Thomas}}, \ and\
  \bibinfo {author} {\bibfnamefont {D.~J.}\ \bibnamefont {Wilson}} (\bibinfo
  {collaboration} {Hadron Spectrum}),\ }\href {\doibase
  10.1103/PhysRevLett.113.182001} {\bibfield  {journal} {\bibinfo  {journal}
  {Phys. Rev. Lett.}\ }\textbf {\bibinfo {volume} {113}},\ \bibinfo {pages}
  {182001} (\bibinfo {year} {2014})},\ \Eprint {http://arxiv.org/abs/1406.4158}
  {arXiv:1406.4158 [hep-ph]} \BibitemShut {NoStop}%
%%CITATION = ARXIV:1406.4158;%%
\bibitem [{\citenamefont {Wilson}\ \emph
  {et~al.}(2015{\natexlab{b}})\citenamefont {Wilson}, \citenamefont {Dudek},
  \citenamefont {Edwards},\ and\ \citenamefont {Thomas}}]{Wilson:2014cna}%
  \BibitemOpen
  \bibfield  {author} {\bibinfo {author} {\bibfnamefont {D.~J.}\ \bibnamefont
  {Wilson}}, \bibinfo {author} {\bibfnamefont {J.~J.}\ \bibnamefont {Dudek}},
  \bibinfo {author} {\bibfnamefont {R.~G.}\ \bibnamefont {Edwards}}, \ and\
  \bibinfo {author} {\bibfnamefont {C.~E.}\ \bibnamefont {Thomas}},\ }\href
  {\doibase 10.1103/PhysRevD.91.054008} {\bibfield  {journal} {\bibinfo
  {journal} {Phys. Rev.}\ }\textbf {\bibinfo {volume} {D91}},\ \bibinfo {pages}
  {054008} (\bibinfo {year} {2015}{\natexlab{b}})},\ \Eprint
  {http://arxiv.org/abs/1411.2004} {arXiv:1411.2004 [hep-ph]} \BibitemShut
  {NoStop}%
%%CITATION = ARXIV:1411.2004;%%
\bibitem [{\citenamefont {Dudek}\ \emph {et~al.}(2016)\citenamefont {Dudek},
  \citenamefont {Edwards},\ and\ \citenamefont {Wilson}}]{Dudek:2016cru}%
  \BibitemOpen
  \bibfield  {author} {\bibinfo {author} {\bibfnamefont {J.~J.}\ \bibnamefont
  {Dudek}}, \bibinfo {author} {\bibfnamefont {R.~G.}\ \bibnamefont {Edwards}},
  \ and\ \bibinfo {author} {\bibfnamefont {D.~J.}\ \bibnamefont {Wilson}}
  (\bibinfo {collaboration} {Hadron Spectrum}),\ }\href {\doibase
  10.1103/PhysRevD.93.094506} {\bibfield  {journal} {\bibinfo  {journal} {Phys.
  Rev.}\ }\textbf {\bibinfo {volume} {D93}},\ \bibinfo {pages} {094506}
  (\bibinfo {year} {2016})},\ \Eprint {http://arxiv.org/abs/1602.05122}
  {arXiv:1602.05122 [hep-ph]} \BibitemShut {NoStop}%
%%CITATION = ARXIV:1602.05122;%%
\bibitem [{\citenamefont {Moir}\ \emph {et~al.}(2016)\citenamefont {Moir},
  \citenamefont {Peardon}, \citenamefont {Ryan}, \citenamefont {Thomas},\ and\
  \citenamefont {Wilson}}]{Moir:2016srx}%
  \BibitemOpen
  \bibfield  {author} {\bibinfo {author} {\bibfnamefont {G.}~\bibnamefont
  {Moir}}, \bibinfo {author} {\bibfnamefont {M.}~\bibnamefont {Peardon}},
  \bibinfo {author} {\bibfnamefont {S.~M.}\ \bibnamefont {Ryan}}, \bibinfo
  {author} {\bibfnamefont {C.~E.}\ \bibnamefont {Thomas}}, \ and\ \bibinfo
  {author} {\bibfnamefont {D.~J.}\ \bibnamefont {Wilson}},\ }\href {\doibase
  10.1007/JHEP10(2016)011} {\bibfield  {journal} {\bibinfo  {journal} {JHEP}\
  }\textbf {\bibinfo {volume} {10}},\ \bibinfo {pages} {011} (\bibinfo {year}
  {2016})},\ \Eprint {http://arxiv.org/abs/1607.07093} {arXiv:1607.07093
  [hep-lat]} \BibitemShut {NoStop}%
%%CITATION = ARXIV:1607.07093;%%
\bibitem [{\citenamefont {Briceno}\ \emph
  {et~al.}(2018{\natexlab{a}})\citenamefont {Briceno}, \citenamefont {Dudek},
  \citenamefont {Edwards},\ and\ \citenamefont {Wilson}}]{Briceno:2017qmb}%
  \BibitemOpen
  \bibfield  {author} {\bibinfo {author} {\bibfnamefont {R.~A.}\ \bibnamefont
  {Briceno}}, \bibinfo {author} {\bibfnamefont {J.~J.}\ \bibnamefont {Dudek}},
  \bibinfo {author} {\bibfnamefont {R.~G.}\ \bibnamefont {Edwards}}, \ and\
  \bibinfo {author} {\bibfnamefont {D.~J.}\ \bibnamefont {Wilson}},\ }\href
  {\doibase 10.1103/PhysRevD.97.054513} {\bibfield  {journal} {\bibinfo
  {journal} {Phys. Rev.}\ }\textbf {\bibinfo {volume} {D97}},\ \bibinfo {pages}
  {054513} (\bibinfo {year} {2018}{\natexlab{a}})},\ \Eprint
  {http://arxiv.org/abs/1708.06667} {arXiv:1708.06667 [hep-lat]} \BibitemShut
  {NoStop}%
%%CITATION = ARXIV:1708.06667;%%
\bibitem [{\citenamefont {Woss}\ \emph {et~al.}(2018)\citenamefont {Woss},
  \citenamefont {Thomas}, \citenamefont {Dudek}, \citenamefont {Edwards},\ and\
  \citenamefont {Wilson}}]{Woss:2018irj}%
  \BibitemOpen
  \bibfield  {author} {\bibinfo {author} {\bibfnamefont {A.}~\bibnamefont
  {Woss}}, \bibinfo {author} {\bibfnamefont {C.~E.}\ \bibnamefont {Thomas}},
  \bibinfo {author} {\bibfnamefont {J.~J.}\ \bibnamefont {Dudek}}, \bibinfo
  {author} {\bibfnamefont {R.~G.}\ \bibnamefont {Edwards}}, \ and\ \bibinfo
  {author} {\bibfnamefont {D.~J.}\ \bibnamefont {Wilson}},\ }\href {\doibase
  10.1007/JHEP07(2018)043} {\bibfield  {journal} {\bibinfo  {journal} {JHEP}\
  }\textbf {\bibinfo {volume} {07}},\ \bibinfo {pages} {043} (\bibinfo {year}
  {2018})},\ \Eprint {http://arxiv.org/abs/1802.05580} {arXiv:1802.05580
  [hep-lat]} \BibitemShut {NoStop}%
%%CITATION = ARXIV:1802.05580;%%
\bibitem [{\citenamefont {Woss}\ \emph {et~al.}(2019)\citenamefont {Woss},
  \citenamefont {Thomas}, \citenamefont {Dudek}, \citenamefont {Edwards},\ and\
  \citenamefont {Wilson}}]{Woss:2019hse}%
  \BibitemOpen
  \bibfield  {author} {\bibinfo {author} {\bibfnamefont {A.~J.}\ \bibnamefont
  {Woss}}, \bibinfo {author} {\bibfnamefont {C.~E.}\ \bibnamefont {Thomas}},
  \bibinfo {author} {\bibfnamefont {J.~J.}\ \bibnamefont {Dudek}}, \bibinfo
  {author} {\bibfnamefont {R.~G.}\ \bibnamefont {Edwards}}, \ and\ \bibinfo
  {author} {\bibfnamefont {D.~J.}\ \bibnamefont {Wilson}},\ }\href {\doibase
  10.1103/PhysRevD.100.054506} {\bibfield  {journal} {\bibinfo  {journal}
  {Phys. Rev.}\ }\textbf {\bibinfo {volume} {D100}},\ \bibinfo {pages} {054506}
  (\bibinfo {year} {2019})},\ \Eprint {http://arxiv.org/abs/1904.04136}
  {arXiv:1904.04136 [hep-lat]} \BibitemShut {NoStop}%
%%CITATION = ARXIV:1904.04136;%%
\bibitem [{\citenamefont {Briceno}\ \emph
  {et~al.}(2018{\natexlab{b}})\citenamefont {Briceno}, \citenamefont {Dudek},\
  and\ \citenamefont {Young}}]{Briceno:2017max}%
  \BibitemOpen
  \bibfield  {author} {\bibinfo {author} {\bibfnamefont {R.~A.}\ \bibnamefont
  {Briceno}}, \bibinfo {author} {\bibfnamefont {J.~J.}\ \bibnamefont {Dudek}},
  \ and\ \bibinfo {author} {\bibfnamefont {R.~D.}\ \bibnamefont {Young}},\
  }\href {\doibase 10.1103/RevModPhys.90.025001} {\bibfield  {journal}
  {\bibinfo  {journal} {Rev. Mod. Phys.}\ }\textbf {\bibinfo {volume} {90}},\
  \bibinfo {pages} {025001} (\bibinfo {year} {2018}{\natexlab{b}})},\ \Eprint
  {http://arxiv.org/abs/1706.06223} {arXiv:1706.06223 [hep-lat]} \BibitemShut
  {NoStop}%
%%CITATION = ARXIV:1706.06223;%%
\bibitem [{\citenamefont {Michael}(1985)}]{Michael:1985ne}%
  \BibitemOpen
  \bibfield  {author} {\bibinfo {author} {\bibfnamefont {C.}~\bibnamefont
  {Michael}},\ }\href {\doibase 10.1016/0550-3213(85)90297-4} {\bibfield
  {journal} {\bibinfo  {journal} {Nucl. Phys.}\ }\textbf {\bibinfo {volume}
  {B259}},\ \bibinfo {pages} {58} (\bibinfo {year} {1985})}\BibitemShut
  {NoStop}%
%%CITATION = NUPHA,B259,58;%%
\bibitem [{\citenamefont {Luscher}\ and\ \citenamefont
  {Wolff}(1990)}]{Luscher:1990ck}%
  \BibitemOpen
  \bibfield  {author} {\bibinfo {author} {\bibfnamefont {M.}~\bibnamefont
  {Luscher}}\ and\ \bibinfo {author} {\bibfnamefont {U.}~\bibnamefont
  {Wolff}},\ }\href {\doibase 10.1016/0550-3213(90)90540-T} {\bibfield
  {journal} {\bibinfo  {journal} {Nucl. Phys.}\ }\textbf {\bibinfo {volume}
  {B339}},\ \bibinfo {pages} {222} (\bibinfo {year} {1990})}\BibitemShut
  {NoStop}%
%%CITATION = NUPHA,B339,222;%%
\bibitem [{\citenamefont {Blossier}\ \emph {et~al.}(2009)\citenamefont
  {Blossier}, \citenamefont {Della~Morte}, \citenamefont {von Hippel},
  \citenamefont {Mendes},\ and\ \citenamefont {Sommer}}]{Blossier:2009kd}%
  \BibitemOpen
  \bibfield  {author} {\bibinfo {author} {\bibfnamefont {B.}~\bibnamefont
  {Blossier}}, \bibinfo {author} {\bibfnamefont {M.}~\bibnamefont
  {Della~Morte}}, \bibinfo {author} {\bibfnamefont {G.}~\bibnamefont {von
  Hippel}}, \bibinfo {author} {\bibfnamefont {T.}~\bibnamefont {Mendes}}, \
  and\ \bibinfo {author} {\bibfnamefont {R.}~\bibnamefont {Sommer}},\ }\href
  {\doibase 10.1088/1126-6708/2009/04/094} {\bibfield  {journal} {\bibinfo
  {journal} {JHEP}\ }\textbf {\bibinfo {volume} {04}},\ \bibinfo {pages} {094}
  (\bibinfo {year} {2009})},\ \Eprint {http://arxiv.org/abs/0902.1265}
  {arXiv:0902.1265 [hep-lat]} \BibitemShut {NoStop}%
%%CITATION = ARXIV:0902.1265;%%
\bibitem [{\citenamefont {Dudek}\ \emph {et~al.}(2008)\citenamefont {Dudek},
  \citenamefont {Edwards}, \citenamefont {Mathur},\ and\ \citenamefont
  {Richards}}]{Dudek:2007wv}%
  \BibitemOpen
  \bibfield  {author} {\bibinfo {author} {\bibfnamefont {J.~J.}\ \bibnamefont
  {Dudek}}, \bibinfo {author} {\bibfnamefont {R.~G.}\ \bibnamefont {Edwards}},
  \bibinfo {author} {\bibfnamefont {N.}~\bibnamefont {Mathur}}, \ and\ \bibinfo
  {author} {\bibfnamefont {D.~G.}\ \bibnamefont {Richards}},\ }\href {\doibase
  10.1103/PhysRevD.77.034501} {\bibfield  {journal} {\bibinfo  {journal} {Phys.
  Rev.}\ }\textbf {\bibinfo {volume} {D77}},\ \bibinfo {pages} {034501}
  (\bibinfo {year} {2008})},\ \Eprint {http://arxiv.org/abs/0707.4162}
  {arXiv:0707.4162 [hep-lat]} \BibitemShut {NoStop}%
%%CITATION = ARXIV:0707.4162;%%
\bibitem [{\citenamefont {Dudek}\ \emph {et~al.}(2012)\citenamefont {Dudek},
  \citenamefont {Edwards},\ and\ \citenamefont {Thomas}}]{Dudek:2012gj}%
  \BibitemOpen
  \bibfield  {author} {\bibinfo {author} {\bibfnamefont {J.~J.}\ \bibnamefont
  {Dudek}}, \bibinfo {author} {\bibfnamefont {R.~G.}\ \bibnamefont {Edwards}},
  \ and\ \bibinfo {author} {\bibfnamefont {C.~E.}\ \bibnamefont {Thomas}},\
  }\href {\doibase 10.1103/PhysRevD.86.034031} {\bibfield  {journal} {\bibinfo
  {journal} {Phys. Rev.}\ }\textbf {\bibinfo {volume} {D86}},\ \bibinfo {pages}
  {034031} (\bibinfo {year} {2012})},\ \Eprint {http://arxiv.org/abs/1203.6041}
  {arXiv:1203.6041 [hep-ph]} \BibitemShut {NoStop}%
%%CITATION = ARXIV:1203.6041;%%
\bibitem [{\citenamefont {Shultz}\ \emph {et~al.}(2015)\citenamefont {Shultz},
  \citenamefont {Dudek},\ and\ \citenamefont {Edwards}}]{Shultz:2015pfa}%
  \BibitemOpen
  \bibfield  {author} {\bibinfo {author} {\bibfnamefont {C.~J.}\ \bibnamefont
  {Shultz}}, \bibinfo {author} {\bibfnamefont {J.~J.}\ \bibnamefont {Dudek}}, \
  and\ \bibinfo {author} {\bibfnamefont {R.~G.}\ \bibnamefont {Edwards}},\
  }\href {\doibase 10.1103/PhysRevD.91.114501} {\bibfield  {journal} {\bibinfo
  {journal} {Phys. Rev.}\ }\textbf {\bibinfo {volume} {D91}},\ \bibinfo {pages}
  {114501} (\bibinfo {year} {2015})},\ \Eprint
  {http://arxiv.org/abs/1501.07457} {arXiv:1501.07457 [hep-lat]} \BibitemShut
  {NoStop}%
%%CITATION = ARXIV:1501.07457;%%
\bibitem [{\citenamefont {Johnson}(1982)}]{Johnson:1982yq}%
  \BibitemOpen
  \bibfield  {author} {\bibinfo {author} {\bibfnamefont {R.~C.}\ \bibnamefont
  {Johnson}},\ }\href {\doibase 10.1016/0370-2693(82)90134-4} {\bibfield
  {journal} {\bibinfo  {journal} {Phys. Lett.}\ }\textbf {\bibinfo {volume}
  {B114}},\ \bibinfo {pages} {147} (\bibinfo {year} {1982})}\BibitemShut
  {NoStop}%
%%CITATION = PHLTA,B114,147;%%
\bibitem [{\citenamefont {Moore}\ and\ \citenamefont
  {Fleming}(2006)}]{Moore:2005dw}%
  \BibitemOpen
  \bibfield  {author} {\bibinfo {author} {\bibfnamefont {D.~C.}\ \bibnamefont
  {Moore}}\ and\ \bibinfo {author} {\bibfnamefont {G.~T.}\ \bibnamefont
  {Fleming}},\ }\href {\doibase 10.1103/PhysRevD.73.014504,
  10.1103/PhysRevD.74.079905} {\bibfield  {journal} {\bibinfo  {journal} {Phys.
  Rev.}\ }\textbf {\bibinfo {volume} {D73}},\ \bibinfo {pages} {014504}
  (\bibinfo {year} {2006})},\ \bibinfo {note} {[Erratum: Phys.
  Rev.D74,079905(2006)]},\ \Eprint {http://arxiv.org/abs/hep-lat/0507018}
  {arXiv:hep-lat/0507018 [hep-lat]} \BibitemShut {NoStop}%
%%CITATION = HEP-LAT/0507018;%%
\bibitem [{\citenamefont {Peardon}\ \emph {et~al.}(2009)\citenamefont
  {Peardon}, \citenamefont {Bulava}, \citenamefont {Foley}, \citenamefont
  {Morningstar}, \citenamefont {Dudek}, \citenamefont {Edwards}, \citenamefont
  {Joo}, \citenamefont {Lin}, \citenamefont {Richards},\ and\ \citenamefont
  {Juge}}]{Peardon:2009gh}%
  \BibitemOpen
  \bibfield  {author} {\bibinfo {author} {\bibfnamefont {M.}~\bibnamefont
  {Peardon}}, \bibinfo {author} {\bibfnamefont {J.}~\bibnamefont {Bulava}},
  \bibinfo {author} {\bibfnamefont {J.}~\bibnamefont {Foley}}, \bibinfo
  {author} {\bibfnamefont {C.}~\bibnamefont {Morningstar}}, \bibinfo {author}
  {\bibfnamefont {J.}~\bibnamefont {Dudek}}, \bibinfo {author} {\bibfnamefont
  {R.~G.}\ \bibnamefont {Edwards}}, \bibinfo {author} {\bibfnamefont
  {B.}~\bibnamefont {Joo}}, \bibinfo {author} {\bibfnamefont {H.-W.}\
  \bibnamefont {Lin}}, \bibinfo {author} {\bibfnamefont {D.~G.}\ \bibnamefont
  {Richards}}, \ and\ \bibinfo {author} {\bibfnamefont {K.~J.}\ \bibnamefont
  {Juge}} (\bibinfo {collaboration} {Hadron Spectrum}),\ }\href {\doibase
  10.1103/PhysRevD.80.054506} {\bibfield  {journal} {\bibinfo  {journal} {Phys.
  Rev.}\ }\textbf {\bibinfo {volume} {D80}},\ \bibinfo {pages} {054506}
  (\bibinfo {year} {2009})},\ \Eprint {http://arxiv.org/abs/0905.2160}
  {arXiv:0905.2160 [hep-lat]} \BibitemShut {NoStop}%
%%CITATION = ARXIV:0905.2160;%%
\bibitem [{\citenamefont {Woss}\ \emph {et~al.}(2020)\citenamefont {Woss},
  \citenamefont {Wilson},\ and\ \citenamefont {Dudek}}]{Woss:2020cmp}%
  \BibitemOpen
  \bibfield  {author} {\bibinfo {author} {\bibfnamefont {A.~J.}\ \bibnamefont
  {Woss}}, \bibinfo {author} {\bibfnamefont {D.~J.}\ \bibnamefont {Wilson}}, \
  and\ \bibinfo {author} {\bibfnamefont {J.~J.}\ \bibnamefont {Dudek}}
  (\bibinfo {collaboration} {Hadron Spectrum}),\ }\href {\doibase
  10.1103/PhysRevD.101.114505} {\bibfield  {journal} {\bibinfo  {journal}
  {Phys. Rev. D}\ }\textbf {\bibinfo {volume} {101}},\ \bibinfo {pages}
  {114505} (\bibinfo {year} {2020})},\ \Eprint
  {http://arxiv.org/abs/2001.08474} {arXiv:2001.08474 [hep-lat]} \BibitemShut
  {NoStop}%
\bibitem [{\citenamefont {Hansen}\ and\ \citenamefont
  {Sharpe}(2019)}]{Hansen:2019nir}%
  \BibitemOpen
  \bibfield  {author} {\bibinfo {author} {\bibfnamefont {M.~T.}\ \bibnamefont
  {Hansen}}\ and\ \bibinfo {author} {\bibfnamefont {S.~R.}\ \bibnamefont
  {Sharpe}},\ }\href {\doibase 10.1146/annurev-nucl-101918-023723} {\bibfield
  {journal} {\bibinfo  {journal} {Ann. Rev. Nucl. Part. Sci.}\ }\textbf
  {\bibinfo {volume} {69}},\ \bibinfo {pages} {65} (\bibinfo {year} {2019})},\
  \Eprint {http://arxiv.org/abs/1901.00483} {arXiv:1901.00483 [hep-lat]}
  \BibitemShut {NoStop}%
\bibitem [{\citenamefont {Rusetsky}(2019)}]{Rusetsky:2019gyk}%
  \BibitemOpen
  \bibfield  {author} {\bibinfo {author} {\bibfnamefont {A.}~\bibnamefont
  {Rusetsky}},\ }\href {\doibase 10.22323/1.363.0281} {\bibfield  {journal}
  {\bibinfo  {journal} {PoS}\ }\textbf {\bibinfo {volume} {LATTICE2019}},\
  \bibinfo {pages} {281} (\bibinfo {year} {2019})},\ \Eprint
  {http://arxiv.org/abs/1911.01253} {arXiv:1911.01253 [hep-lat]} \BibitemShut
  {NoStop}%
\bibitem [{\citenamefont {Dudek}\ \emph
  {et~al.}(2011{\natexlab{b}})\citenamefont {Dudek}, \citenamefont {Edwards},
  \citenamefont {Peardon}, \citenamefont {Richards},\ and\ \citenamefont
  {Thomas}}]{Dudek:2010ew}%
  \BibitemOpen
  \bibfield  {author} {\bibinfo {author} {\bibfnamefont {J.~J.}\ \bibnamefont
  {Dudek}}, \bibinfo {author} {\bibfnamefont {R.~G.}\ \bibnamefont {Edwards}},
  \bibinfo {author} {\bibfnamefont {M.~J.}\ \bibnamefont {Peardon}}, \bibinfo
  {author} {\bibfnamefont {D.~G.}\ \bibnamefont {Richards}}, \ and\ \bibinfo
  {author} {\bibfnamefont {C.~E.}\ \bibnamefont {Thomas}},\ }\href {\doibase
  10.1103/PhysRevD.83.071504} {\bibfield  {journal} {\bibinfo  {journal} {Phys.
  Rev.}\ }\textbf {\bibinfo {volume} {D83}},\ \bibinfo {pages} {071504}
  (\bibinfo {year} {2011}{\natexlab{b}})},\ \Eprint
  {http://arxiv.org/abs/1011.6352} {arXiv:1011.6352 [hep-ph]} \BibitemShut
  {NoStop}%
%%CITATION = ARXIV:1011.6352;%%
\bibitem [{\citenamefont {de~Swart}(1963)}]{deSwart:1963pdg}%
  \BibitemOpen
  \bibfield  {author} {\bibinfo {author} {\bibfnamefont {J.~J.}\ \bibnamefont
  {de~Swart}},\ }\href {\doibase 10.1103/RevModPhys.35.916} {\bibfield
  {journal} {\bibinfo  {journal} {Rev. Mod. Phys.}\ }\textbf {\bibinfo {volume}
  {35}},\ \bibinfo {pages} {916} (\bibinfo {year} {1963})},\ \bibinfo {note}
  {[Erratum: Rev. Mod. Phys.37,326(1965)]}\BibitemShut {NoStop}%
%%CITATION = RMPHA,35,916;%%
\bibitem [{\citenamefont {Edwards}\ \emph {et~al.}(2008)\citenamefont
  {Edwards}, \citenamefont {Joo},\ and\ \citenamefont {Lin}}]{Edwards:2008ja}%
  \BibitemOpen
  \bibfield  {author} {\bibinfo {author} {\bibfnamefont {R.~G.}\ \bibnamefont
  {Edwards}}, \bibinfo {author} {\bibfnamefont {B.}~\bibnamefont {Joo}}, \ and\
  \bibinfo {author} {\bibfnamefont {H.-W.}\ \bibnamefont {Lin}},\ }\href
  {\doibase 10.1103/PhysRevD.78.054501} {\bibfield  {journal} {\bibinfo
  {journal} {Phys. Rev.}\ }\textbf {\bibinfo {volume} {D78}},\ \bibinfo {pages}
  {054501} (\bibinfo {year} {2008})},\ \Eprint {http://arxiv.org/abs/0803.3960}
  {arXiv:0803.3960 [hep-lat]} \BibitemShut {NoStop}%
%%CITATION = ARXIV:0803.3960;%%
\bibitem [{\citenamefont {Lin}\ \emph {et~al.}(2009)\citenamefont {Lin} \emph
  {et~al.}}]{Lin:2008pr}%
  \BibitemOpen
  \bibfield  {author} {\bibinfo {author} {\bibfnamefont {H.-W.}\ \bibnamefont
  {Lin}} \emph {et~al.} (\bibinfo {collaboration} {Hadron Spectrum}),\ }\href
  {\doibase 10.1103/PhysRevD.79.034502} {\bibfield  {journal} {\bibinfo
  {journal} {Phys. Rev.}\ }\textbf {\bibinfo {volume} {D79}},\ \bibinfo {pages}
  {034502} (\bibinfo {year} {2009})},\ \Eprint {http://arxiv.org/abs/0810.3588}
  {arXiv:0810.3588 [hep-lat]} \BibitemShut {NoStop}%
%%CITATION = ARXIV:0810.3588;%%
\bibitem [{\citenamefont {Flatte}(1976{\natexlab{a}})}]{Flatte:1976xv}%
  \BibitemOpen
  \bibfield  {author} {\bibinfo {author} {\bibfnamefont {S.~M.}\ \bibnamefont
  {Flatte}},\ }\href {\doibase 10.1016/0370-2693(76)90655-9} {\bibfield
  {journal} {\bibinfo  {journal} {Phys. Lett. B}\ }\textbf {\bibinfo {volume}
  {63}},\ \bibinfo {pages} {228} (\bibinfo {year}
  {1976}{\natexlab{a}})}\BibitemShut {NoStop}%
\bibitem [{\citenamefont {Flatte}(1976{\natexlab{b}})}]{Flatte:1976xu}%
  \BibitemOpen
  \bibfield  {author} {\bibinfo {author} {\bibfnamefont {S.~M.}\ \bibnamefont
  {Flatte}},\ }\href {\doibase 10.1016/0370-2693(76)90654-7} {\bibfield
  {journal} {\bibinfo  {journal} {Phys. Lett. B}\ }\textbf {\bibinfo {volume}
  {63}},\ \bibinfo {pages} {224} (\bibinfo {year}
  {1976}{\natexlab{b}})}\BibitemShut {NoStop}%
\bibitem [{\citenamefont {Edwards}\ \emph {et~al.}(2013)\citenamefont
  {Edwards}, \citenamefont {Mathur}, \citenamefont {Richards},\ and\
  \citenamefont {Wallace}}]{Edwards:2012fx}%
  \BibitemOpen
  \bibfield  {author} {\bibinfo {author} {\bibfnamefont {R.~G.}\ \bibnamefont
  {Edwards}}, \bibinfo {author} {\bibfnamefont {N.}~\bibnamefont {Mathur}},
  \bibinfo {author} {\bibfnamefont {D.~G.}\ \bibnamefont {Richards}}, \ and\
  \bibinfo {author} {\bibfnamefont {S.~J.}\ \bibnamefont {Wallace}} (\bibinfo
  {collaboration} {Hadron Spectrum}),\ }\href {\doibase
  10.1103/PhysRevD.87.054506} {\bibfield  {journal} {\bibinfo  {journal} {Phys.
  Rev.}\ }\textbf {\bibinfo {volume} {D87}},\ \bibinfo {pages} {054506}
  (\bibinfo {year} {2013})},\ \Eprint {http://arxiv.org/abs/1212.5236}
  {arXiv:1212.5236 [hep-ph]} \BibitemShut {NoStop}%
%%CITATION = ARXIV:1212.5236;%%
\bibitem [{\citenamefont {Zyla}\ \emph {et~al.}(2020)\citenamefont {Zyla} \emph
  {et~al.}}]{Zyla:2020zbs}%
  \BibitemOpen
  \bibfield  {author} {\bibinfo {author} {\bibfnamefont {P.}~\bibnamefont
  {Zyla}} \emph {et~al.},\ }\href {\doibase 10.1093/ptep/ptaa104} {\bibfield
  {journal} {\bibinfo  {journal} {PTEP}\ }\textbf {\bibinfo {volume} {2020}},\
  \bibinfo {pages} {083C01} (\bibinfo {year} {2020})}\BibitemShut {NoStop}%
\bibitem [{\citenamefont {Lipkin}(1989)}]{Lipkin:1988um}%
  \BibitemOpen
  \bibfield  {author} {\bibinfo {author} {\bibfnamefont {H.~J.}\ \bibnamefont
  {Lipkin}},\ }\href {\doibase 10.1016/0370-2693(89)90846-0} {\bibfield
  {journal} {\bibinfo  {journal} {Phys. Lett. B}\ }\textbf {\bibinfo {volume}
  {219}},\ \bibinfo {pages} {99} (\bibinfo {year} {1989})}\BibitemShut
  {NoStop}%
\bibitem [{\citenamefont {Page}(2001)}]{Page:1999ak}%
  \BibitemOpen
  \bibfield  {author} {\bibinfo {author} {\bibfnamefont {P.~R.}\ \bibnamefont
  {Page}},\ }\href {\doibase 10.1103/PhysRevD.64.056009} {\bibfield  {journal}
  {\bibinfo  {journal} {Phys. Rev.}\ }\textbf {\bibinfo {volume} {D64}},\
  \bibinfo {pages} {056009} (\bibinfo {year} {2001})},\ \Eprint
  {http://arxiv.org/abs/hep-ph/9911301} {arXiv:hep-ph/9911301 [hep-ph]}
  \BibitemShut {NoStop}%
%%CITATION = HEP-PH/9911301;%%
\bibitem [{\citenamefont {Wilson}\ \emph {et~al.}(2019)\citenamefont {Wilson},
  \citenamefont {Briceno}, \citenamefont {Dudek}, \citenamefont {Edwards},\
  and\ \citenamefont {Thomas}}]{Wilson:2019wfr}%
  \BibitemOpen
  \bibfield  {author} {\bibinfo {author} {\bibfnamefont {D.~J.}\ \bibnamefont
  {Wilson}}, \bibinfo {author} {\bibfnamefont {R.~A.}\ \bibnamefont {Briceno}},
  \bibinfo {author} {\bibfnamefont {J.~J.}\ \bibnamefont {Dudek}}, \bibinfo
  {author} {\bibfnamefont {R.~G.}\ \bibnamefont {Edwards}}, \ and\ \bibinfo
  {author} {\bibfnamefont {C.~E.}\ \bibnamefont {Thomas}},\ }\href {\doibase
  10.1103/PhysRevLett.123.042002} {\bibfield  {journal} {\bibinfo  {journal}
  {Phys. Rev. Lett.}\ }\textbf {\bibinfo {volume} {123}},\ \bibinfo {pages}
  {042002} (\bibinfo {year} {2019})},\ \Eprint
  {http://arxiv.org/abs/1904.03188} {arXiv:1904.03188 [hep-lat]} \BibitemShut
  {NoStop}%
%%CITATION = ARXIV:1904.03188;%%
\bibitem [{\citenamefont {Burns}\ and\ \citenamefont
  {Close}(2006)}]{Burns:2006wz}%
  \BibitemOpen
  \bibfield  {author} {\bibinfo {author} {\bibfnamefont {T.}~\bibnamefont
  {Burns}}\ and\ \bibinfo {author} {\bibfnamefont {F.}~\bibnamefont {Close}},\
  }\href {\doibase 10.1103/PhysRevD.74.034003} {\bibfield  {journal} {\bibinfo
  {journal} {Phys. Rev. D}\ }\textbf {\bibinfo {volume} {74}},\ \bibinfo
  {pages} {034003} (\bibinfo {year} {2006})},\ \Eprint
  {http://arxiv.org/abs/hep-ph/0604161} {arXiv:hep-ph/0604161} \BibitemShut
  {NoStop}%
\bibitem [{\citenamefont {Feldmann}(2000)}]{Feldmann:1999uf}%
  \BibitemOpen
  \bibfield  {author} {\bibinfo {author} {\bibfnamefont {T.}~\bibnamefont
  {Feldmann}},\ }\href {\doibase 10.1142/S0217751X00000082} {\bibfield
  {journal} {\bibinfo  {journal} {Int. J. Mod. Phys. A}\ }\textbf {\bibinfo
  {volume} {15}},\ \bibinfo {pages} {159} (\bibinfo {year} {2000})},\ \Eprint
  {http://arxiv.org/abs/hep-ph/9907491} {arXiv:hep-ph/9907491} \BibitemShut
  {NoStop}%
\bibitem [{\citenamefont {Escribano}\ and\ \citenamefont
  {Nadal}(2007)}]{Escribano:2007cd}%
  \BibitemOpen
  \bibfield  {author} {\bibinfo {author} {\bibfnamefont {R.}~\bibnamefont
  {Escribano}}\ and\ \bibinfo {author} {\bibfnamefont {J.}~\bibnamefont
  {Nadal}},\ }\href {\doibase 10.1088/1126-6708/2007/05/006} {\bibfield
  {journal} {\bibinfo  {journal} {JHEP}\ }\textbf {\bibinfo {volume} {05}},\
  \bibinfo {pages} {006} (\bibinfo {year} {2007})},\ \Eprint
  {http://arxiv.org/abs/hep-ph/0703187} {arXiv:hep-ph/0703187} \BibitemShut
  {NoStop}%
\bibitem [{\citenamefont {Thomas}(2007)}]{Thomas:2007uy}%
  \BibitemOpen
  \bibfield  {author} {\bibinfo {author} {\bibfnamefont {C.}~\bibnamefont
  {Thomas}},\ }\href {\doibase 10.1088/1126-6708/2007/10/026} {\bibfield
  {journal} {\bibinfo  {journal} {JHEP}\ }\textbf {\bibinfo {volume} {10}},\
  \bibinfo {pages} {026} (\bibinfo {year} {2007})},\ \Eprint
  {http://arxiv.org/abs/0705.1500} {arXiv:0705.1500 [hep-ph]} \BibitemShut
  {NoStop}%
\bibitem [{\citenamefont {Close}\ and\ \citenamefont
  {Kirk}(1997)}]{Close:1997nm}%
  \BibitemOpen
  \bibfield  {author} {\bibinfo {author} {\bibfnamefont {F.~E.}\ \bibnamefont
  {Close}}\ and\ \bibinfo {author} {\bibfnamefont {A.}~\bibnamefont {Kirk}},\
  }\href {\doibase 10.1007/s002880050569} {\bibfield  {journal} {\bibinfo
  {journal} {Z. Phys. C}\ }\textbf {\bibinfo {volume} {76}},\ \bibinfo {pages}
  {469} (\bibinfo {year} {1997})},\ \Eprint
  {http://arxiv.org/abs/hep-ph/9706543} {arXiv:hep-ph/9706543} \BibitemShut
  {NoStop}%
\bibitem [{\citenamefont {Close}\ and\ \citenamefont
  {Kirk}(2015)}]{Close:2015rza}%
  \BibitemOpen
  \bibfield  {author} {\bibinfo {author} {\bibfnamefont {F.}~\bibnamefont
  {Close}}\ and\ \bibinfo {author} {\bibfnamefont {A.}~\bibnamefont {Kirk}},\
  }\href {\doibase 10.1103/PhysRevD.91.114015} {\bibfield  {journal} {\bibinfo
  {journal} {Phys. Rev. D}\ }\textbf {\bibinfo {volume} {91}},\ \bibinfo
  {pages} {114015} (\bibinfo {year} {2015})},\ \Eprint
  {http://arxiv.org/abs/1503.06942} {arXiv:1503.06942 [hep-ex]} \BibitemShut
  {NoStop}%
\bibitem [{\citenamefont {Barnes}\ \emph {et~al.}(2003)\citenamefont {Barnes},
  \citenamefont {Black},\ and\ \citenamefont {Page}}]{Barnes:2002mu}%
  \BibitemOpen
  \bibfield  {author} {\bibinfo {author} {\bibfnamefont {T.}~\bibnamefont
  {Barnes}}, \bibinfo {author} {\bibfnamefont {N.}~\bibnamefont {Black}}, \
  and\ \bibinfo {author} {\bibfnamefont {P.}~\bibnamefont {Page}},\ }\href
  {\doibase 10.1103/PhysRevD.68.054014} {\bibfield  {journal} {\bibinfo
  {journal} {Phys. Rev. D}\ }\textbf {\bibinfo {volume} {68}},\ \bibinfo
  {pages} {054014} (\bibinfo {year} {2003})},\ \Eprint
  {http://arxiv.org/abs/nucl-th/0208072} {arXiv:nucl-th/0208072} \BibitemShut
  {NoStop}%
\bibitem [{\citenamefont {McNeile}\ and\ \citenamefont
  {Michael}(2006)}]{McNeile:2006bz}%
  \BibitemOpen
  \bibfield  {author} {\bibinfo {author} {\bibfnamefont {C.}~\bibnamefont
  {McNeile}}\ and\ \bibinfo {author} {\bibfnamefont {C.}~\bibnamefont
  {Michael}} (\bibinfo {collaboration} {UKQCD}),\ }\href {\doibase
  10.1103/PhysRevD.73.074506} {\bibfield  {journal} {\bibinfo  {journal} {Phys.
  Rev.}\ }\textbf {\bibinfo {volume} {D73}},\ \bibinfo {pages} {074506}
  (\bibinfo {year} {2006})},\ \Eprint {http://arxiv.org/abs/hep-lat/0603007}
  {arXiv:hep-lat/0603007 [hep-lat]} \BibitemShut {NoStop}%
%%CITATION = HEP-LAT/0603007;%%
\bibitem [{\citenamefont {Edwards}\ and\ \citenamefont
  {Joo}(2005)}]{Edwards:2004sx}%
  \BibitemOpen
  \bibfield  {author} {\bibinfo {author} {\bibfnamefont {R.~G.}\ \bibnamefont
  {Edwards}}\ and\ \bibinfo {author} {\bibfnamefont {B.}~\bibnamefont {Joo}}
  (\bibinfo {collaboration} {SciDAC, LHPC, UKQCD}),\ }\bibfield  {booktitle}
  {\emph {\bibinfo {booktitle} {{Lattice field theory. Proceedings, 22nd
  International Symposium, Lattice 2004, Batavia, USA, June 21-26, 2004}}},\
  }\href {\doibase 10.1016/j.nuclphysbps.2004.11.254} {\bibfield  {journal}
  {\bibinfo  {journal} {Nucl. Phys. Proc. Suppl.}\ }\textbf {\bibinfo {volume}
  {140}},\ \bibinfo {pages} {832} (\bibinfo {year} {2005})},\ \Eprint
  {http://arxiv.org/abs/hep-lat/0409003} {arXiv:hep-lat/0409003 [hep-lat]}
  \BibitemShut {NoStop}%
%%CITATION = HEP-LAT/0409003;%%
\bibitem [{\citenamefont {Clark}\ \emph {et~al.}(2010)\citenamefont {Clark},
  \citenamefont {Babich}, \citenamefont {Barros}, \citenamefont {Brower},\ and\
  \citenamefont {Rebbi}}]{Clark:2009wm}%
  \BibitemOpen
  \bibfield  {author} {\bibinfo {author} {\bibfnamefont {M.~A.}\ \bibnamefont
  {Clark}}, \bibinfo {author} {\bibfnamefont {R.}~\bibnamefont {Babich}},
  \bibinfo {author} {\bibfnamefont {K.}~\bibnamefont {Barros}}, \bibinfo
  {author} {\bibfnamefont {R.~C.}\ \bibnamefont {Brower}}, \ and\ \bibinfo
  {author} {\bibfnamefont {C.}~\bibnamefont {Rebbi}},\ }\href {\doibase
  10.1016/j.cpc.2010.05.002} {\bibfield  {journal} {\bibinfo  {journal}
  {Comput. Phys. Commun.}\ }\textbf {\bibinfo {volume} {181}},\ \bibinfo
  {pages} {1517} (\bibinfo {year} {2010})},\ \Eprint
  {http://arxiv.org/abs/0911.3191} {arXiv:0911.3191 [hep-lat]} \BibitemShut
  {NoStop}%
%%CITATION = ARXIV:0911.3191;%%
\bibitem [{\citenamefont {Babich}\ \emph {et~al.}(2010)\citenamefont {Babich},
  \citenamefont {Clark},\ and\ \citenamefont {Joo}}]{Babich:2010mu}%
  \BibitemOpen
  \bibfield  {author} {\bibinfo {author} {\bibfnamefont {R.}~\bibnamefont
  {Babich}}, \bibinfo {author} {\bibfnamefont {M.~A.}\ \bibnamefont {Clark}}, \
  and\ \bibinfo {author} {\bibfnamefont {B.}~\bibnamefont {Joo}},\ }in\ \href
  {http://www1.jlab.org/Ul/publications/view_pub.cfm?pub_id=10186} {\emph
  {\bibinfo {booktitle} {{SC 10 (Supercomputing 2010) New Orleans, Louisiana,
  November 13-19, 2010}}}}\ (\bibinfo {year} {2010})\ \Eprint
  {http://arxiv.org/abs/1011.0024} {arXiv:1011.0024 [hep-lat]} \BibitemShut
  {NoStop}%
%%CITATION = ARXIV:1011.0024;%%
\bibitem [{\citenamefont {Clark}\ \emph {et~al.}(2016)\citenamefont {Clark},
  \citenamefont {Joó}, \citenamefont {Strelchenko}, \citenamefont {Cheng},
  \citenamefont {Gambhir},\ and\ \citenamefont {Brower}}]{Clark:2016rdz}%
  \BibitemOpen
  \bibfield  {author} {\bibinfo {author} {\bibfnamefont {M.}~\bibnamefont
  {Clark}}, \bibinfo {author} {\bibfnamefont {B.}~\bibnamefont {Joó}},
  \bibinfo {author} {\bibfnamefont {A.}~\bibnamefont {Strelchenko}}, \bibinfo
  {author} {\bibfnamefont {M.}~\bibnamefont {Cheng}}, \bibinfo {author}
  {\bibfnamefont {A.}~\bibnamefont {Gambhir}}, \ and\ \bibinfo {author}
  {\bibfnamefont {R.}~\bibnamefont {Brower}},\ }in\ \href@noop {} {\emph
  {\bibinfo {booktitle} {SC '16: Proceedings of the International Conference
  for High Performance Computing, Networking, Storage and Analysis}}}\
  (\bibinfo {year} {2016})\ pp.\ \bibinfo {pages} {{795--806}},\ \Eprint
  {http://arxiv.org/abs/1612.07873} {arXiv:1612.07873 [hep-lat]} \BibitemShut
  {NoStop}%
\bibitem [{\citenamefont {Chung}(1971)}]{Chung:1971ri}%
  \BibitemOpen
  \bibfield  {author} {\bibinfo {author} {\bibfnamefont {S.~U.}\ \bibnamefont
  {Chung}},\ }\href {\doibase 10.5170/CERN-1971-008} {\  (\bibinfo {year}
  {1971}),\ 10.5170/CERN-1971-008}\BibitemShut {NoStop}%
%%CITATION = CERN-71-08;%%
\end{thebibliography}%
